\documentclass[aps,pre,twocolumn,superscriptaddress,epsf]{revtex4-1}

\usepackage{latexsym}
\usepackage{amssymb,amsmath,amsbsy}
\usepackage{graphicx,color}
\usepackage{bm}
\usepackage{longtable}
\usepackage{epsf}
\usepackage[utf8]{inputenc}
\usepackage{enumerate}
\usepackage[english]{babel}               
\usepackage{natbib}

\def\eps{\varepsilon}

\def\S{\mathcal{S}}
\def\D{\mathcal{D}}

\def\eRM{{\mathrm e}}
\def\dRM{{\mathrm d}}

\def\mf{{\bm f}}

\def\mk{{\bm k}}

\def\mpp{{\bm p}}
\def\mq{{\bm q}}

\def\mv{{\bm v}}
\def\mV{{\bm V}}
\def\mx{{\bm x}}

\def\G{\Gamma}

\def\boldnabla{{\bm \nabla}}
\allowdisplaybreaks

\begin{document}


\title {Influence of turbulent mixing on critical behavior of directed percolation process : effect of
 compressibility}
\author{J. Honkonen} 
\affiliation{National Defence University, 00861 Helsinki, Finland}
\author{T.~Lu\v{c}ivjansk\'y}
\affiliation{Faculty of Sciences, P.J. \v{S}af\'arik University, 04154 Ko\v{s}ice, Slovakia}
\affiliation{Peoples’ Friendship University of Russia (RUDN University), 6 Miklukho-Maklaya St, Moscow, 117198, Russian Federation} 
\author{V.~\v{S}kult\'ety}
\affiliation{Department of Physics, Stockholm University, AlbaNova University Center, SE-106 91 Stockholm, Sweden}
\date{\today}
\begin{abstract}
Universal behavior is a typical emergent feature of critical systems. A paramount model of the
non-equilibrium critical behavior is the directed bond percolation process that exhibits an 
active-to-absorbing state phase transition in the vicinity of a percolation threshold. 
 Fluctuations of the ambient environment might affect or destroy
the universality properties completely. In this work we assume that the random environment
can be described by means of compressible velocity fluctuations. Using 
field-theoretic models and renormalization group methods
we investigate large-scale and long-time behavior. Altogether eleven universality classes
 are found, out of which four are stable in the infrared limit and
thus macroscopically accessible. In contrast to the model without velocity fluctuations
 a possible candidate for a realistic three-dimensional case, a regime with relevant
short-range noise, is identified.
 Depending on the dimensionality of space
and the structure of the turbulent flow we calculate critical exponents of the directed percolation
process. In the limit of the purely transversal
velocity field random force critical exponents comply with the incompressible results 
obtained by previous authors. We have found intriguing
non-universal behavior related to the mutual effect of compressibility and advection.
\end{abstract}

\maketitle

\section{Introduction \label{sec:intro}}
The non-equilibrium systems are a fascinating  branch of physics, which
 encompasses many natural phenomena.
 In last decades a lot of effort has been put into study of different aspects, but
  still a general theory is lacking \cite{Zia95,krapivsky,Tauber}.

 A paradigmatic example is the directed bond percolation (DP) process  also known as Gribov process, in
 which 
  an absorbing phase transition between an active (fluctuating) and 
 an inactive (absorbing) state occurs. At this transition
 vigorous spatio-temporal fluctuations of an order parameter dominate and the resulting collective
behavior is  analogous to equilibrium phase transitions \cite{Amit,Zinn,ZinnRG,Vasiliev}. The main difference
is in the scaling of the time variable different from that of the spatial variables 
 \cite{HHL08,Tauber}. The DP process describes creation of fractal percolation structures
 \cite{Cardy80,obukhov80}.
In high energy physics the DP process was developed in a different context of Reggeon field theory 
with aim to describe behavior of hadrons. Later it became clear that DP and Reggeon field theory are just
 different versions of the same critical theory.
 In complex non-equilibrium models non-linearities
  pose a crucial challenge for a theoretical description. In order
  to make a model mathematically amenable one can either exploit a certain
 special feature or implement a sophisticated numerical scheme. The former approach is realized in the
 critical domain of DP, where correlated regions of microscopic degrees of freedom
 can be conveniently described by means of continuous fields.
 
 As was conjectured by Janssen and Grassberger
 \cite{Janssen81,Grassberger82},  necessary
conditions for a system to be in DP universality class are: i) a single absorbing state,
  ii) short-ranged interactions,
 iii) a positive order parameter and iv) no additional symmetry or coupling with other
 slow variables.
 Several models have been identified and their adherence to the DP class has been shown, e.g. 
  reaction-diffusion problems \cite{Odor04}, percolation processes \cite{JT04}, hadron
 interactions \cite{Cardy80}.
 
Of the conditions of the DP universality class the item (iv) is very subtle from the experimental point of view.
 In realistic situations impurities and defects,  which are not taken into account
 in the original DP formulation, are expected to induce violations of 
 universal properties of the model. 
 This is believed to be one of the reasons  why there are
not so many direct experimental realizations \cite{RRR03,TKCS07,Sano2016,LSAJAH16} of 
the percolation process itself.
 A study of deviations from the ideal situation could proceed in different routes and this 
 still constitutes a topic of an ongoing debate \cite{HHL08}.
A substantial effort has been made  in studying  a long-range interaction using
L{\'e}vy flights \cite{Jan99,Hin06,Hin07}, effects
 of immunization \cite{Hin00,JT04}, or in the presence of spatially quenched
 disorder \cite{Janssen97}.
 In general, a novel behavior is observed  with a  possibility that the critical
 behavior is lost.
 For instance, the presence of quenched disorder in the latter case causes a shift
 of the critical fixed point to the unphysical region.
 This leads to such interesting phenomena as activated
 dynamical scaling or Griffiths singularities \cite{MorDic96,CGM98,Vojta05,Vojta06}.

\begin{figure}[h!]
	\centering
	\includegraphics[width=8.6cm]{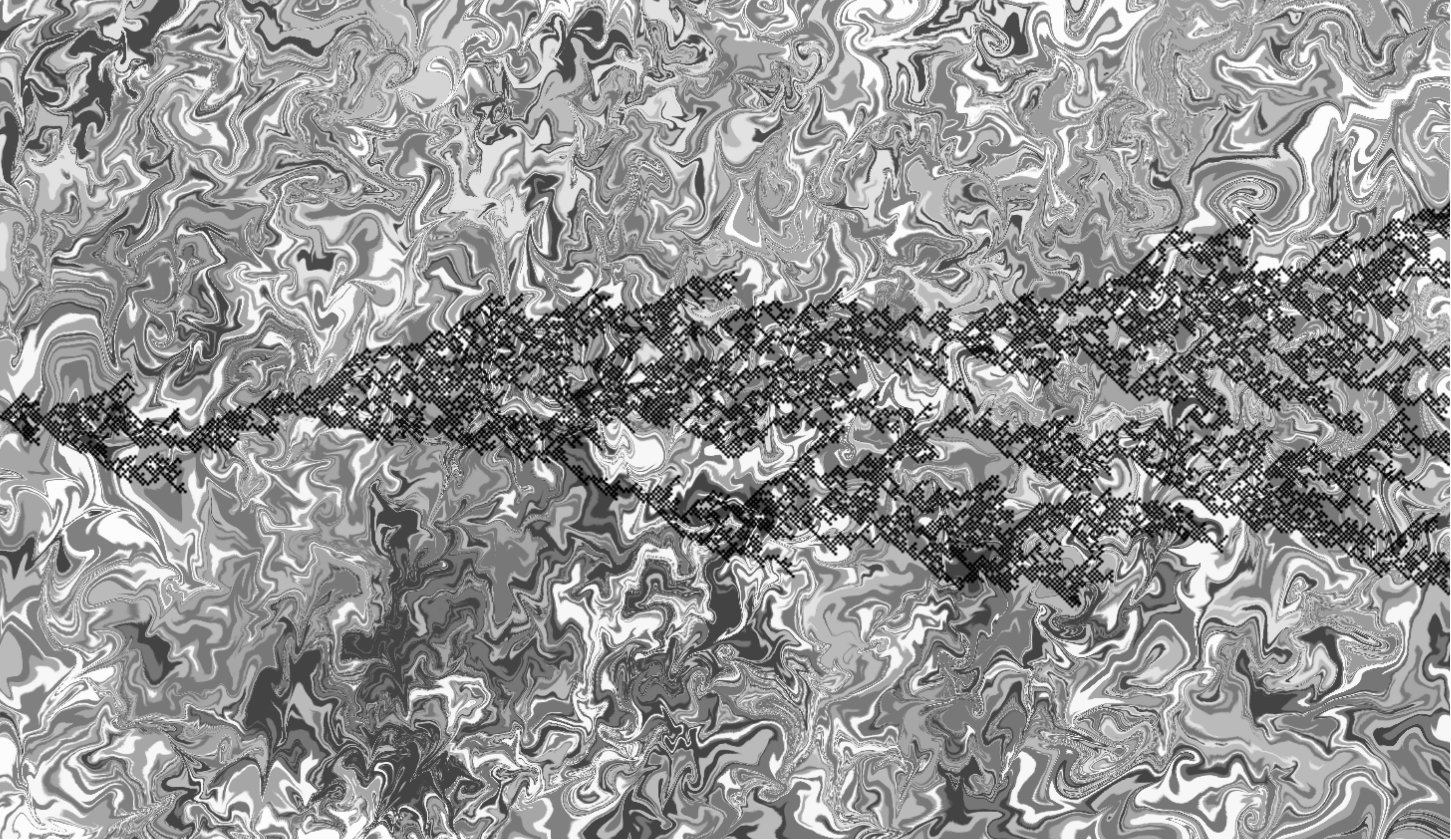}
	\caption{Schematic visualization of the DP process in the presence of the fully
	developed turbulent flow. The black path represents the DP process (starting from the
	left and going rightwards) and the fully developed turbulence is displayed in the
	background.}
	\label{fig:artist_viktor}
\end{figure}

In this paper, we address the question how DP is affected by velocity fluctuations
 of an ambient environment in which DP takes place (qualitatively displayed
 in Fig.~\ref{fig:artist_viktor}). 
 Velocity fluctuations are hardly avoidable
in any of laboratory experiments. For instance, a vast majority of chemical reactions occurs
at finite temperature, which is inevitably accompanied with the presence of a thermal noise. 
Furthermore, disease spreading and chemical reactions may be affected heavily by turbulent
advection \cite{AdzAnt98,Ant00}.
In general, turbulence is a rule rather than an exception 
and many physical phenomena cannot be properly explained without turbulence
\cite{Frisch,davidson,chapters,voyage}.
Here, our aim is
to estimate how strong compressibility of the ambient fluid can affect the DP process and what are the 
main differences
from the incompressible case 
 \cite{AIK10,AIM10,AKM11,DHLM13}.

For analytic description of steady turbulent flow it is customary to use randomly forced (stochastic)
Navier-Stokes equation \cite{Kraichnan68,FGV01,turbo}.
In this framework an important question is how the properties of the random force affect
the turbulence. In case of incompressible turbulence, the random force is chosen to
contain only transversal modes. Longitudinal modes generated by the nonlinearity of the Navier-Stokes equation are 
absorbed in fluctuations of pressure. In turbulence of compressible fluid longitudinal modes of
the velocity field are always present and
have to be incorporated in a proper fashion.
 In our model there is no influence of the percolating
 field on the velocity fluctuations. In other words, our model corresponds to
 the passive advection of the  reacting scalar field.
 
Recently, there has been an increased interest in different advection problems in 
turbulent flows \cite{Benzi09,Pig12,Volk14,depietro15}. These studies have shown that
compressibility plays a decisive r\^{o}le for population dynamics or chaotic mixing of colloids.

Our main aim in this paper is to elucidate to what extent strong compressible modes change
the critical behavior of the  DP universality class.
To this end we use functional presentation of the stochastic problem with the subsequent
application of the field-theoretic renormalization group (RG) \cite{Amit,Zinn,Vasiliev}. 
This theoretical framework allows
to examine asymptotic scaling behavior 
and infer quantitative
predictions about universal
quantities in a controllable fashion.

This paper is organized as follows. In 
 Sec.~\ref{subsec:DP}, we introduce a
coarse-grained formulation of the DP problem, which we reformulate
 into the field-theoretic model. Also we introduce relevant quantities that we want
 to analyze. Next in Sec.~\ref{subsec:turbul} we give a brief overview of the
  compressible Navier-Stokes equation and main differences to the incompressible case.
 Sec.~\ref{sec:RG} is reserved for the  main steps of the
  renormalization group procedure. In Sec.~\ref{sec:results},  we present 
 an analysis of possible regimes involved in the model. We analyze numerically
 and to some extent analytically  fixed points' structure.
 In Sec.~\ref{sec:concl} we give  a concluding summary. 
 Technical details concerning a calculation of the RG constants
 and functions are presented in Appendix \ref{app:const} and Appendix \ref{app:RG}. 
 Certain explicit expressions are summarized in Appendix \ref{app:explicit}.

{\section{The model}  \label{sec:model} }
{\subsection{Directed percolation} \label{subsec:DP}}
The field-theoretic formulation of the DP  
process may be obtained with the use of the Langevin equation \cite{JT04,Tauber}
\begin{align}   
  \partial_{t} \psi = D_{0}(\partial^{2} - R[\psi]) \psi +\sqrt{\psi N[\psi]}f,
  \label{eq:DP_pheLE}
\end{align}
where $ \psi = \psi(x) $ is the order parameter field (e.g. the density of a species), 
$ x = \{ \mx,t \}, \ D_{0} $ is the microscopic (bare in the parlance of RG) diffusion 
constant, $ \partial^{2}=(\partial/\partial x_i) (\partial/\partial x_i)\equiv\partial_{i}\partial_{i} $ is
the Laplace operator, $ R[\psi] $
is a given time-local reaction functional which will be specified later, $ N[\psi]$ is a local 
noise functional and
$ f= f(x) $ is the random force with the following properties
\begin{align} 
  \langle f(x) \rangle & = 0, 
  \label{eq:DP_pheNoise1}
  \\ 
  \langle f(x) f(x') \rangle & = \delta(x-x'),
  \label{eq:DP_pheNoise2}
\end{align}
Note that Eq.~(\ref{eq:DP_pheLE}) is a coarse grained model that captures the essential (universal) 
properties of DP only in the critical domain. 

Further, it is important that the field $ \psi(x) $ has
been taken out from the functional $ N[\psi] $  in Eq.~(\ref{eq:DP_pheLE})
in order to obtain a multiplicative noise. This 
type of noise ensures that fluctuations vanish in the absorbing state $ \psi = 0 $, which is a 
fundamental property of models undergoing an active-to-absorbing phase transition \cite{JT04,HHL08}. 
In the critical domain, the density $ \psi $ is a slow variable and
the reaction and the noise 
functional may be expanded as follows 
\begin{align}
   R[\psi] & = \tau_{0} +  \lambda_{0} \psi/2 + \cdots,\\
   N[\psi] & = D_{0} \lambda_{0} + \cdots, 
\end{align}
where in the last expression we have extracted the diffusion constant $D_0$ due to dimensional reasons.
The Langevin equation~(\ref{eq:DP_pheLE})
then assumes the form 
\begin{align}
  \partial_{t} \psi &= D_{0}(\partial^{2} - \tau_{0}) \psi + D_{0}\lambda_{0} \psi^{2}/2 + \sqrt{ D_{0} \lambda_{0}\psi(x)}f\,.
  \label{eq:DPlangevin}
\end{align} 
In Eq.~(\ref{eq:DPlangevin}) the parameter $ \tau_{0} $ can be interpreted as the deviation from the percolation
threshold
($\tau_0$ is thus analogous to $\tau\propto (T-T_c)$, the deviation from the critical temperature in
the $\varphi^4-$theory of static critical 
phenomena \cite{Amit,Zinn}) and $ \lambda_{0} $ plays the r\^ole of
a coupling constant. The subscript $ ``0`` $ denotes
bare quantities for the future use of the RG method. 

The field-theoretic formulation of the DP is given by the De Dominicis-Janssen 
action functional \cite{DeDM76,Janssen76}
\begin{align} 
  \S^{\psi}[\Phi_\text{DP}] =&\ \psi'\{ \partial_{t} + D_{0}(\tau_{0} - \partial^{2}) \} \psi +
  \frac{\lambda_{0} D_{0}}{2} \{ \psi' - \psi \} \psi' \psi,
  \label{eq:DP_action}
\end{align}
where $ \psi'=\psi'(x) $ is Martin-Siggia-Rose response field \cite{MSR}, 
 $ \Phi_\text{DP} \equiv \{\psi',\psi\} $
is the set of all
 DP related fields in the action functional and the integration over all temporal and spatial variables 
 is assumed, e.g.
 the first term on the right-hand side of Eq.~(\ref{eq:DP_action}) stands for
\begin{align}
   \psi'\partial_{t}\psi = \int \dRM t \int \dRM^{d}x \ \psi(\mx,t) \partial_{t} \psi(\mx,t).
\end{align}
The dynamic action (\ref{eq:DP_action}) corresponds to the It\^o interpretation of the stochastic
differential equation (\ref{eq:DPlangevin}), (\ref{eq:DP_pheNoise2}).
The field-theoretic formulation means that all correlations and response functions are represented
as functional integrals with the functional measure $ \D\psi\exp(- \S^{\psi}[\Phi_\text{DP}]) $.
For instance, the pair connectedness function \cite{HHL08} is given by
\begin{align}
  \langle \psi'(x') \psi(x) \rangle = \int \D\psi'\D\psi \ \psi'(x') \psi(x)
  \exp ( -  \S^{\psi}[\Phi_\text{DP}] ). 
  \label{eq:DP_pathInt}
\end{align}
Quantities of primary importance are the number of particles $ N(t) $, radius of gyration $ R^{2}(t) $,
density of species $ \rho(t) $ and the survival probability $ P(t) $ defined
as follows \cite{JT04,HHL08}
\begin{align}
  \mathcal{N}(t) &= \int \dRM^{d} x \  \langle \psi'({\bm 0},0)\psi(\mx,t) \rangle, 
  \label{eq:CExtp1} \\
  \mathcal{R}^{2}(t) &= N^{-1} \int \dRM^{d}x\, \mx^{2} \langle \psi'({\bm 0},0)\psi(\mx,t) \rangle, 
  \label{eq:CExtpN} \\
  \rho(t) &= \langle \psi(\mx,t) \rangle, \\
  \mathcal{P}(t) &= -\lim\limits_{k\rightarrow\infty} \langle \psi'(\mx,-t) 
  \eRM^{-k\int \dRM^{d} x \, \psi(\mx,0)} \rangle. 
  \label{eq:CExtp2}
\end{align}
The asymptotic long-time behavior is governed by the following universal
power laws
\begin{align}
  N(t) &\sim t^{\Theta}, \quad \rho(t) \sim t^{-\delta}, \\
  R^{2}(t) &\sim t^{\tilde{z}}, \quad P(t) \sim t^{-\delta'}.
  \label{eq:DP_exponents}
\end{align}
Numerical values of the corresponding critical exponents in the mean-field approximation are
\begin{equation}
  \Theta = 0, \quad \tilde{z} = 1,\quad \delta = \delta' = 1,
\end{equation}
where the last equality follows from the rapidity symmetry \cite{JT04}, i.e., from the fact that the 
action functional (\ref{eq:DP_action}) is invariant with respect to the transformation
\begin{equation} 
  \psi(\mx,t) \leftrightarrow -\psi'(\mx,-t).
  \label{eq:DP_RS}
\end{equation}
{\subsection{Turbulent advection}  \label{subsec:turbul}}
In order to study the advection of DP by the random velocity field let us recall
 that we have to replace the time derivative 
with the generalized covariant derivative \cite{Landau_fluid,Monin}
\begin{align}
  \partial_{t} \rightarrow \nabla_{t} + a_{0} 
  (\boldnabla \cdot \mv),
\end{align}
where 
$ \nabla_{t} \equiv \partial_{t} + (\mv\cdot\boldnabla) $
is the standard convective derivative and the 
parameter $ a_{0} $ has to be introduced only in the case of compressible velocity field \cite{AK10}.
The permissible physical (microscopic/bare) values of this parameter are $ a_{0} = 0 $ and
$ a_{0} = 1 $, where the corresponding Langevin equation describes either an advection of the tracer
field or an advection of the density field, respectively \cite{Landau_fluid}. 
Effectively this discussion leads to an additional term in the action functional~(\ref{eq:DP_action})
of the following form
\begin{equation}
  \S^\text{adv}[\psi',\psi,\mv] = \psi'(\mv\cdot\boldnabla)\psi + a_0 \psi' (\boldnabla\cdot\mv)\psi.
  \label{eq:adv_action}
\end{equation}
Introduction of the velocity
field in Eq.~(\ref{eq:DP_action}) may generally break down the rapidity symmetry (\ref{eq:DP_RS})
which increases the number of independent critical exponents to four. However, as was shown previously
in the case of compressible Kraichnan-velocity ensemble \cite{AK10}, this symmetry has to be modified
\begin{align}
  \psi(\mx,t) \leftrightarrow \psi'(-\mx,-t), \quad a_{0} \rightarrow (1-a_{0}), \quad \lambda_{0}
  \rightarrow -\lambda_{0},
\end{align}
in order to ensure the number of independent exponents to remain three. The
sign of the coupling constant $ \lambda $ also appears to be unimportant, since the parameter of the
perturbation expansion is rather $ \lambda^{2} $ than $ \lambda $.

In the present case, the velocity field is generated by the compressible NS (cNS) equation 
 \cite{SYKO90,ANU97,AGKL17}, written in the component form as
\begin{equation}
  \rho \nabla_{t} v_{i} = \mu_{0} (\delta_{ij}\partial^{2} - \partial_{i} \partial_{j}/3) v_{j} + 
  \zeta_{0} \partial_{i} \partial_{j} v_{j} - \partial_{i} p_{i} + \rho f_{i}^{\mv}, 
  \label{eq:cNS_cNS} 
\end{equation}
and
the continuity equation
\begin{equation}  
  \partial_{t} \rho = - \partial_{i}(\rho v_{i}), 
  \label{eq:cNS_contEq}
\end{equation}
where $ \rho = \rho({x}) $ is the density, $ v = v({x}) $ is velocity and $ p = p({x}) $ 
is the pressure of the fluid, $ \mu_{0} $ and $ \zeta_{0} $ are the dynamical and the bulk viscosity.
 Density of the random force per unit mass $ f_{i}^{\mv} = f_{i}^{\mv}(x) $  mimics
the energy input into the system \cite{ANU97,Vasiliev}, which is necessary to compensate
loss of energy due to viscous forces. In order to obtain a closed set of
equations, we further assume the isothermal condition to hold that relates the density and pressure of the fluid
 in the following way
\begin{equation}
  (p - \overline{p}) = c_{0}^{2} (\rho - \overline{\rho}). 
  \label{eq:cNS_AC}
\end{equation}
Here, $ \overline{p},\overline{\rho} $ are the mean pressure and the mean density and $ c_{0} $ is
the
speed of sound. Eqs.~(\ref{eq:cNS_cNS}) and (\ref{eq:cNS_contEq}) can be then 
cast into a more convenient form 
\begin{align}
  \nabla_{t} v_{i} &= \nu_{0} (\delta_{ij}\partial^{2} - \partial_{i} \partial_{j}) v_{j} 
  + \nu_{0} u_{0} \partial_{i} \partial_{j} v_{j} - \partial_{i}\phi + f_{i}^{\mv}, 
  \label{eq:cNS_cNS2} \\
  \nabla_{t} \phi &= -c_{0}^{2}( \partial_{i}v_{i}), 
  \label{eq:cNS_contEq2}
\end{align}
where $ \nu_{0} = \eta_{0}/\overline{\rho} $ is the kinematic viscosity (see \cite{ANU97} for
more details), $ u_{0} $ is a new parameter related to the bulk viscosity via relation 
$ \nu_{0} (u_{0}-1) = -\nu_{0}/3 + \zeta_{0}/\overline{\rho} $ and we have introduced a new density-related field $ \phi = c_{0}^{2} \ln \rho/\overline{\rho} $. In Eq.~(\ref{eq:cNS_cNS2})
the specific random force $ f_{i}^{\mv} $ obeys 
Gaussian statistics with zero mean and the two-point correlator \cite{AGKL17}
\begin{align}
  \langle f_{i}^{\mv}( x ) f_{j}^{\mv}( x' ) \rangle =&\ 
  \frac{\delta(t-t')}{(2\pi)^{d}} \int_{k> m} \dRM^{d} k \ D_{ij}^{\mv}(\mk) 
  \mathrm{e}^{i\mk(\mx-\mx')}, 
  \label{eq:cNS_randForce}
\end{align}
where $ m $ plays a r\^ole of  the infrared (IR) cut-off and the spectrum $ D_{ij}^{v}(\mk) $ is in
adopted in the form
\begin{align}
  D_{ij}^{\mv}(\mk) = g_{10} \nu_{0}^{3} k^{4-d-y}(P_{ij}(\mk) + \alpha Q_{ij}(\mk)) +
  g_{20} \nu_{0}^{3}. 
  \label{eq:vRFcorr}
\end{align}
Here, $ g_{10},g_{20} $ are coupling constants, $ P_{ij}(\mk) = \delta_{ij} - k_{i}k_{j}/k^{2} $ 
and $ Q_{ij}(\mk) = k_{i}k_{j}/k^{2} $ are transversal and longitudinal projection operators, 
$ d $ is the dimension of the space and $ y $ is an analytic regulator that serves as an
expansion parameter in the perturbative RG \cite{Vasiliev,HHL16}. The parameter $ y $ is analogous to the classical
$ \varepsilon = 4 - d $ in the theory of critical phenomena which is introduced in order to
regularize the ultraviolet (UV) divergences in the Feynman diagrams of the perturbative expansion.
This procedure is also referred to as an analytic regularization \cite{Vasiliev}. The most relevant 
value of $ y $ is $ 4 $, where the trace of the non-local part of Eq.~(\ref{eq:cNS_randForce}) becomes 
proportional to $ \delta(\mk) $ which mimics the energy input from 
 the largest spatial scales $\mk\rightarrow 0 $ \cite{turbo}.

The first term on the right hand side in Eq.~(\ref{eq:vRFcorr}) represents a classical way of introducing the 
random force in the perturbative RG theory of turbulence, whereas the second term has to be added in 
order to ensure the multiplicative renormalization of the model around $ d = 4 $ \cite{AGKL17}.
This point is discussed further in Sec.~\ref{sec:Renormalization}. The local part of the 
random-force correlation functions can be interpreted as a term responsible for thermal fluctuations (since 
in the real space it represents a delta correlated term that mimics the energy input from
all spatial scales including the smallest).

An important point to discuss here is the meaning of the parameter $ \alpha $, which is lacking
in previous literature \cite{ANU97,AK14,AGKL17}.
Let us consider the stochastic NS Eq.~(\ref{eq:cNS_cNS}) in a different form \cite{Landau_fluid}
\begin{align}
  \partial_{t}(\rho v_{i}) + \partial_{j}(\rho v_{i} v_{j}) = \partial_{j} (\sigma_{ij}' - 
  p \delta_{ij}) + \rho f_{i}^{\mv}, 
  \label{eq:cNS_cNS_compact}
\end{align}
where $ \sigma_{ij}' $ is the viscous stress tensor responsible for the energy dissipation 
\cite{Landau_fluid}, whose exact form is unimportant at this stage. By taking the divergence of
 Eq.~(\ref{eq:cNS_cNS_compact}) with the subsequent insertion into the time derivative of the continuity equation
 (\ref{eq:cNS_contEq}) (together with the adiabatic condition (\ref{eq:cNS_AC})) we arrive at
\begin{align} 
  \partial_{tt}^{2} \rho - c_{0}^{2} \partial^{2} \rho = \partial_{i} \partial_{j} (\rho v_{j} v_{i}
  - \sigma_{ij}') - f_{i}^{\mv} \partial_{i} \rho - \rho \partial_{i} f_{i}^{\mv}.
  \label{eq:LightEq}
\end{align}
This is nothing else (let alone the random force) than the Lighthill equation of aero-acoustics 
\cite{Monin}. 
 From Eq.~(\ref{eq:LightEq}) it is obvious that the longitudinal part of the random force is
responsible for generation of sound waves. However, in the case of purely solenoidal random
force $ ( \boldnabla\cdot {\mf}^{\mv} = 0) $ the sound waves may still be generated due to the
non-linearities on the right hand side o Eq.~(\ref{eq:LightEq}). This implies that the 
sole limit $ \alpha \rightarrow 0 $ 
in Eq.~(\ref{eq:cNS_randForce}) does not correspond to the incompressible limit. On 
the other hand, it has been shown within the one-loop approximation \cite{ANU97} that in the 
limit $ \alpha \rightarrow 0 $ the energy spectrum of the fully developed turbulence 
coincides with the Kolmogorov $ -5/3 $ law for the incompressible turbulence.

Using standard procedures \cite{DeDM76,Vasiliev,Tauber} we finally
obtain the De Dominicis-Janssen action functional
\begin{align}
  \S^{\mv}[ \Phi_\text{vel} ] =&\ - \frac{v_{i}' D_{ij}^{\mv} v_{j}'}{2} + 
  v_{i}' \big\{ \nabla_{t} v_{i}
  - \nu_{0} [\delta_{ij} \partial^{2} - \partial_{i} \partial_{k}] v_{k}  
  \nonumber \\
  &\ - u_{0} \nu_{0} \partial_{i} \partial_{j} v_{j} + \partial_{i} \phi \big\} 
  \nonumber \\
  &\ + \phi' \left\{ \nabla_{t} \phi - \tilde{v}_{0} \nu_{0} \partial^{2} \phi + c_{0}^2 
  \partial_{i} v_{i} \right\}, 
  \label{eq:cNS_action}
\end{align}
where $ \Phi_\text{vel} \equiv \{ \mv', \mv,\phi',\phi \}$ is a full set of velocity-related fields, 
$ D_{ij}^{\mv} $ is the velocity field 
 random force correlator 
 (\ref{eq:cNS_randForce}) and the term proportional to $ \tilde{v} $ has been added in order to
ensure the multiplicative renormalization \cite{ANU97}.

The model for compressible turbulence based on compressible NS equation was firstly proposed
in \cite{SYKO90}. However, as mentioned in \cite{ANS95,VN96} authors did not pay attention to the 
multiplicatively renormalization and the model they obtained was not multiplicatively 
renormalizable as well as the term proportional to $ \tilde{v} $ was missing.

In contrast to the synthetic model for velocity field and
its variations \cite{Ant99,Ant00,A06} the model (\ref{eq:cNS_action})
is Galilean invariant, which is specified in the next section. This 
leads to a restriction of possible terms that can be
generated during the
RG procedure and pose certain conditions on renormalization constants \cite{DHLM13,AHKLM16}.

{\section{Field-theoretic renormalization group} \label{sec:RG}}
{\subsection{Perturbation theory} \label{subsec:pertur}}
The entire model describing the advection of the DP process in presence of compressible fully 
developed turbulence is given by the sum of action functionals
 (\ref{eq:DP_action}), (\ref{eq:adv_action}) and (\ref{eq:cNS_action}), briefly written as
\begin{equation}   
   \S[\Phi] = \S^{\psi} + \S^{\mv} + \S^\text{adv},  
   \label{eq:total_action}
\end{equation}  
where $\Phi = \Phi_\text{DP} \cup \Phi_\text{vel}$ is the set of all fields.
  Main objects 
of a practical interest are connected correlation functions $ W_{\varphi\dots\varphi} $
with $ \varphi \equiv \varphi( x ) $ being any permissible field from the set $\Phi$. 
The generating functional $ \mathcal{W} $ for 
connected correlation functions is defined as \cite{Zinn,Amit,Vasiliev}
\begin{align}
  \mathcal{W}[A] = \ln \mathcal{Z}[A], \quad  W_{\varphi\dots\varphi} =
  \frac{\delta \mathcal{W}}{\delta \varphi\dots\varphi} \bigg|_{A = 0}\,,
\end{align}
where $A$ stands for the set of source fields corresponding to $\Phi$.

\begin{figure}[t]
	\includegraphics[width=8.5cm]{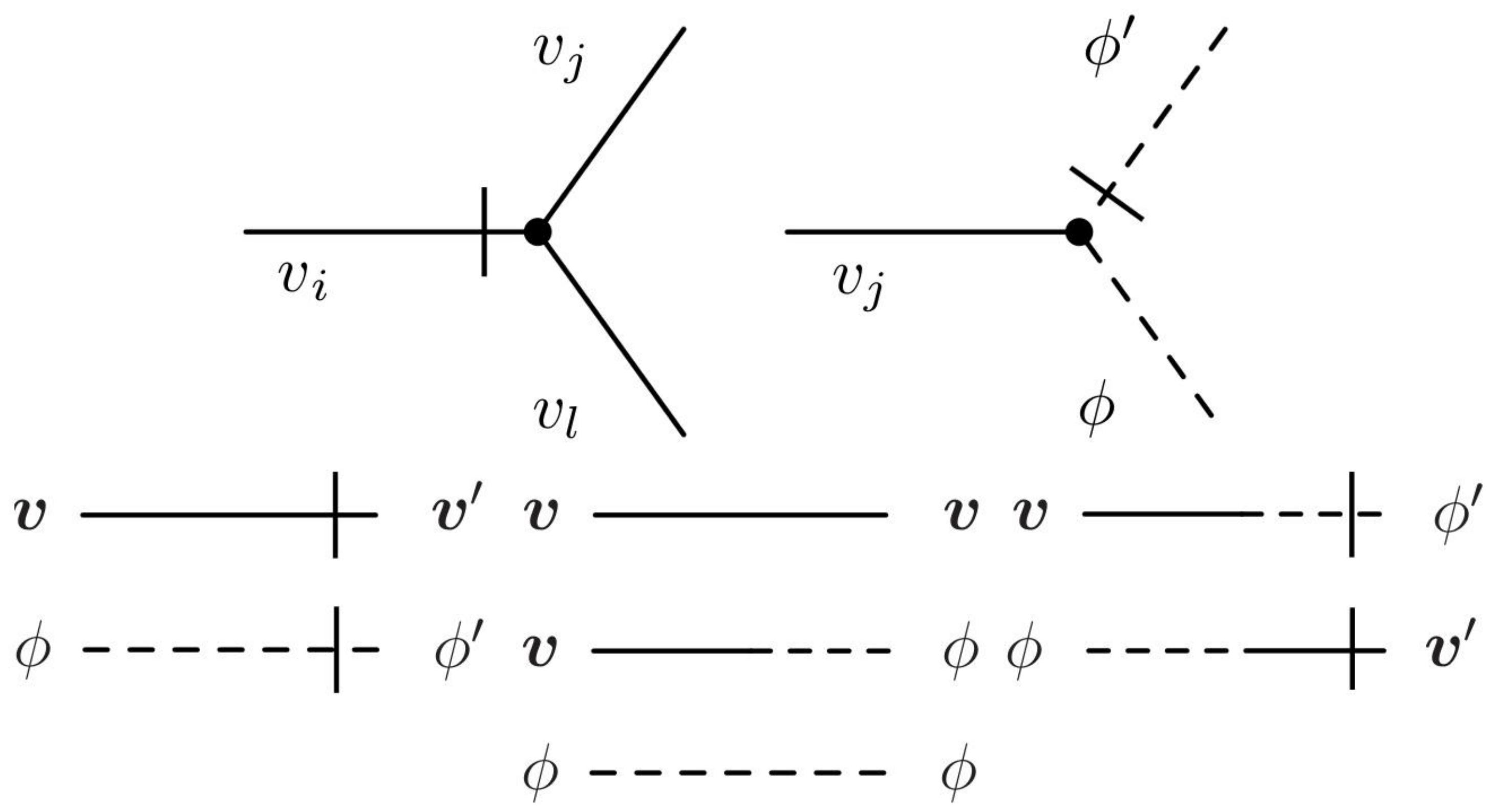}
	\caption{Graphical representation of the propagators of model
	 (\ref{eq:total_action}).} 
	\label{fig:cNS_FDC}
	\includegraphics[width=8.5cm]{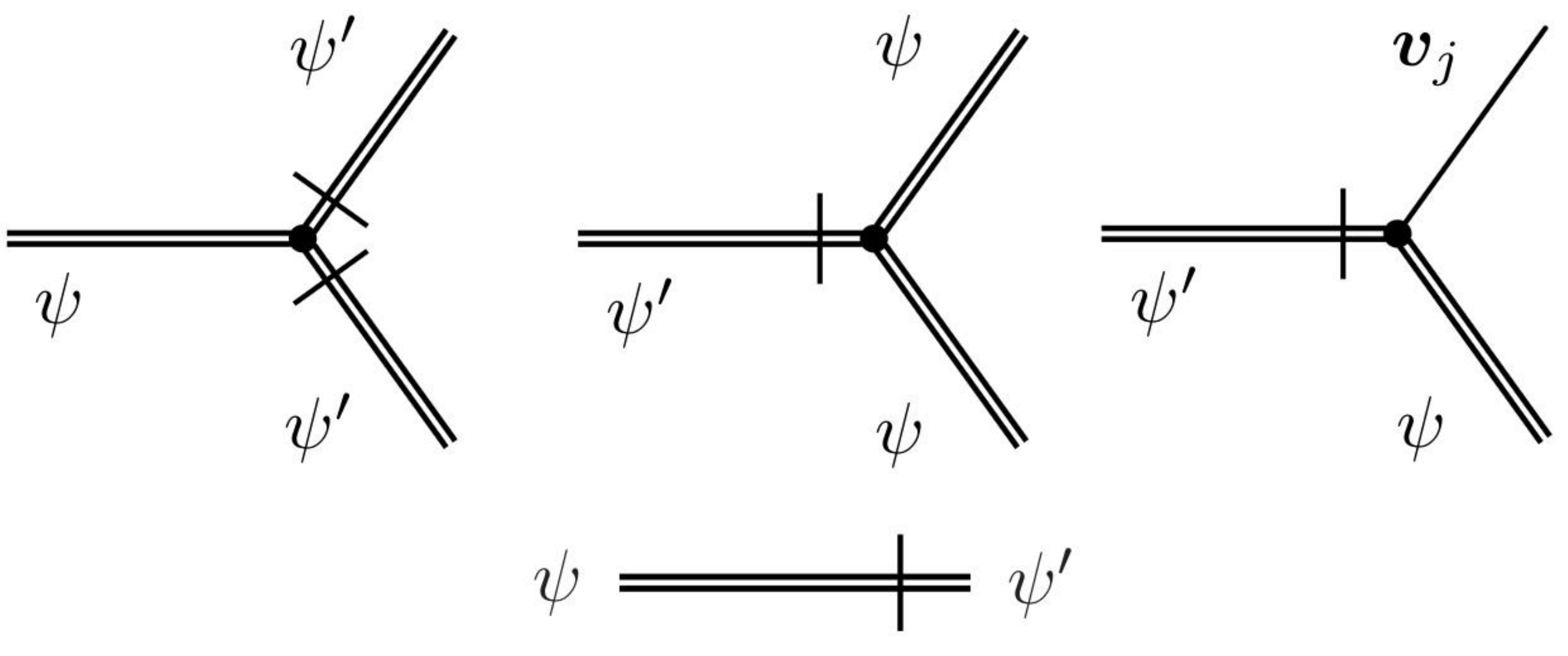}
	\caption{Graphical representation of the 
	interaction part of model (\ref{eq:total_action}).} 
	\label{fig:DP_FDC}
\end{figure}

The building blocks of the perturbation theory
are propagators and vertex factors.
Propagators are acquired from the inverse of the quadratic part of the action
functional (\ref{eq:total_action})
\begin{align}
  \langle v_{i}v_{j} \rangle_0 &= \frac{d_{1}^{f}}{|\epsilon_{1}|^{2}} P_{ij} + d_{2}^{f} 
  \left| \frac{\epsilon_{3}}{R} \right|^{2} Q_{ij} , \hspace{0.2cm} \langle \phi' \phi 
  \rangle_0 = \frac{\epsilon_{2}^{*}}{R^{*}}, 
  \label{eq:prop1}   \\
  \langle v_{i}'v_{j} \rangle_0 &= \frac{1}{\epsilon_{1}^{*}} P_{ij} + 
  \frac{\epsilon_{3}^{*}}{R^{*}} Q_{ij} , \hspace{1.35cm} \langle v_{i}'\phi \rangle_0 =
  \frac{ic_{0}^{2} k_{i}}{R^{*}},
  \label{eq:prop2}   \\   
  \langle \phi \phi \rangle_0 &= \frac{c_{0}^{4} k^{2}}{|R|^{2}} d_{2}^{f},  \hspace{2.5cm} 
  \langle v_{i}\phi' \rangle_0 = \frac{-i k_{i}}{R} , 
  \label{eq:prop3}   \\
  \langle v_{i} \phi \rangle_0 &= \frac{i c_{0}^{2} \epsilon_{3} k_{i} }{|R|^{2}} d_{2}^{f} , 
  \hspace{2.15cm} \langle \psi \psi' \rangle_{0} = \frac{1}{L}, 
  \label{eq:prop4} 
\end{align}
where we have introduced the shorthand notation
\begin{align}
  \epsilon_{1} &= -i \omega + \nu_{0}k^{2}, \hspace{0.68cm} d_{1}^{f} = 
  \nu_{0}^{3}( g_{10} k^{d-4-y} + g_{20}), \\
  \epsilon_{2} &= -i \omega + \nu_{0} u_{0} k^{2}, \quad d_{2}^{f} = 
  \nu_{0}^{3}(\alpha g_{10} k^{d-4-y} + g_{20}), \\
  \epsilon_{3} &= -i \omega + \nu_{0} \tilde{v}_{0} k^{2}, \hspace{0.53cm} L =
  - i \omega + \nu_{0} w_{0} (k^{2} + \tau_{0}), \\
  R &= \epsilon_{1} \epsilon_{2} + c_{0}^{2} k^{4}.
\end{align}
The complex conjugate counterparts $ \langle \varphi \varphi' \rangle
= \langle \varphi' \varphi \rangle^{*},  \varphi \in\Phi $ are not displayed. All
other free-field correlation functions are zero. The vertex factors $ V_{\varphi\dots\varphi} $ are extracted from the 
interaction part of the action functionals (\ref{eq:DP_action}) and (\ref{eq:cNS_action}):
\begin{align}
  V_{v_{i}'v_{j}(\mq)v_{l}(\mk)} &= -i(k_{j}\delta_{il}+q_{l}\delta_{ij}), 
  \label{eq:ver1} \\
  V_{\phi'v_{j}\phi(\mk)} &= -ik_{j}, \\
  V_{\psi'\psi'\psi} &= -V_{\psi'\psi\psi} = \lambda_{0} D_{0}, \\
  V_{\psi'v_{i}(\mq)\psi(\mk)} &= -i(k_{i} + a_{0}q_{i}). 
  \label{eq:ver2}
\end{align}
 A graphical representation of the Feynman rules is depicted 
 in Fig.~\ref{fig:cNS_FDC} and Fig.~\ref{fig:DP_FDC}.
Apparently, theory (\ref{eq:total_action}) is translation-invariant. For such theories
 it is convenient to work with the
 effective potential $ \G $, defined
as the Legendre transform of $ \mathcal{W} $ \cite{Vasiliev_blue,Zinn}
\begin{align}
  \mathcal{W}[A] = \G[\alpha] + A \alpha, \quad 
  \alpha = \frac{\delta \mathcal{W}[A]}{\delta A( x )} 
  \bigg|_{A( x ) = 0}.
\end{align}
The effective action $ \G $ represents also the generating functional 
for vertex functions $ \G_{\alpha\dots\alpha} $. It can be shown \cite{Zinn,Amit,Vasiliev}
that after the relabeling $\alpha \rightarrow \Phi$
 the relation between the effective potential and the original action functional takes the
 simple form
\begin{align}
  \G[\Phi] = -\S[\Phi] + (\text{1P loop corrections}).  
  \label{eq:effectivePot}
\end{align}
1P (one-irreducible)
loop corrections in (\ref{eq:effectivePot}) are Feynman diagrams, which
remain connected if one of the lines is removed.
 At the tree level we have $ \G_{0}[\Phi] = -\S_{0}[\Phi] $.
The next-to-leading order requires calculation of all one-loop Feynman diagrams.
Such calculations are usually plagued with divergences, what can be taken
care of with a proper renormalization scheme \cite{Vasiliev,Zinn}.

{\subsection{UV divergences and renormalization procedure} \label{sec:Renormalization} }
Renormalization of model (\ref{eq:cNS_action}) has been carried out
both directly at $ d = 3 $ in \cite{ANU97} and with the use of
 a double expansion scheme in the vicinity of the space dimension $ d = 4 $ in \cite{AGKL17}.
Since the upper critical dimension of
the DP model is four (the coupling constant of the DP becomes dimensionless, see below) we have to
 renormalize the cNS model around $ d = 4 $ as well. In field-theoretic models
 of passive turbulent
advection or advection of models such as (\ref{eq:DP_action})  the
velocity field is not renormalized at all, what simplifies the RG procedure
\cite{turbo,AIK10,AK10,AK14}. 
The same situation occurs also in the present case. 
Because the velocity model has been discussed in detail elsewhere \cite{AGKL17} and
it is not our main aim here, 
we do not dwell on the full renormalization procedure of the cNS model
but discuss only parts relevant to our model.

The initial part of the RG procedure consists of
 the analysis of the UV divergences based on a calculation of canonical 
dimensions \cite{Zinn,Amit,Vasiliev}. Dynamical models such as (\ref{eq:DP_action}) and (\ref{eq:cNS_action})
have two independent length scales - spatial and temporal length scale.
 We introduce $ d_{Q}^{k} $ and $ d_{Q}^{\omega} $ as the
 momentum and a frequency canonical dimension, respectively.
 Their linear combination
\begin{align}
  d_{Q} = d_{Q}^{k} + 2 d_{Q}^{\omega}
  \label{eq:totalCD}
\end{align}
denotes the total canonical dimension $ d_{Q} $ of the quantity $ Q $.
 The total canonical dimension (\ref{eq:totalCD}) plays the same role as a 
  standard canonical dimension in static theories \cite{Zinn,Amit}.
  A list of all canonical dimensions for the current model is shown
  in Tab.~\ref{tab:CD}.
\begin{table*}
	\centering
	\def\arraystretch{1.5}
\begin{tabular}{| c || c  c  c  c  c  c  c  c  c  c  c  c  c |}
			\hline \hline
			$Q$ & $ \mv_{0}' $  & $ \mv_{0} $ & $ \phi_{0}' $ & $ \phi_{0} $ & $ \psi' $
			& $ \psi $ & $ m, \mu, \Lambda $ & $ \tau $ & $\nu_0,\nu,D_{0},D $ & $c_{0},c$
			& $g_{10}$ & $g_{20},g_{30}=\lambda_{0}^{2}$ 
			& $ u_0,v_0,w_{0},u,v,w,g_1,g_2,g_{3},\alpha_{0},\alpha $ \\ 
			\hline \hline
			$d_Q^k$ & $d+1$ & $-1$ & $d+2$ & $-2$ & $ d/2 $ & $ d/2 $ & $1$ & $2$ & $-2$ &
			$-1$  & $y$ & $\varepsilon$ & $0$ \\
			 
			$d^\omega_Q$ & $-1$ & $1$ & $-2$ & $2$ & $0$ & $0$ & $0$ & $0$ & $1$ & $1$  &
			$0$ & $0$ & $0$ \\
			 
			$d_Q$ & $d-1$ & $1$ & $d-2$ & $2$ & $ d/2 $ & $ d/2 $ & $1$ & $2$ & $0$ & $1$ 
			& $y$ & $\varepsilon$ & $0$ \\ 
			\hline \hline
\end{tabular}
	\caption{Canonical dimensions of all the bare fields and bare parameters for the model of velocity 
	fluctuations. Parameters $ m $ and $ \Lambda $ are the IR and UV cut-off, respectively,
	and $ \mu $ is
	the scale-setting parameter. } 
	\label{tab:CD}
\end{table*}

Superficial divergences are present in such 1I functions $ \G $
for which the UV exponent
\begin{align}
  \delta_{\G} = d_{\G}|_{\varepsilon=y=0} = 6 - \sum_{\varphi} N_{\varphi} d_{\varphi},
  \label{eq:super_degree}
\end{align}
is non-negative. The sum in Eq.~(\ref{eq:super_degree}) runs over
all field arguments of the function $ \G $. 
 
To sort out 1I functions $ \G $ with real UV divergences the following properties of the
model are used.
\begin{enumerate}[(a)]
	\item 1I functions without at least one response field $ v',\phi' $ and $ \psi' $
	necessarily contain
	a closed loop of retarded propagators and therefore no such counterterm can appear. 
	Moreover, structures with at least one field $ \psi' $ must contain at least one
	field $ \psi $. Otherwise we obtain again a closed loop. A detailed
	technical exposition can be found in \cite{Vasiliev}.
	\item Fields $ v $ and $ \phi $ appear in interaction
	vertices of the action (\ref{eq:cNS_action}) 
	together with their derivatives and hence the real UV exponent is reduced according 
	to 
	\begin{align}
	  \delta_{\G}' = \delta_{\G} - n_{v} - n_{\phi} .
	\end{align}
	For instance,
	1I function
	$ \G^{\psi'\psi\phi} $ has $ \delta_{\G} = 0 $ but $ \delta_{\G}' = -1 $ 
	and therefore it is a finite function (does not require renormalization).
	\item The Galilean invariance \cite{turbo,AGKL17} of model (\ref{eq:cNS_action})
	ensures that the convective
	derivative $ \nabla_{t} $ must enter the counterterms as a single
	object \cite{Vasiliev,ANU97}.
	This 
	implies that structures $ \psi'\partial_{t}\psi $ and 
	$ \psi'(\mv\cdot\boldnabla)\psi $ must 
	be renormalized by  the same counterterm.
	An additional observation which reduces possible types of counterterms is the 
	generalized Galilean invariance under the time-dependent
	transformation (instantaneous) velocity parameter ${\bm V}_i(t)$:
	\begin{align}
	  {\bm v}_{w}(x) & =  {\bm v}(x_{w}) - {\bm V}_i(t), 
	  &x &= (t,{\bm x}),
	  \nonumber \\
	  \Phi_{w}(x)& = \Phi(x_{w});  &x_{w}& =(t,{\bm x}+{\bm V}(t));
	  \nonumber \\
	  {\bm V}(t) & = \int^{t}_{ -\infty} {\bm V}_i(t') dt',
	  \label{GGi}
	\end{align}
	where $\Phi$ stands for any of the three remaining fields~-- $v',\phi',\phi$.
	The crucial point is that {despite} the fact that the action functional
	is {\it not} invariant  with respect to such a transformation, it transforms in the identical way with 
	the generating functional
	of the {1-irreducible} Green functions
	\begin{align}
	 {\cal S}[\Phi_{w}] & =  {\cal S}[\Phi] + \mv'\cdot \partial_{t} \mV_i, \nonumber \\
	  {\Gamma}[\Phi_{w}] & =  {\Gamma}[\Phi] + \mv'\cdot\partial_{t} \mV_i.
	  \label{gGal}
	\end{align}
        The latter expression could be expressed in the form (\ref{eq:effectivePot}).
	In fact, the expressions~\eqref{gGal} mean that the counterterms appear invariant under
	the generalized Galilean transformation~(\ref{GGi}).
	\item From the explicit form of the propagators in Eq. (\ref{eq:prop1})-(\ref{eq:prop2}), we
	observe that $ \langle v'\phi \rangle_{0} $ and $ \langle v\phi \rangle_{0} $ are proportional 
	to $ c_{0}^{2} $ while $ \langle \phi\phi \rangle_{0} $ is proportional to $ c_{0}^{4} $. On 
	the other hand, propagator $ \langle \phi\phi' \rangle_0 $ is not proportional to any power of 
	$ c_{0} $. Since these factors have a positive total canonical dimension 
	(see Tab.~\ref{tab:CD}), they appear as an 
	external factor in a given Feynman diagram. Hence, the real UV exponent is 
	reduced by the number of 
	fields containing this factor. The vertex function with $ N_{\phi'}>N_{\phi} $ contains 
	factor
	$ c_{0}^{2(N_{\phi'}-N_{\phi})} $. For example, the Green function $ \G^{\psi'\psi\phi'} $ is 
	UV finite, since the UV exponent is reduced from $ \delta_{\G} = 0 $ to
	$ \delta_{\G}'=-2 $ \cite{AK14}.	
\end{enumerate}
As a consequence, we arrive at the conclusion
that all UV divergences of the DP model can be removed by addition of the following 
counterterms
\begin{align}
  \psi'\partial_{t} \psi&, \quad \psi' \partial^{2} \psi, \quad \tau \psi'\psi, \quad \psi' 
  ( \mv \cdot\boldnabla ) \psi, \\
  \psi' (\boldnabla\cdot\mv) \psi&, \quad \psi'\psi^{2}, \quad \psi'^{2}\psi \ .
\end{align}
All of these terms are already present in model (\ref{eq:total_action}) and thus the 
model is multiplicatively renormalizable.

In explicit terms renormalization of the DP action functional is accomplished by the following renormalization
of the parameters and fields
\begin{align} 
  g_{30} &= g_{3} \mu^{\varepsilon} Z_{g_{3}}, \quad D_{0} = D Z_{D}, \quad
  \tau_{0} = \tau Z_{\tau}+\tau_c, 
  \label{eq:norm_rel1}
  \\
  \lambda_{0} &= \lambda \mu^{\varepsilon/2} Z_{\lambda}, \quad w_{0} = w Z_{w},
  \quad a_{0} = a Z_{a},  
  \label{eq:norm_rel2}
\end{align}
with the substitution $ \psi' \rightarrow \psi' Z_{\psi'}, \ \psi \rightarrow \psi Z_{\psi} $ and
similarly for the cNS field \cite{ANU97,AGKL17}. 
Note that
the term $\tau_c$ is a non-perturbative effect \cite{Sym73,Schloms89,JT04}, which
is not captured by the  dimensional regularization.

For completeness (details in \cite{AGKL17})
we note that the following renormalization of the velocity part of the action
 (\ref{eq:total_action}) is needed
\begin{align}
g_{10} & =  g_1 \mu^y Z_{g_1}, &u_{0}& = u Z_{u}, &\nu_0& = \nu Z_{\nu},
\nonumber \\
g_{20} & =  g_2 \mu^\varepsilon Z_{g_2}, &v_{0}& = v Z_{v}, &c_{0}& = c Z_{c},
\label{mult}
\end{align}
supplemented with the renormalization of $\phi$ and $\phi'$ fields
\begin{equation}
  \phi \to Z_{\phi}\phi, \quad \phi' \to Z_{\phi'}\phi'.
\end{equation}
The total renormalized action functional of DP advected by compressible turbulent flow
is then $ \S_{R} = \; \S_{R}^{\psi} + \S_{R}^{\mv} + S_R^\text{adv} $ explicitly
given by
\begin{widetext}
  \begin{align} 
  \S_{R}^{\psi}[\Phi] = \; & \psi'\{ Z_{1} \partial_{t} + D(- Z_{2}\partial^{2} + Z_{3}\tau) \}
  \psi - \frac{\lambda D}{2} \{ Z_{4} \psi' - Z_{5} \psi \} \psi' \psi + 
  \psi'\{ Z_{1} v_{i}\partial_{i} + Z_{6} a (\partial_{i}v_{i}) \} \psi ,
  \label{eq:RenormalizedAction}
\end{align}
\end{widetext}
which has to be
 augmented by the relations for the renormalization constants
\begin{align}
  Z_{1} &= Z_{\psi'} Z_{\psi}, \hspace{1.25cm} Z_{2} = Z_{\psi'} Z_{\psi} Z_{D}, 
  \label{eq:NC1} \\
  Z_{3} &= Z_{\psi'} Z_{\psi} Z_{D} Z_{\tau}, \quad Z_{4} = Z_{\psi'}^{2} Z_{\psi} Z_{D} Z_{\lambda},
  \label{eq:NC2} \\
  Z_{5} &= Z_{\psi'} Z_{\psi}^{2} Z_{D} Z_{\lambda}, \quad Z_{6} = Z_{\psi'} Z_{\psi} Z_{a}.
  \label{eq:NC3}
\end{align}
These relations can be easily inverted to express the RG constants 
 of fields and parameters in terms of $Z_i,i=1,\ldots,6$. 
 In the one-loop approximation, the following
diagrams are required for the DP part of the whole model (\ref{eq:total_action})
\begin{align}
  \G_{\psi'\psi} = &\ i\omega Z_{1} - D k^{2} Z_{2} - D \tau Z_{3} + \nonumber \\
  & \hspace{-1.1cm} + \frac{1}{2} \ \raisebox{-1.5ex}{\includegraphics[width=2cm]{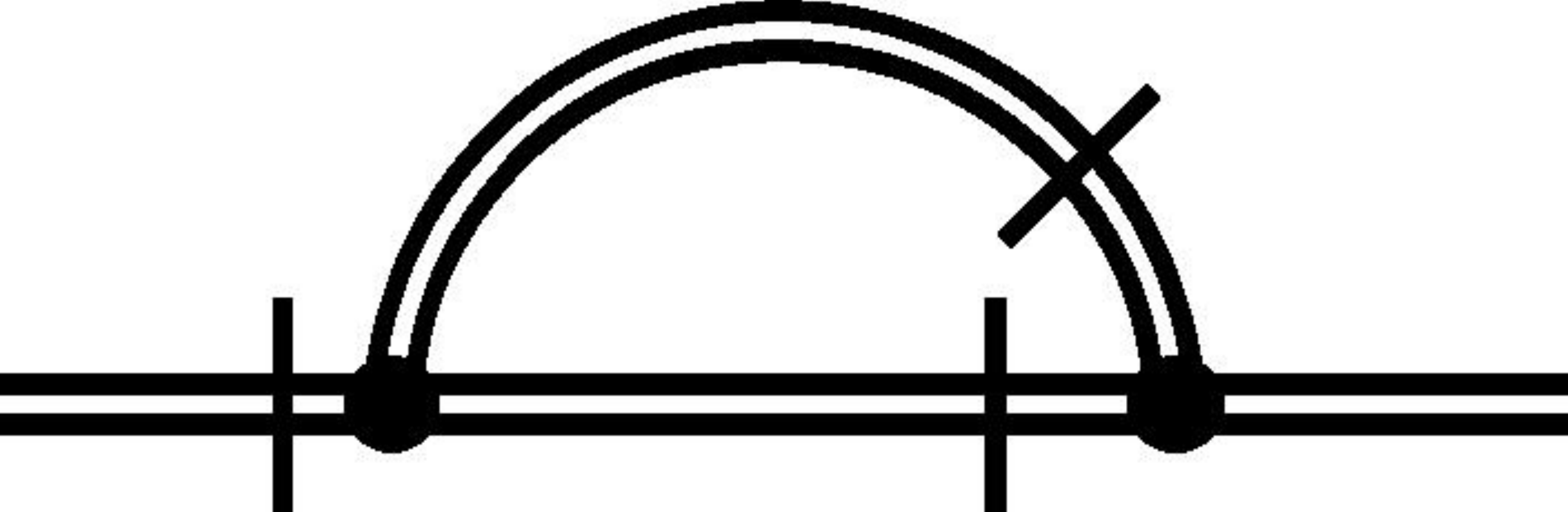}} +
  \raisebox{-1.5ex}{\includegraphics[width=2cm]{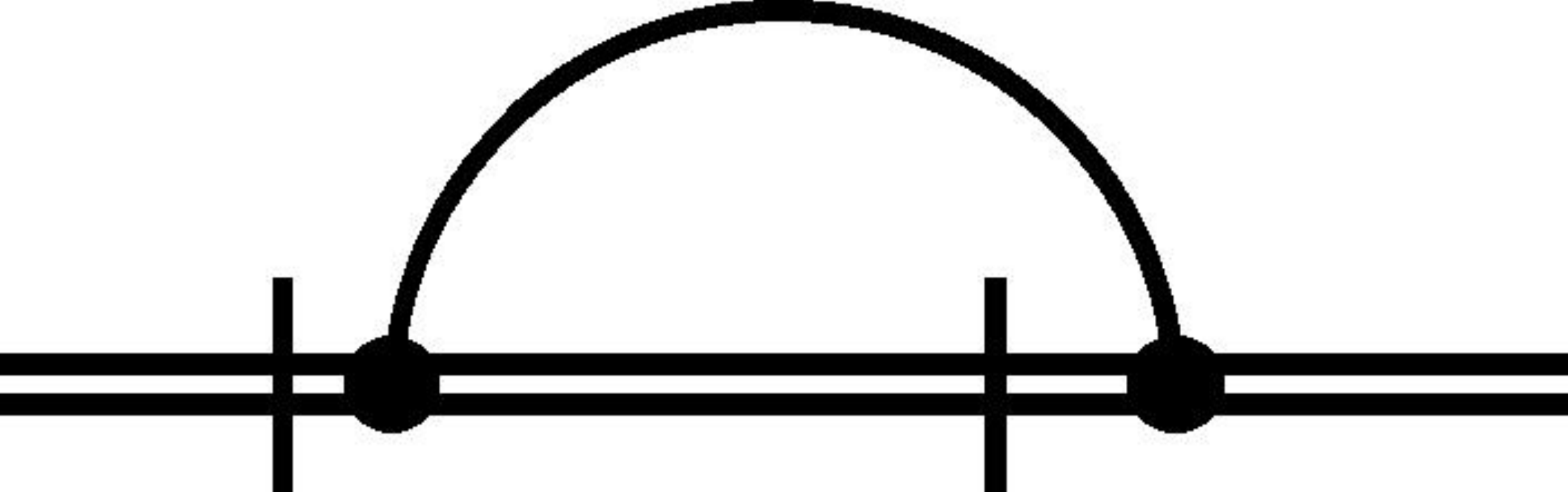}}, \\
  \G_{\psi\psi'\psi'} = &\ D \lambda \mu^{\frac{\varepsilon}{2}} Z_{4} + \nonumber \\
  & \hspace{-1.4cm} + \raisebox{-5ex}{\includegraphics[width=1.9cm]{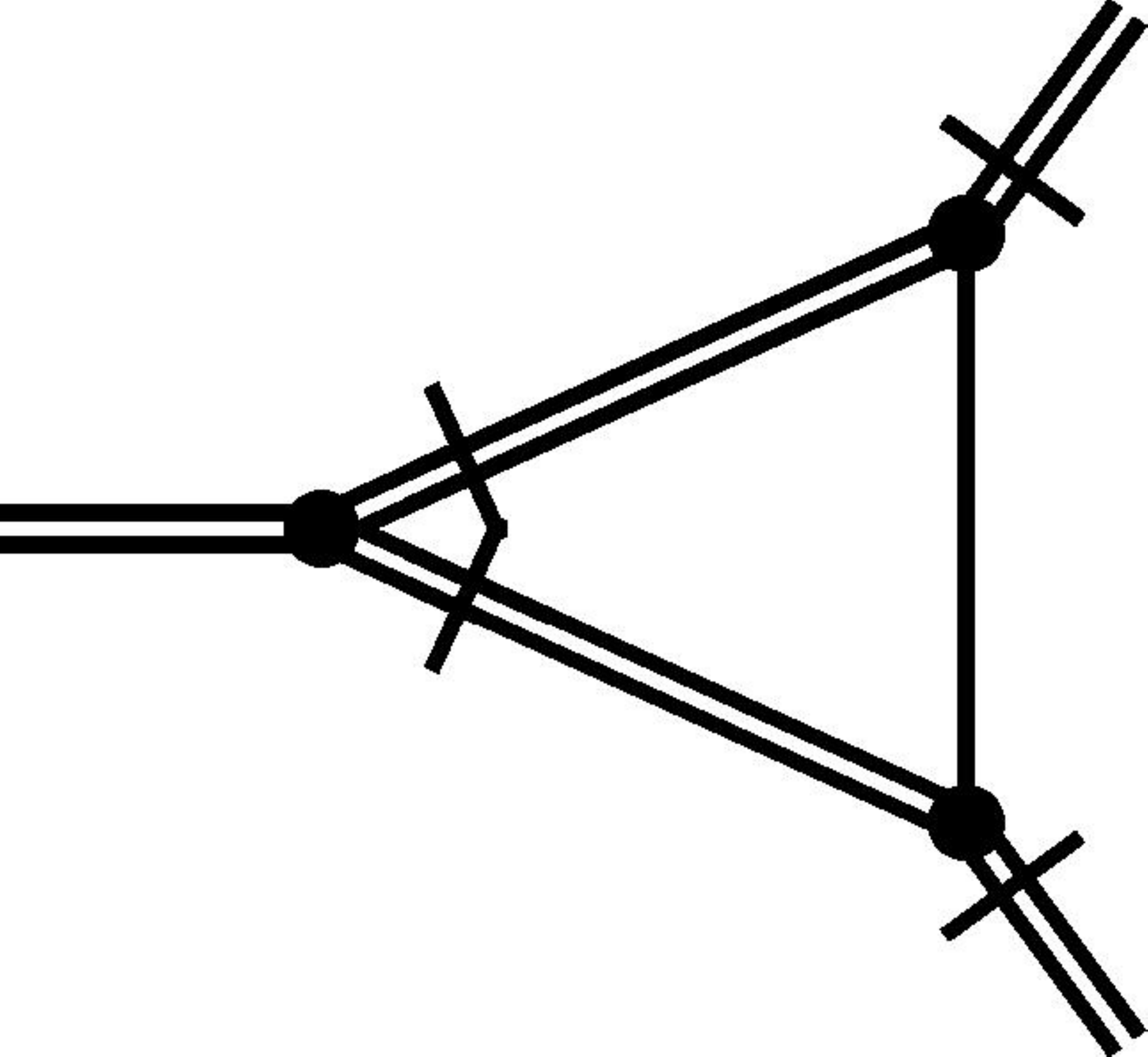}} + 2 \ 
  \raisebox{-5ex}{\includegraphics[width=1.9cm]{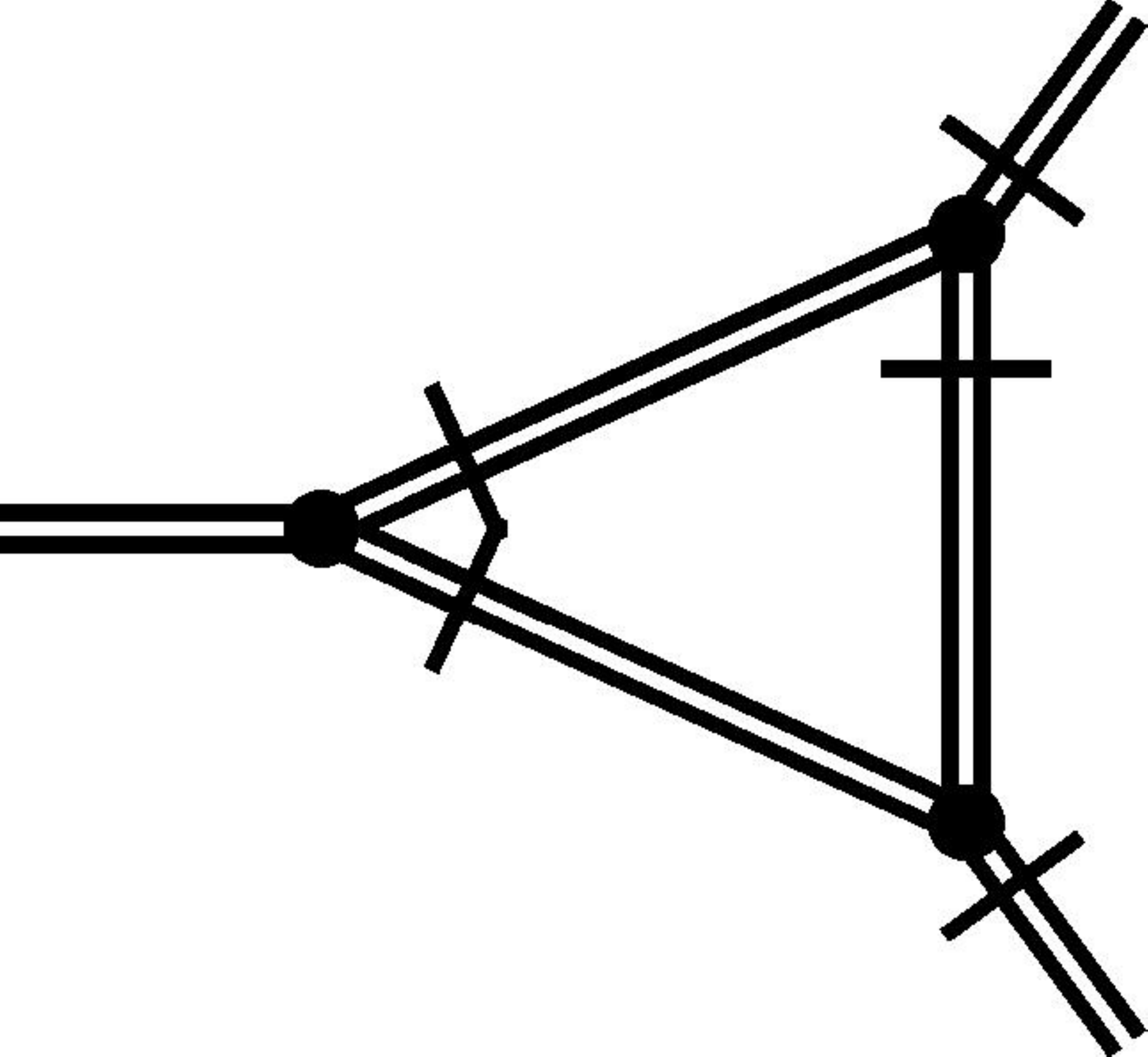}} + 2 \ 
  \raisebox{-5ex}{\includegraphics[width=1.9cm]{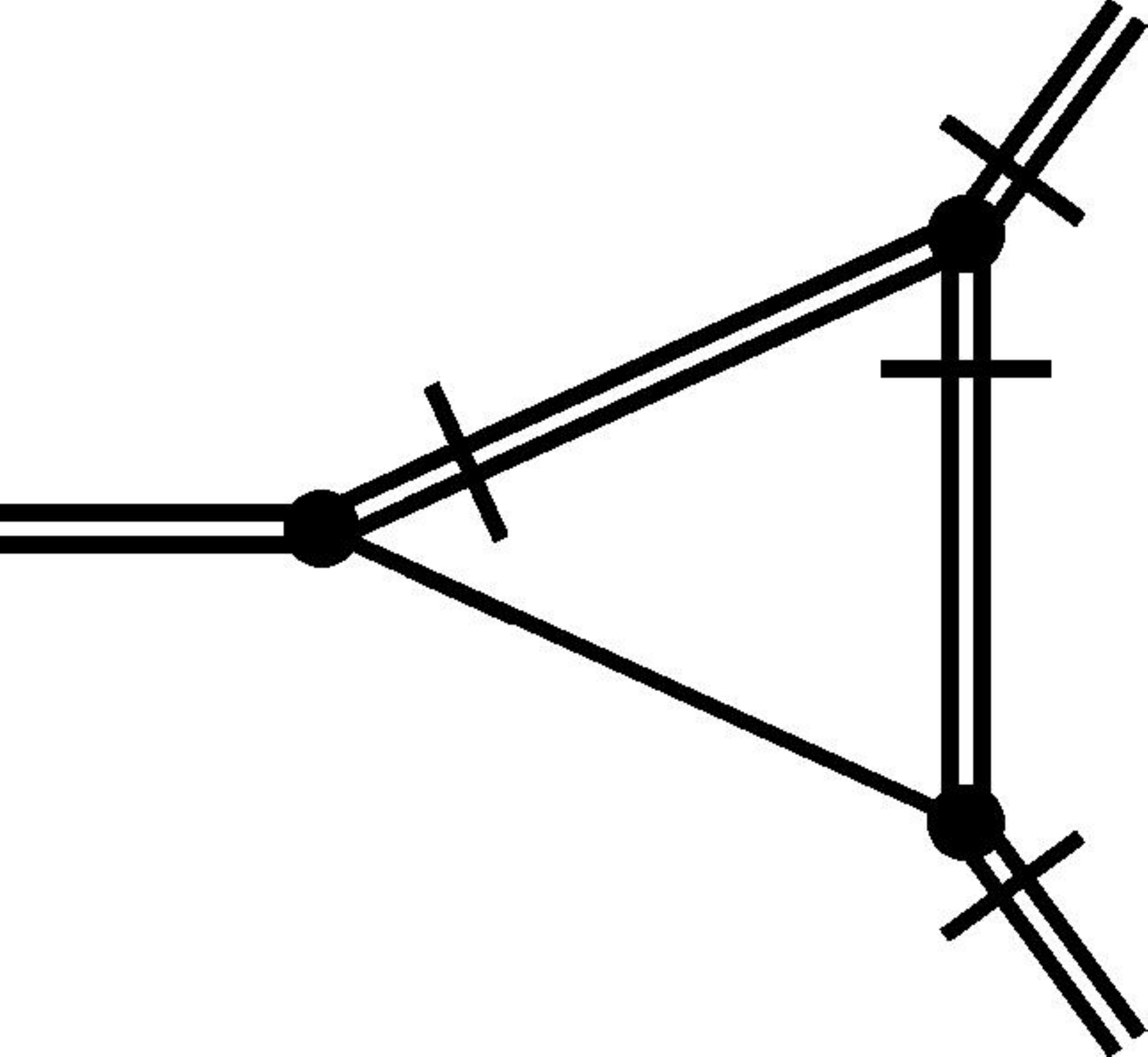}}, \\
  \G_{\psi'\psi\psi} = & - D \lambda \mu^{\frac{\varepsilon}{2}} Z_{5} + \nonumber \\
  & \hspace{-1.4cm} + \raisebox{-5ex}{\includegraphics[width=1.9cm]{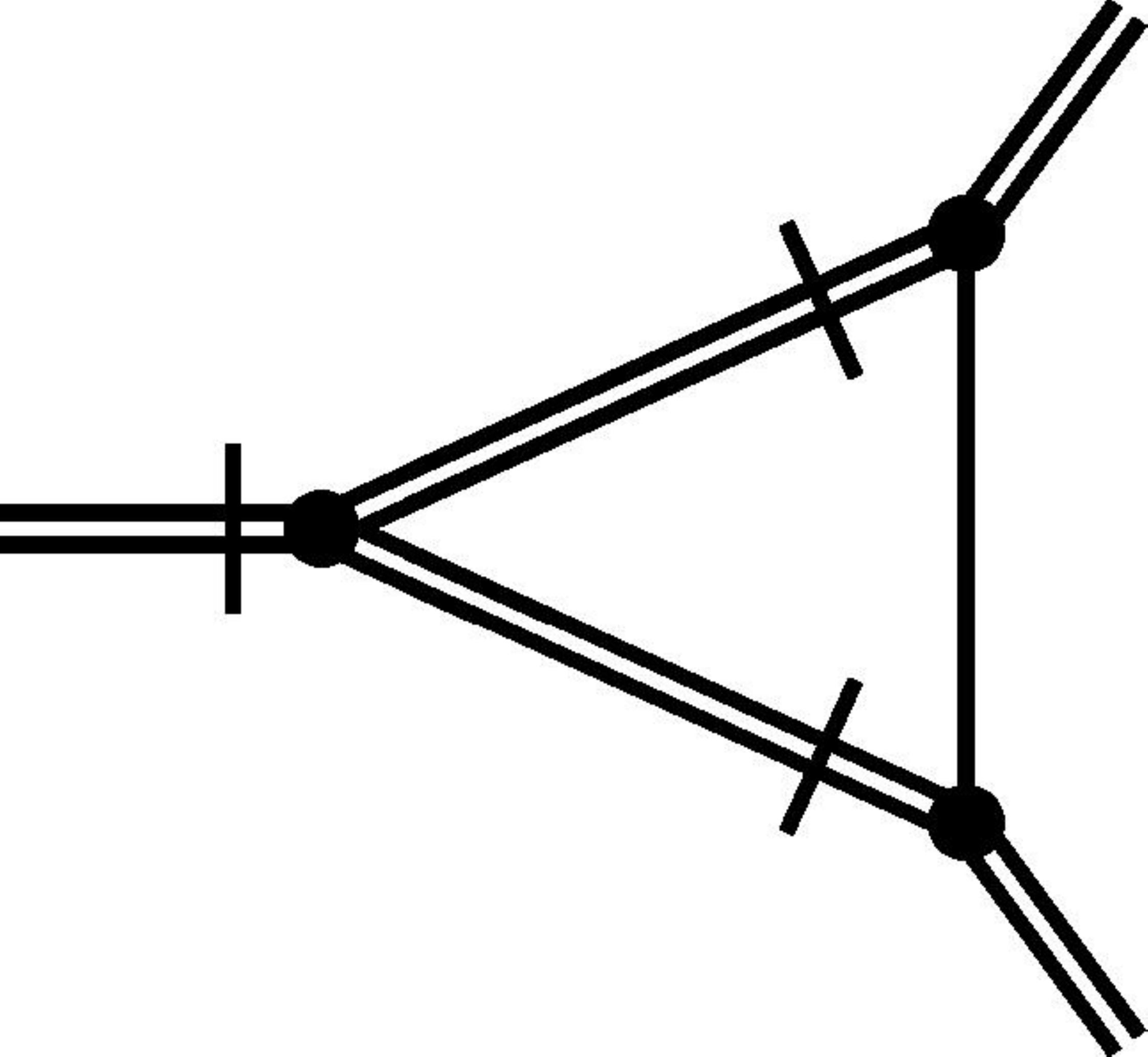}} + 2 \ 
  \raisebox{-5ex}{\includegraphics[width=1.9cm]{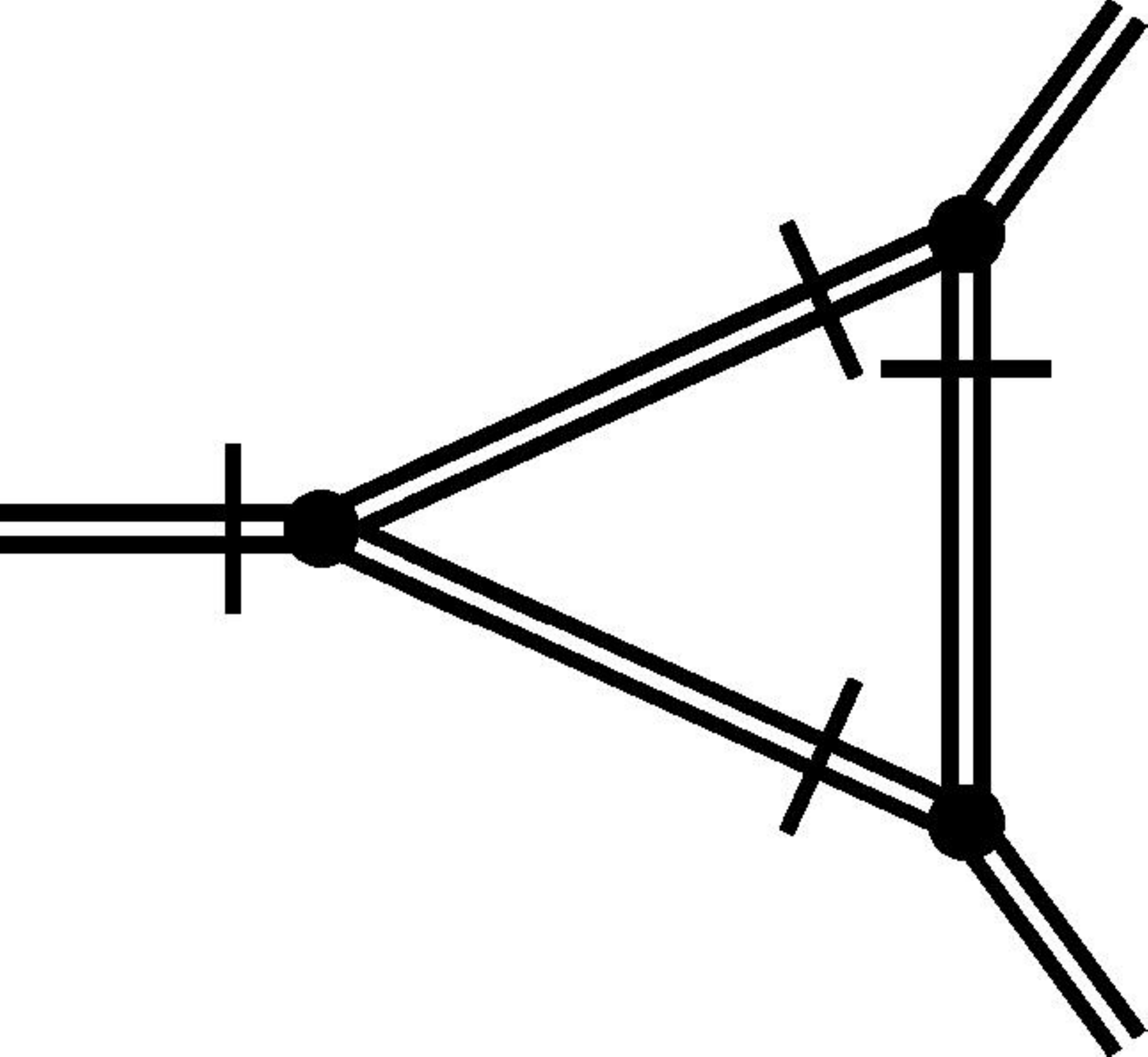}} + 2\ 
  \raisebox{-5ex}{\includegraphics[width=1.9cm]{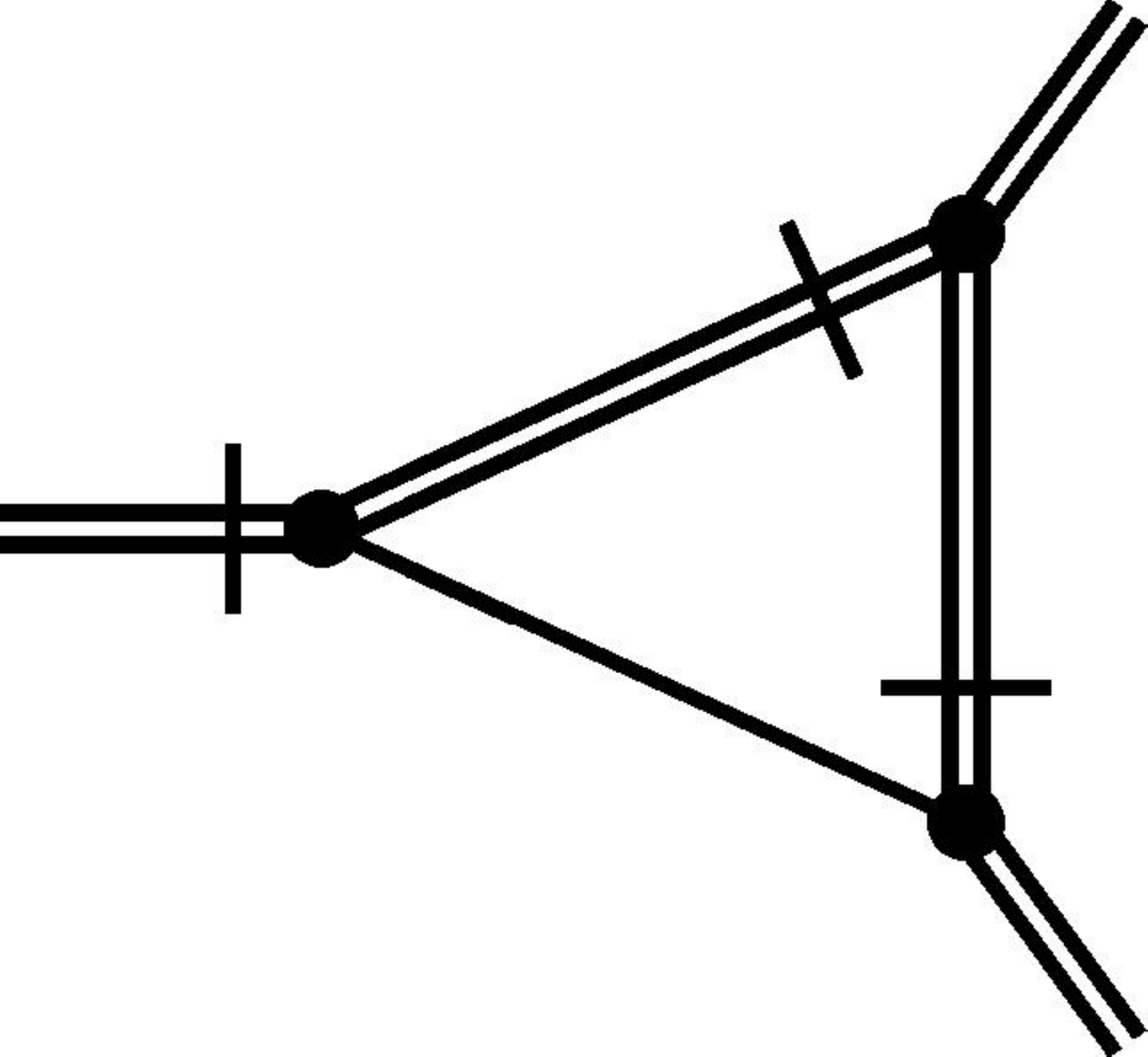}}, \\
  \G_{\psi'\psi v_{i}} = & - i p_{i} Z_{1} - i aq_{i} Z_{6} + \nonumber \\
  & \hspace{-1.4cm} + \raisebox{-5ex}{\includegraphics[width=1.9cm]{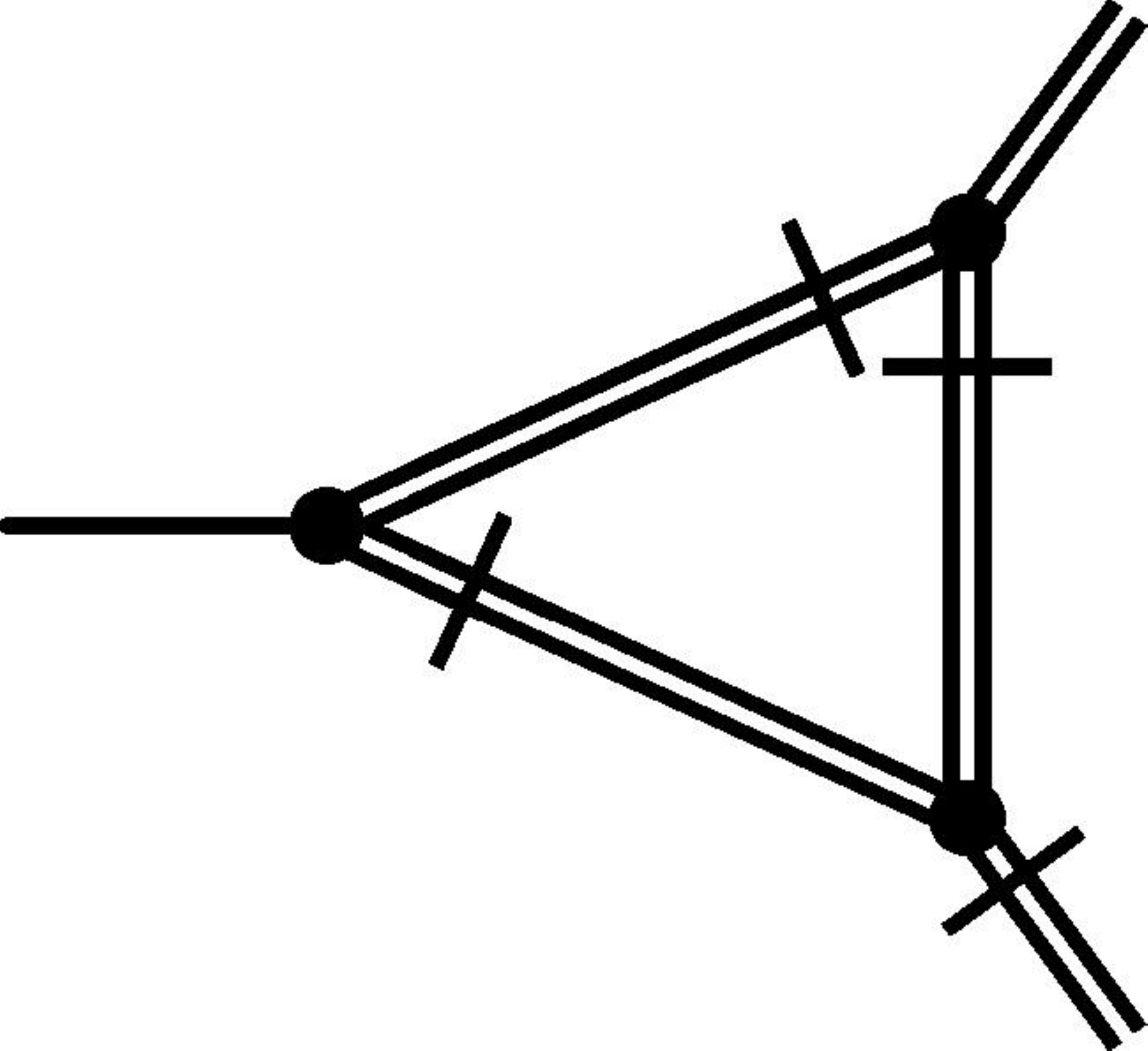}} +
  \raisebox{-5ex}{\includegraphics[width=1.9cm]{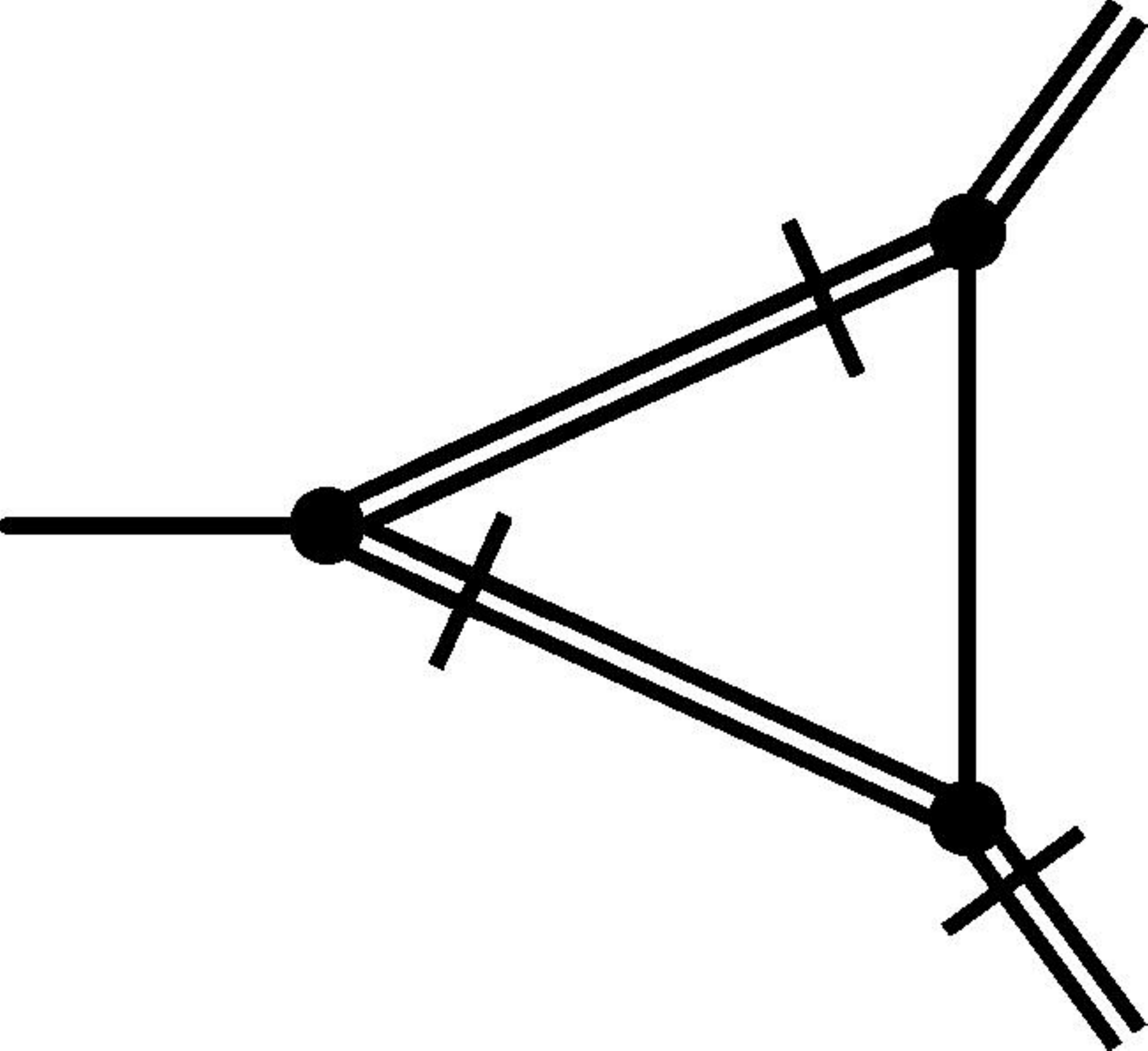}} +
  \raisebox{-5ex}{\includegraphics[width=1.9cm]{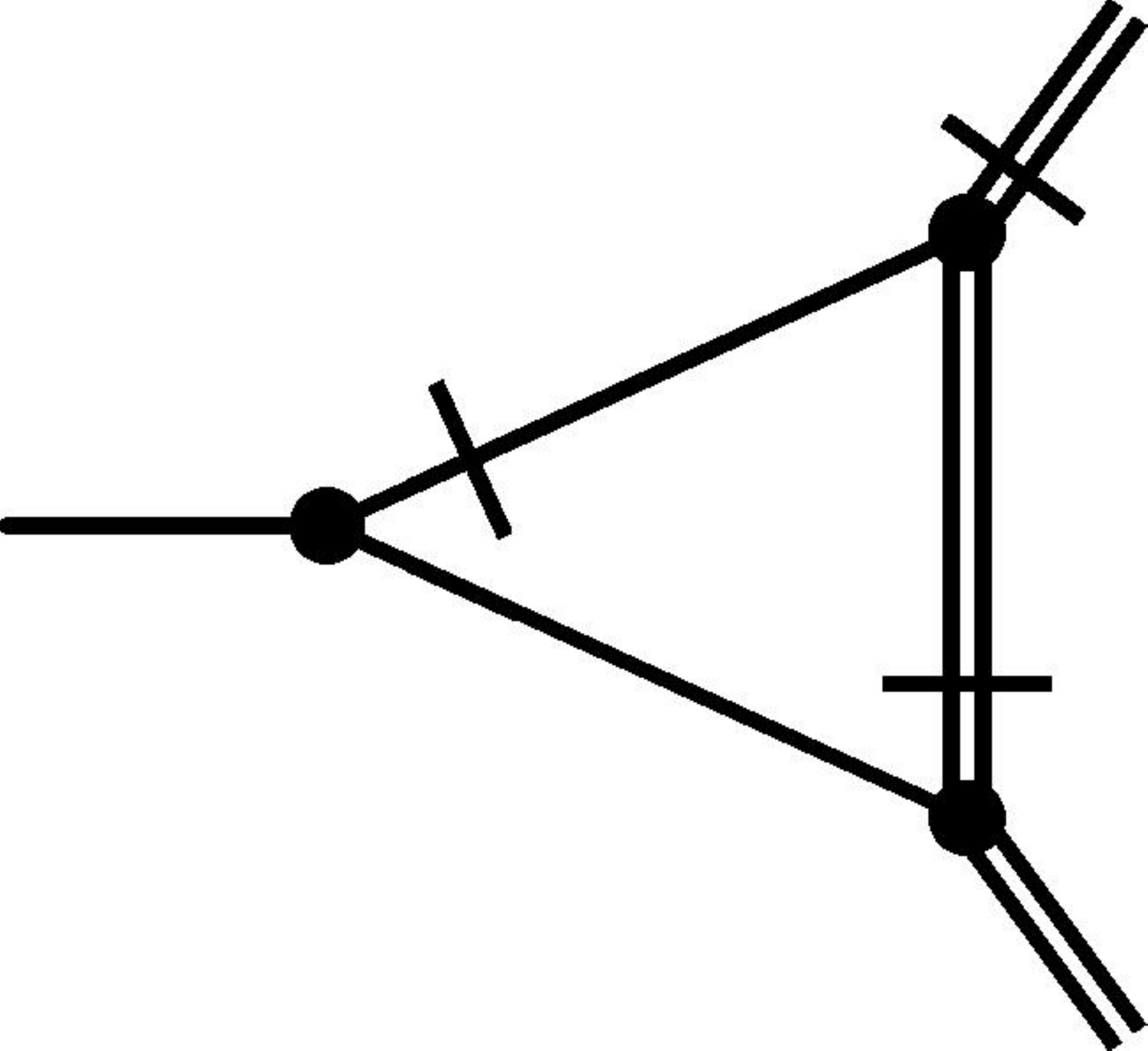}} \nonumber \\
  & \hspace{-1.4cm} + \raisebox{-5ex}{\includegraphics[width=1.9cm]{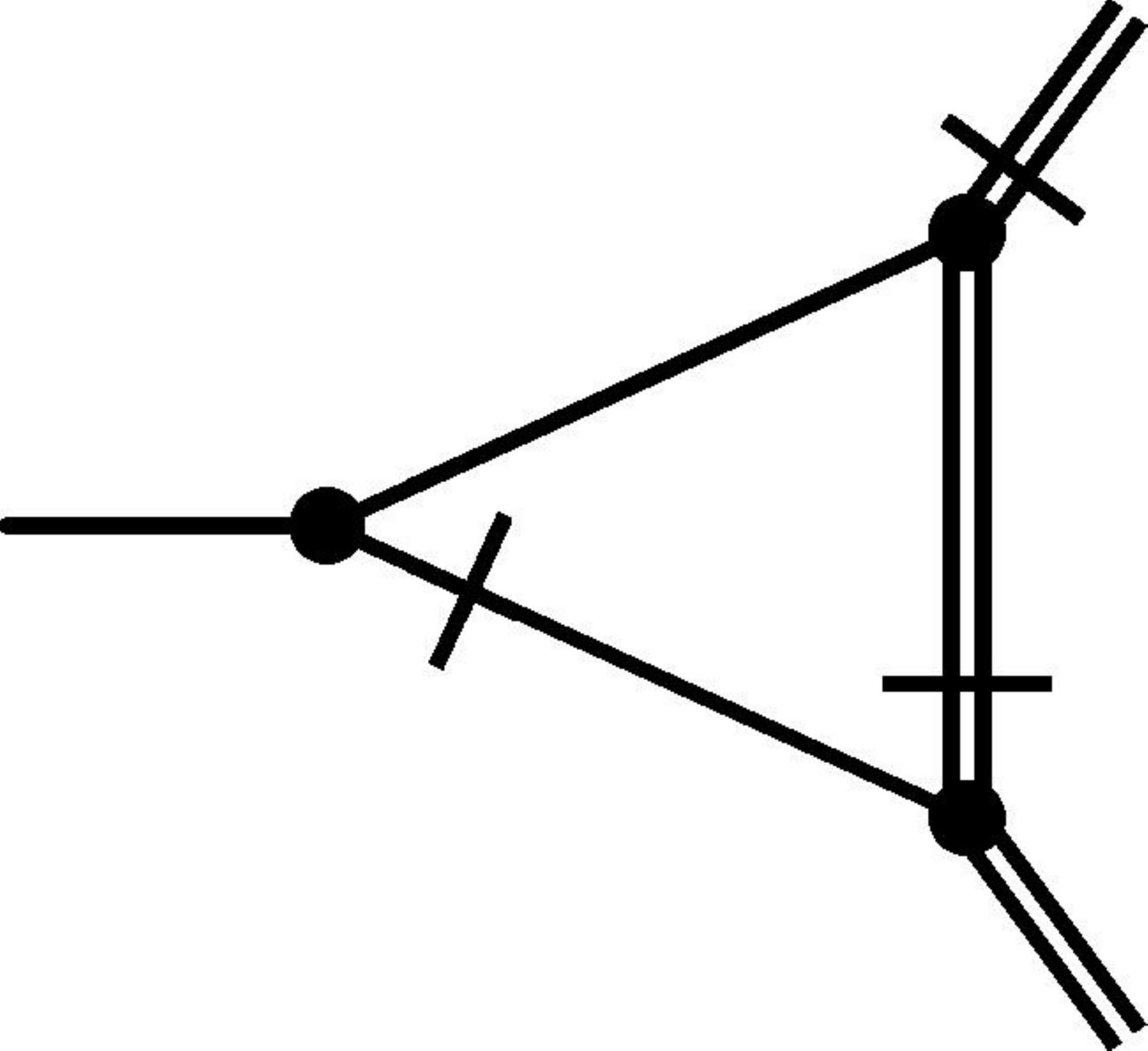}} +
  \raisebox{-5ex}{\includegraphics[width=1.9cm]{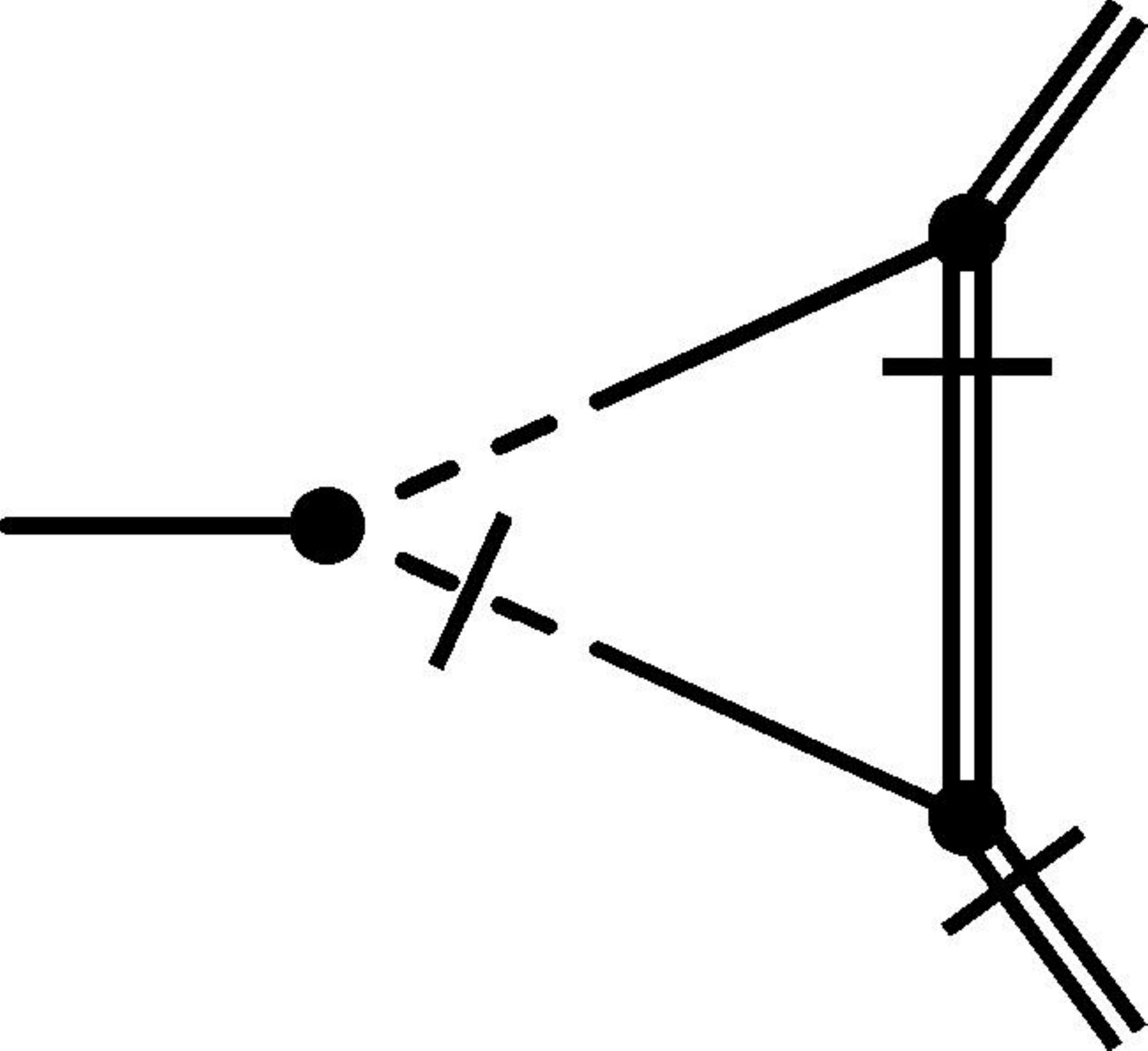}} +
  \raisebox{-5ex}{\includegraphics[width=1.9cm]{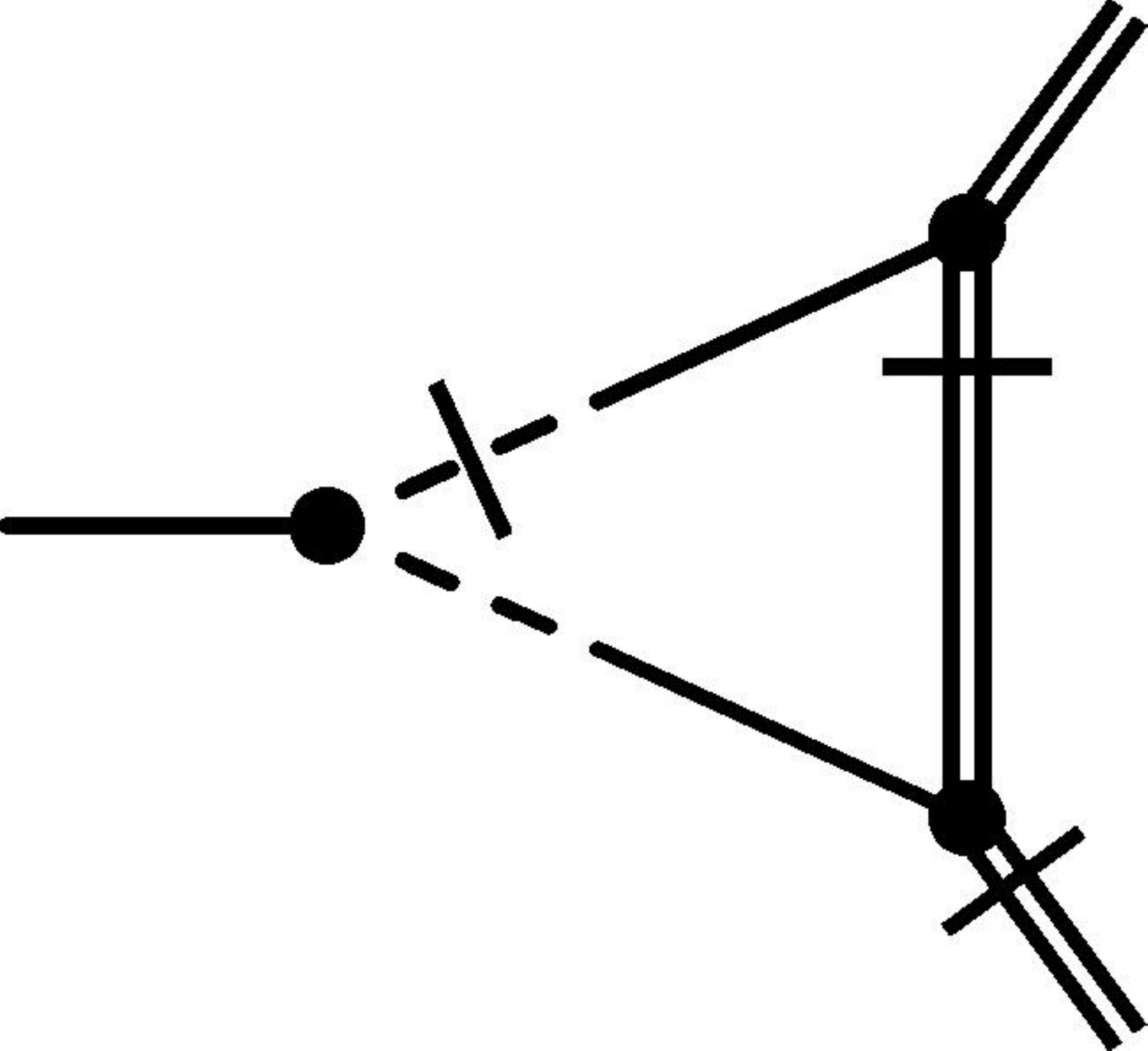}},
\end{align}
 where $ p_{i} $ and $ q_{i} $ in the last equation stand for momenta of fields $ \psi $ and
 $ v_{i} $, respectively. Passivity of the problem directly leads to the absence of additional corrections 
 to cNS model caused by the DP perturbation elements. The explicit expressions for renormalization
 constants are given
 in Appendix \ref{app:const}.

 Investigation of the large-scale and long-time universal properties of the field-theoretic models 
 calls for an analysis of Green functions at different spatio-temporal
  scales. The relation between the renormalized
$ G $ and bare $ G_{0} $ Green functions is the following
\cite{Vasiliev,Amit}
\begin{align}
  G_{0}^{\varphi'\varphi}(\{ k_i \},e_{0}) = Z_{\varphi'}^{N_{\varphi'}}(g)
  Z_{\varphi}^{N_{\varphi}}(g) G^{\varphi'\varphi}(\{ k_i \},e,\mu),
  \label{eq:RG_relation}
\end{align}
where $ \varphi,\varphi' $ stand for any permissible field, 
  and $ k = \{\mk,\omega\}, g = g(\mu) $ is the full set of renormalized charges and $ e_{0},e = e(\mu) $ 
are sets of all bare and renormalized parameters (including masses) with $ \mu $ being the 
reference mass scale. 
 In what follows, we denote the logarithmic derivative with respect to any quantity
 $ x $ as $ \D_{x} \equiv x \partial_{x} $.
Let us denote $ \tilde{\D}_{\mu} $
  the logarithmic derivative with respect to $ \mu $ with fixed bare parameters.
  Application of $\tilde{\D}_{\mu}$ on (\ref{eq:RG_relation})
  yields the fundamental RG equation \cite{Zinn,Amit}
\begin{align}
  \{ \D_\text{RG} + N_{\varphi'} \gamma_{\varphi'} + N_{\varphi} \gamma_{\varphi} \} 
  G^{\varphi'\varphi} (\{ k_i \},e,\mu) = 0 ,
  \label{eq:basicRG}
\end{align}
where $ \gamma_{Q} $ is the anomalous dimension of the quantity $ Q $
\begin{align}
  \gamma_{Q} = \tilde{\D}_{\mu} \ln Z_{Q}.
  \label{eq:AD}
\end{align}
Further, $ \D_\text{RG} $ in Eq.~(\ref{eq:basicRG}) is the $ \tilde{\D}_{\mu} $ operator expressed
in terms of renormalized parameters
\begin{equation}
  \D_\text{RG} =  \D_{\mu} + \beta_{g} \partial_{g}  - \gamma_{\nu} \D_{\nu} - \gamma_{c}\D_{c} -
  \gamma_{\tau}\D_{\tau}, 
\end{equation}
where in the second term summation over all charges $g$ of theory is implied.
For convenience we denote by $g$  the following set
\begin{equation}  
  g = \{ g_{1},g_{2},u,v,g_{3},w,a \}.
  \label{eq:all_charges}
\end{equation}

\begin{table*}
	\centering
	\def\arraystretch{1.75}
	\begin{tabular}{ | c || c c c c c c c | c c c c c c c | }	
		\hline\hline
		FP$\!\!	\biggl/\!\!g^*|\lambda_i$ & $ g_{1}^{*} $ & $ g_{2}^{*} $ & $ u^{*} $ 
		& $ \tilde{v}^{*} $ & $ g_{3}^{*} $ & $ w^{*} $ & $ a^{*} $
		& \multicolumn{4}{c}{$ \lambda_{1}  \dots \lambda_{4} $} & $ \lambda_{5} $ 
		& $ \lambda_{6} $ & $ \lambda_{7} $ \\
		\hline\hline
		FP0 & 0 & 0 & NF & NF & 0 & NF & NF & \multicolumn{4}{c}{$ (y < 0, \varepsilon < 0) $}
		& $ -\varepsilon $ & $ 0 $ & $ 0 $ \\
		
		FPI & 0 & 0 & NF & NF & $ \frac{4}{3}\varepsilon $ & 0 & $ \frac{1}{2} $ &
		\multicolumn{4}{c}{$(y < 0, \varepsilon < 0) $} & $ \varepsilon $ &
		$ -\frac{1}{12} \varepsilon $ & $ \frac{1}{6} \varepsilon $ \\
		
		FPII & 0 & $ \frac{8}{3}\varepsilon $ & $ 1 $ & $ 1 $ & $ 0 $ & $ w^{*}(a^{*}) $ 
		& $ a^{*}(w^{*}) $ & \multicolumn{4}{c}{$(y<\frac{3}{2}\varepsilon, \varepsilon > 0) $}
		& $ \Lambda_{5}(w^{*})\varepsilon $ & $ \Lambda_{6}(w^{*})\varepsilon $ & $ 0 $ \\
		
		FPIII & 0 & $ \frac{8}{3}\varepsilon $ & $ 1 $ & $ 1 $ & $ 0.3505(0) \varepsilon $ 
		& $ 1.0819(3) $ & $ \frac{1}{2} $ & 
		\multicolumn{4}{c}{$(y<\frac{3}{2}\varepsilon, \varepsilon > 0) $} 
		& $  0.0438(1) \varepsilon $ & $ 0.2165(3) \varepsilon $ & $ 0.8083(8) \varepsilon $ \\
		
		FPIV & $ G(\Delta) $ & $ H(\Delta) $ & $ 1 $ & $ 1 $ & $ 0 $ & $ w^{*}(a^{*},\Delta) $ 
		& $ a^{*}(w^{*},\Delta) $ & \multicolumn{4}{c}{$(y>\frac{3}{2}\varepsilon, y > 0) $} 
		& $ \lambda_{5}(w^{*},\Delta) $ & $ \lambda_{6}(w^{*},\Delta) $ & $ 0 $ \\
		
		FPV & $ G(\Delta) $ & $ H(\Delta) $ & $ 1 $ & $ 1 $ & $ g_{3}^{*}(\Delta) $ & $ w^{*}(\Delta) $
		& $ \frac{1}{2} $ & \multicolumn{4}{c}{$(y>\frac{3}{2}\varepsilon, y > 0) $} 
		& $ \lambda_{5}(\Delta) $ & $ \lambda_{6}(\Delta) $ & $ \lambda_{5}(\Delta) $ \\
		\hline
		$ \underset{\alpha \rightarrow 0}{\text{FPIV}} $ & $ \frac{16 y}{9} $ & $ 0 $ & $ 1 $
		& $ 1 $ & $ 0 $ & $ 1 $ & NF & \multicolumn{4}{c}{$(y>\frac{3}{2}\varepsilon, y > 0) $} 
		& $ \frac{2}{3}(y - \frac{3}{2} \varepsilon) $ & $ \frac{1}{2}y $ & $ 0 $ \\
		
		$ \underset{\alpha \rightarrow 0}{\text{FPV}} $ & $ \frac{16 y}{9} $ & $ 0 $ & $ 1 $ 
		& $ 1 $ & $ \frac{16}{15} (\frac{3}{2} \varepsilon - y) $ & 
		$ \frac{1}{2} \left(\sqrt{1 + \frac{40 y}{4 y - \varepsilon}} - 1\right) $ & $ \frac{1}{2} $
		& \multicolumn{4}{c}{$(y>\frac{3}{2}\varepsilon, y > 0) $} & $ \lambda_{5}(y,\varepsilon) $
		& $ \lambda_{6}(y,\varepsilon) $ & $ \frac{2}{15} (\frac{3}{2} \varepsilon - y) $ \\
		\hline
		
		FPVI & $ 0 $ & $ \frac{8 \varepsilon }{3} $ & $ \infty $ & $ C $ & $ 0 $ & $ C $ & $
		\text{NF} $ & \multicolumn{4}{c}{unstable} & $ -\frac{\varepsilon}{3} $ &
		$ \frac{2(13+\sqrt{13})}{3(1+\sqrt{13})^{2}}\varepsilon $ & $ 0 $ \\
		
		FPVII & $ 0 $ & $ \frac{8 \varepsilon }{3} $ & $ \infty $ & $ C $ &
		$ \frac{8 \varepsilon }{15} $ & $ \frac{1}{6} \left(\sqrt{129}-3\right) $
		& $ \frac{1}{2} $ & \multicolumn{4}{c}{unstable} & $ 0.919918\varepsilon $ 
		& $ 0.295449\varepsilon $ & $ \frac{\varepsilon}{15} $ \\
		
		FPVIII & $ \frac{8 y}{3} $ & $ 0 $ & $ \infty $ & $ C $ & $ 0 $ & $ C $ & $ \text{NF} $
		& \multicolumn{4}{c}{unstable} & $ \frac{2}{3}(y - \frac{3}{2} \varepsilon) $ 
		& $ \frac{2 \left(\sqrt{13}+13\right)}{3 \left(\sqrt{13}+1\right)^2} y $ & $ 0 $ \\
		
		FPIX & $ \frac{8 y}{3} $ & $ 0 $ & $ \infty $ & $ C $ 
		& $ \frac{16}{15} (\frac{3}{2} \varepsilon - y) $ & 
		$ \frac{1}{2} \left(\sqrt{1 + \frac{40 y}{4 y - \varepsilon}} - 1\right) $ & 
		$ \frac{1}{2} $ & \multicolumn{4}{c}{unstable} & $ \lambda_{5}(y,\varepsilon) $ 
		& $ \lambda_{6}(y,\varepsilon) $ & $ \frac{2}{15} (\frac{3}{2} \varepsilon - y) $ \\

  		 FPX & \multicolumn{4}{c}{Any from above} & $ \frac{4}{3} \varepsilon $ & $ \infty $
  		& $ \frac{1}{2} $ & & & \multicolumn{4}{c}{unstable}  & 	
  		 \\	
		\hline\hline	
		
	\end{tabular}	
	\caption{List of all fixed points. We use the following abbreviations
	$ (\Delta=\{y,\varepsilon,\alpha\}) $, 
	$ \ G(\Delta) = \frac{16 y (2 y-3 \varepsilon )}{9 ((\alpha +2) y-3 \varepsilon )},
	\ H(\Delta) = \frac{16 \alpha  y^2}{9 ((\alpha +2) y-3 \varepsilon )}, \ C = (\sqrt{13}-1)/2 $ 
	and NF stands for "not fixed", i.e. a coordinate is not determined in 
	unambiguous fashion. Expressions that are not displayed are rather lengthy and a few
	explicit formulas can be found in Appendix \ref{app:explicit}. Coordinates of the
	fixed-points values of
	cNS charges are taken from \cite{AGKL17}. Further comments are found in the main text. 
	Numerical values are rounded to five decimal places, where the last number in
	brackets denotes the last rounded digit.}
	\label{tab:FP_tab1}
\end{table*}
The beta functions $ \beta_{g} $, describing the dependence of charges on the reference mass 
scale $ \mu $, are defined as
\begin{align}
  \beta_{g} = \D_{\mu} g.
\end{align}
For the DP process advected by turbulent flow they are found from
 (\ref{eq:norm_rel1}) and (\ref{eq:norm_rel2})
\begin{align}
  \beta_{g_{3}} = -g_{3} (\varepsilon + \gamma_{g_{3}}), \quad  \beta_{w} = -w \gamma_{w}, \quad 
  \beta_{a} = -a\gamma_{a}, 
  \label{eq:BetaF}
\end{align}
and similarly for charges of the cNS field \cite{AGKL17}
\begin{align}
\beta_{g_1} & =  -g_1\,(y + \gamma_{g_1}),
&\beta_{g_2} & =  -g_2\,(\varepsilon + \gamma_{g_2}),
\nonumber \\
\beta_{u} & =  -u\gamma_{u},
&\beta_{v} & = -v\gamma_{v},
\label{betagw}
\end{align}
which we quote here for completeness.

The explicit form of the RG functions can be
found in Appendix \ref{app:RG}. The asymptotic behavior is described 
by the IR fixed point (FP) $ g^{*} $ at which all the charges satisfy
\begin{equation}
  \forall g:\quad\beta_{g}(g^{*}) = 0.
  \label{eq:zeroes}
\end{equation}
Recall the abbreviations (\ref{eq:all_charges}), so in fact Eq.~(\ref{eq:zeroes}) is
 a system of seven connected equations for seven unknowns.
 Stability of the given fixed point is then determined by eigenvalues of the matrix
 of the first derivatives
\begin{equation}  
  \Omega_{ij} = \frac{\partial \beta_{g_{i}}}{\partial g_{j}} \bigg|_{g=g^{*}}.
  \label{eq:StabilityMatrix}
\end{equation}
In case of the IR attractive stable point, eigenvalues 
of this matrix have to be strictly positive \cite{Vasiliev}.
{\section{Results} \label{sec:results}}
{\subsection{Fixed points} \label{subsec:fixed}}

Apart from the Gaussian (free) fixed point FP0, eleven fixed points have been found, out of which 
four embody qualitatively new universality classes.  A list of all fixed points and eigenvalues
of the corresponding stability matrix (\ref{eq:StabilityMatrix}) are summarized
 in Table~\ref{tab:FP_tab1}. As expected, the trivial fixed point FP0 is  stable 
 for negative values of $ \varepsilon,y $, and for any $ \alpha $. The first non-trivial 
 fixed point FPI, represents the bare DP process with an irrelevant velocity field, which
 corresponds to the standard DP regime \cite{JT04,HHL08}. In 
 contrast to previous work \cite{AIK10,AKM11} this fixed point has been found unstable to
 any permissible values of $ y,\varepsilon,\alpha $. The following two fixed points FPII and 
 FPIII correspond to universality classes of the passive scalar and DP 
 advected by the thermal fluctuations of the velocity field. By a passive regime
 we henceforth have always in mind a regime for which DP interactions are irrelevant, i.e.
  $g_3^*=0$.
 For FPII and FPIII only a local part of the 
  random force for velocity is relevant. For fixed point FPII with irrelevant DP 
 interactions, we find that parameters $ w^{*} $ and $ a^{*} $ are not fixed. However,
 they are related to each other.
 Consequently, we do not have a fixed point, but rather a fixed line constrained by 
 the relation $ a^{*} = a^{*}(w^{*}) $. The plot of the function $ w^{*}(a^{*}) $ can be
 seen in Fig.~\ref{fig:wa} and several explicit results together with eigenvalues of the
 stability matrix are given in Appendix \ref{app:FPII}. If we restrict ourselves
 to the interval $ a^{*} \in \langle 0,1 \rangle $,  
 parameter $ w^{*}$ attains values from the interval $\langle 1, 1.0518(8) \rangle $ and the
 maximum value $w^{*}$ is reached at $ a^{*} = 1/2 $ (see Fig.~\ref{fig:wa}). We have 
 checked numerically that for any 
 accessible value of $ w^{*}(a^{*}) $ the eigenvalue $ \lambda_{6} $ is
 negative for $ \varepsilon > 0 $ and therefore we infer that this fixed 
 point is unstable. FPIII is stable in the region $ y<3\varepsilon/2 $ and
 $ \varepsilon > 0 $.

In the case with $ g_{1}^{*} > 0 $ and $ g_{2}^{*} > 0 $, two fixed points have been found. The
regime FPIV describes turbulent advection of a passive scalar  with irrelevant DP interactions.
 In a fashion similar to the fixed point FPII, parameters $ a^{*} $ and $ w^{*} $ 
 cannot be determined unambiguously, but again they are related and we have a
 whole line of fixed points. In contrast to FPII, this line depends on parameters
$ \Delta = \{y,\varepsilon,\alpha\} $. The explicit expression for $ a^{*}(w^{*}) $ can be found 
in Appendix \ref{app:FPIV}. The plot for $ (y,\varepsilon)=(4,1) $ and $ \alpha\in\{0,1,\infty\} $
is shown in Fig.~\ref{fig:wa} below. In the case of $ \alpha = 0 $, parameter $ w^{*} = 1 $, and
 parameter $ a^{*} $ remain undetermined. This is in accordance with results obtained in \cite{AGKL17}.
However, for $ \alpha > 0 $ we have $ w^{*} > 1 $ if $ a^{*} \in \langle 0,1 \rangle $.
This means that the universality of the fixed point FPIV changes with $ \alpha > 0 $. 
 Although it still corresponds to the turbulent advection by compressible turbulent flow, 
it is quantitatively
different from that obtained in \cite{AGKL17}. The plot for $ \alpha > 0 $ is symmetric 
around $ a^{*} = 1/2 $ as in FPII with maxima at $ w^{*} = 1.0518(8) $ for $ \alpha = 1 $
and $ w^{*} = 1.1085(4)$ for $ \alpha = \infty $. The most non-trivial fixed point FPV 
represents the universality class of DP advected by the compressible turbulent flow. 
In the case $ \alpha = 0 $, FPV becomes unstable for all $ (y,\varepsilon) $ and FPIV is stable 
for $ y>3\varepsilon/2 $ and $ \varepsilon>0 $. However, if $ \alpha > 0 $ the exact structure 
for $ g_{3}^{*}(\Delta)$, $w^{*}(\Delta) $ as well as eigenvalues for FPV and FPIV are too cumbersome 
 to carry out a direct analysis. This problem requires numerical solution of a complicated non-linear set
of connected differential equations \cite{wolfram}. Before we turn our attention
to numerical results, let us discuss other analytical results. 
\begin{figure}[h!]
	\includegraphics[width=8.cm]{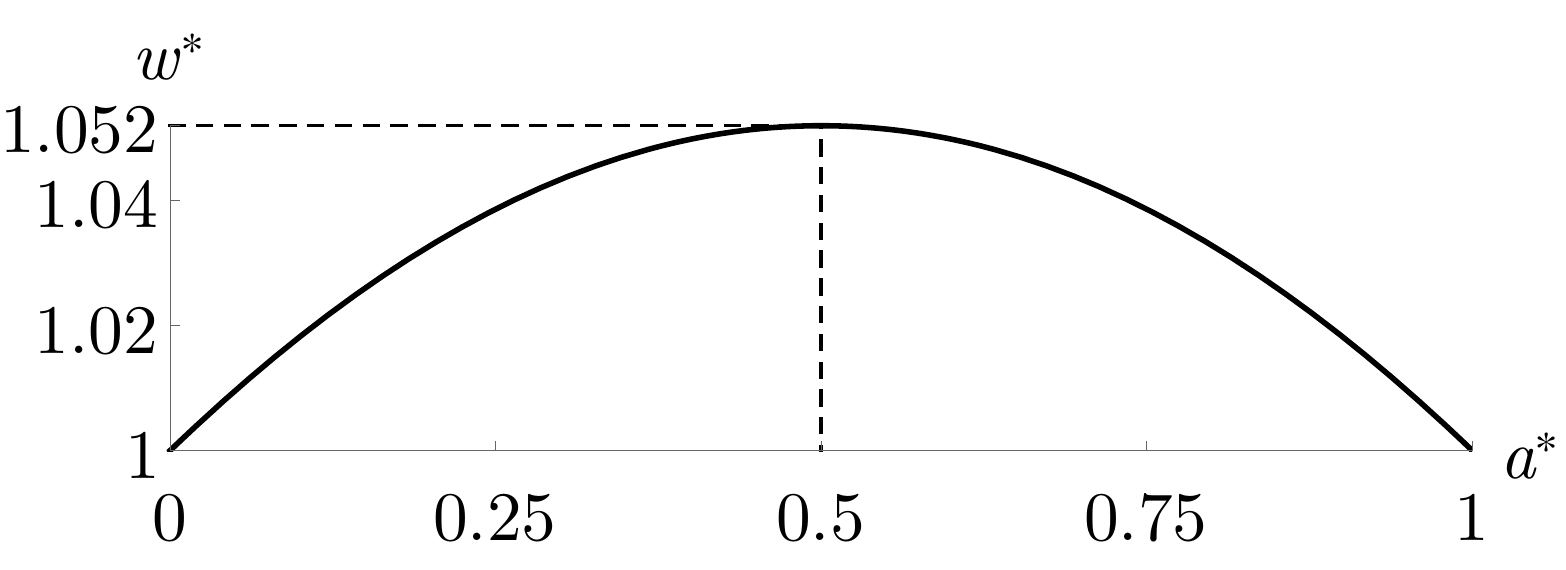}
	\includegraphics[width=8.cm]{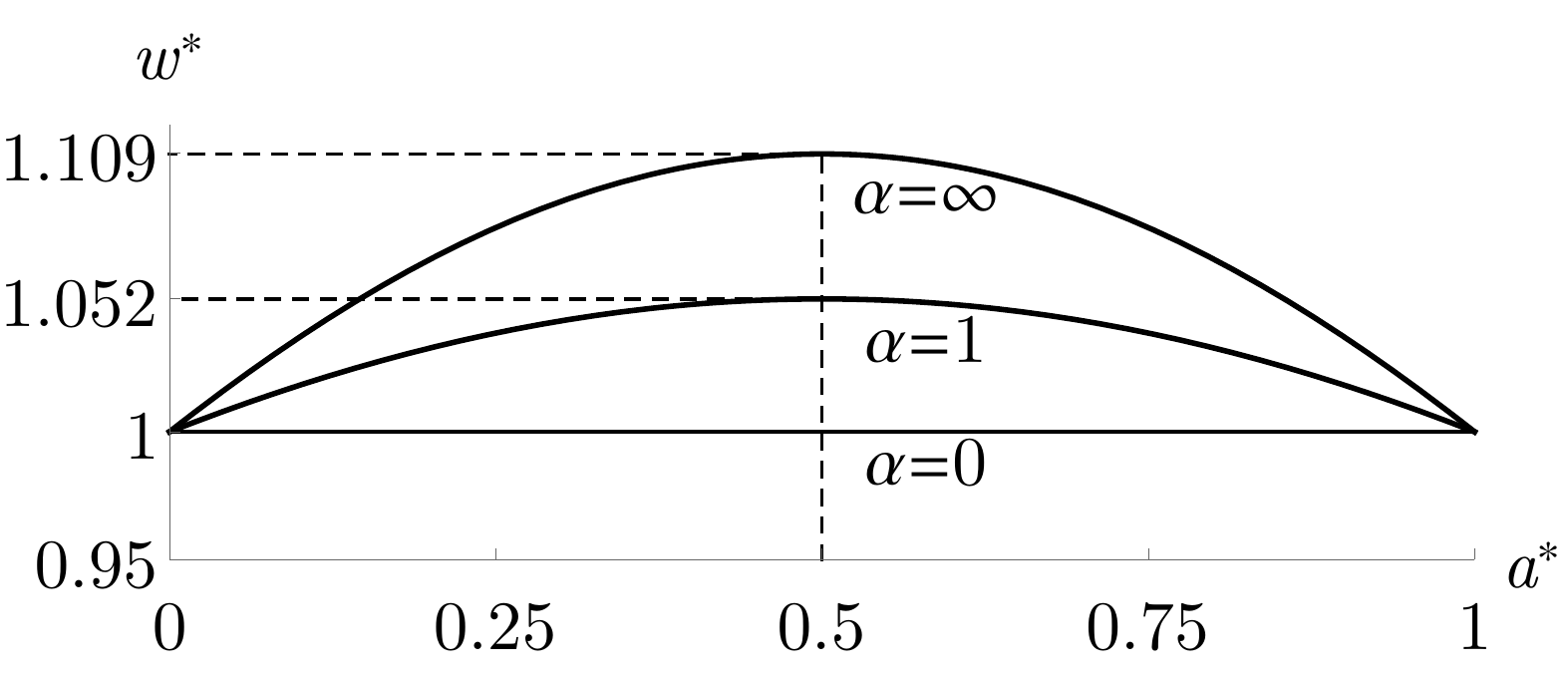}
	\caption{Relation between $ w^{*} $ and $ a^{*} $ for FPII (upper figure) and
	FPIV (lower figure), respectively. For FPII, the parameter
	$ w^{*} \in \langle 0,1.0518(8) \rangle $ 
	for any $ a^{*} \in \langle 0,1 \rangle $. The same situation occurs in the case of 
	FPIV, where the corresponding maximum depends on $ \Delta $. The lower plot is made for
	the physically relevant choice $ (y,\varepsilon)=(4,1) $
	and three different values of $ \alpha $. Note that all plots are symmetric around the
	point $ a^{*} = 1/2. $} 
	\label{fig:wa}
\end{figure}

In order to obtain the full set of fixed points, one has to analyze limiting cases $ \{ u,v,w \} 
\rightarrow \infty $ as well. 
As can be easily seen from (\ref{eq:prop1}) and (\ref{eq:prop2})
 in the limit $ u\rightarrow\infty $, the propagators $ \langle v_{i}v_{j} \rangle_{0} $ 
and $ \langle v_{i}v_{j}' \rangle_{0} $ become purely transversal and the propagators $ \langle v_{i}\phi_{j} 
\rangle_{0} $ and $ \langle v_{i}\phi_{j}' \rangle_{0} $ vanish. Hence, $ u \rightarrow \infty $ describes the
incompressible limit. In this case, four fixed points FPVI-FPIX are found. The first two, FPVI and 
FPVII, describe new universality classes of a passive scalar and DP advected by thermal 
fluctuations of the incompressible velocity field. These two fixed points are not present in the
previous studies, where the velocity field is generated by the incompressible NS equation \cite{AKM11}. 
The incompressible NS model does not possess divergence around $ d = 4 $ and therefore no fixed point 
with only $ g_{2} $ relevant (in our notation) can appear. Fixed points FPVIII and FPIX belong to
a  universality class similar to FPVI and FPVII except that the velocity field now describes the fully 
developed incompressible turbulent flow. It can be shown, however, that all fixed
points of the velocity field in the incompressible limit are unstable \cite{AGKL17}. 

The limiting case $ \tilde{v} \rightarrow \infty $ is uninteresting, since in the one-loop
approximation the parameter $ \tilde{v} $ enters only the $ \beta_{\tilde{v}} $ function and 
therefore fixed point values of other parameters are identical to the ones already mentioned 
above (see \cite{AGKL17}). Similarly to the previous case, 
 corresponding fixed points in the limit $ u \rightarrow \infty $  are unstable.

\begin{figure}[h!]
	\includegraphics[width=8.cm]{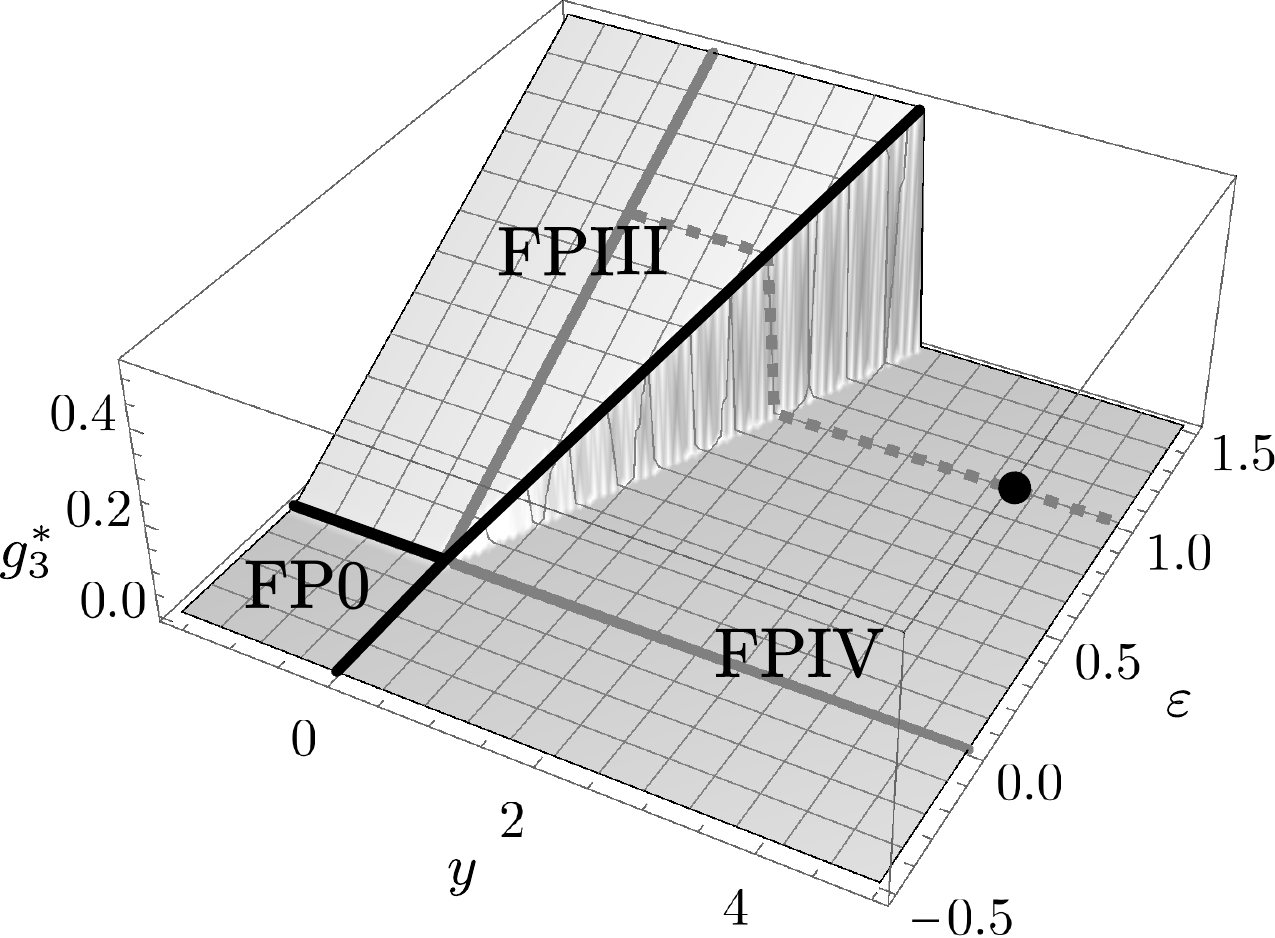} \\
	\includegraphics[width=8.cm]{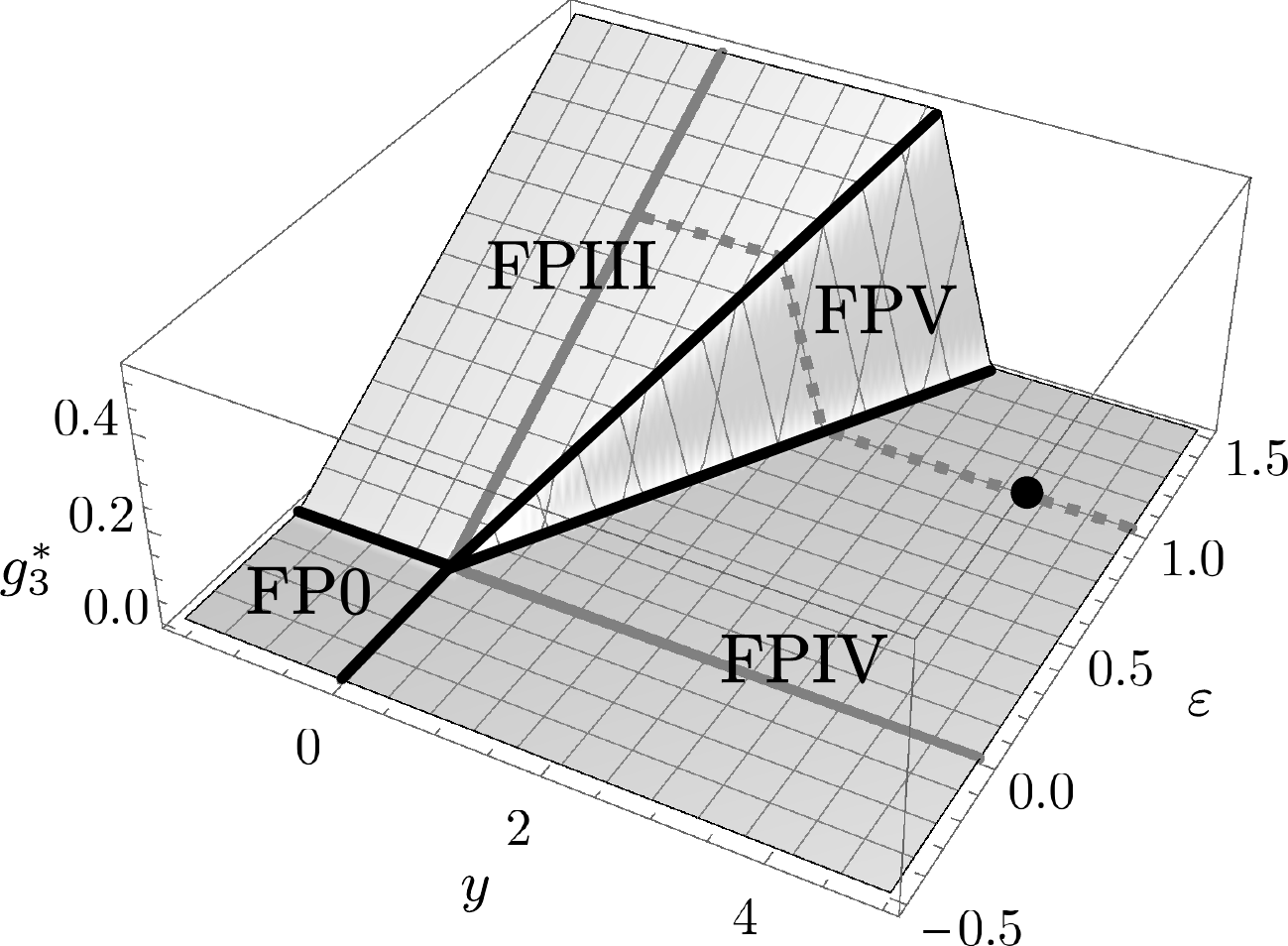} \\
	\includegraphics[width=8.cm]{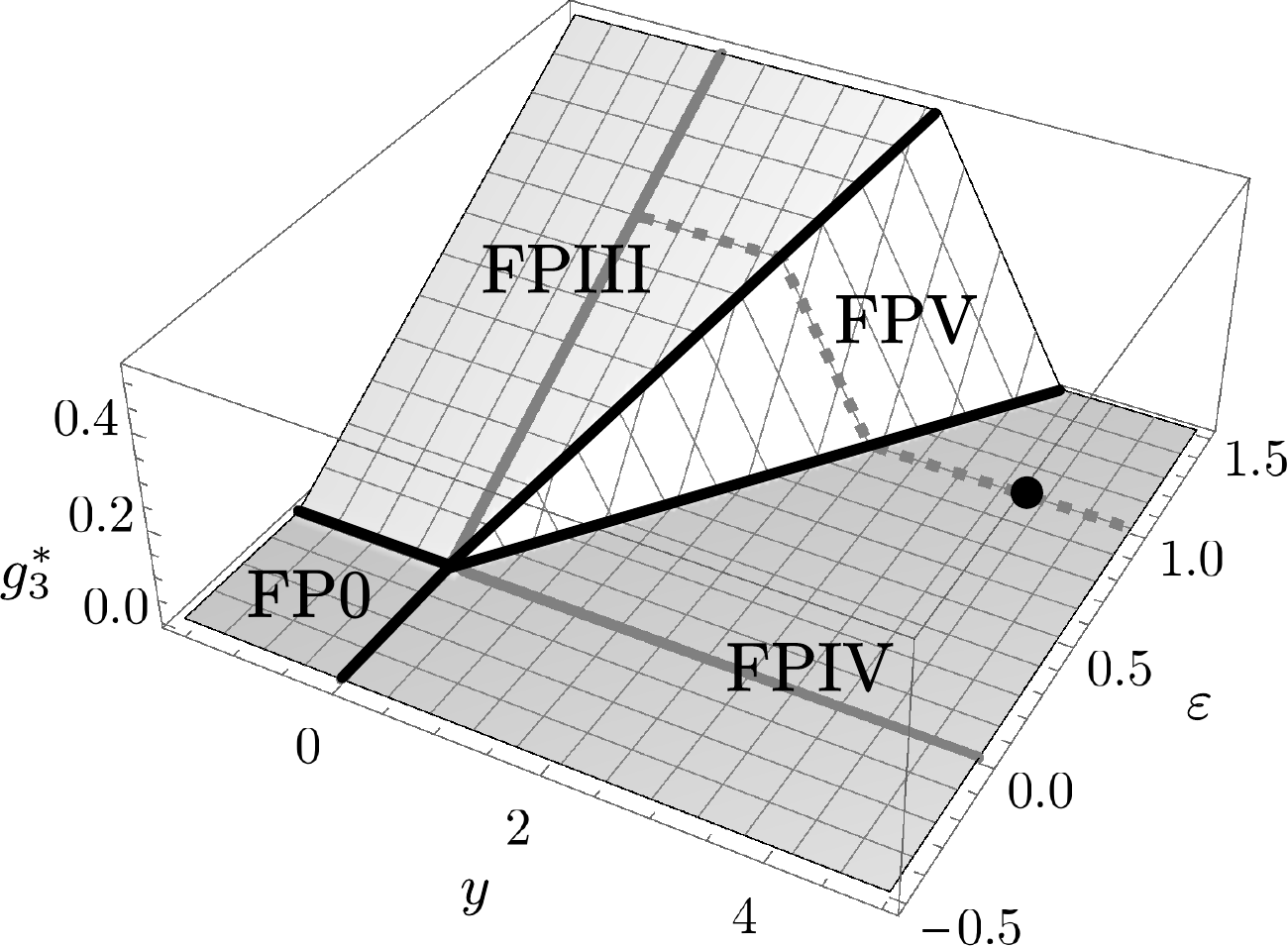}
	\caption{Numerical solution for the fixed point's coordinate of the charge 
	$ g_{3}^{*} $ for three different values of parameter $ \alpha \in \{ 0,1,\infty \} $ 
	(depicted in a given order from top to
	 bottom). Distinct regions of stability are separated by the solid black line, the 
	line $ \varepsilon = 1\, (d=3) $ is represented by the gray dashed line and black dot represents
	the case $ (y,\varepsilon)=(4,1) $. Technical details concerning
	boundary between regions of stability for FPIV and FPV can be found in Appendix~\ref{subsec:boundary}.} 
	\label{fig:g3}
\end{figure}
Finally, let us discuss the limit $ w \rightarrow \infty $. The
contributions to the DP renormalization constants
from the velocity field (terms proportional to $ g_{1} $ and $ g_{2} $) vanish
   and accordingly the process 
belongs to the universality class of DP with irrelevant velocity field (see Appendix \ref{app:const}). 
We have checked that for any fixed point values for the velocity field the fixed point FPX is unstable 
for any $ \Delta $.
\begin{figure}
	\includegraphics[width=8.cm]{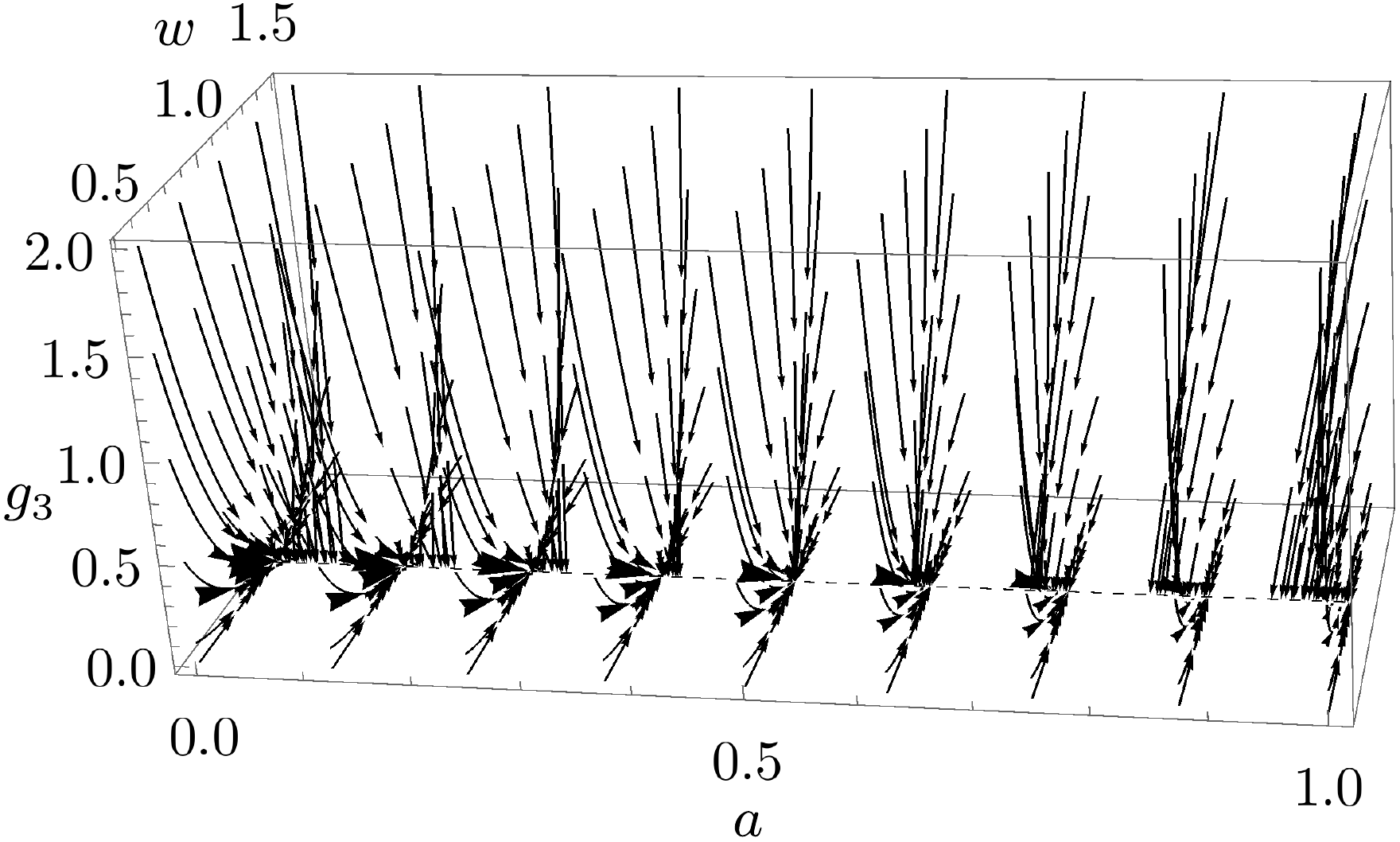} \\
	\includegraphics[width=8.cm]{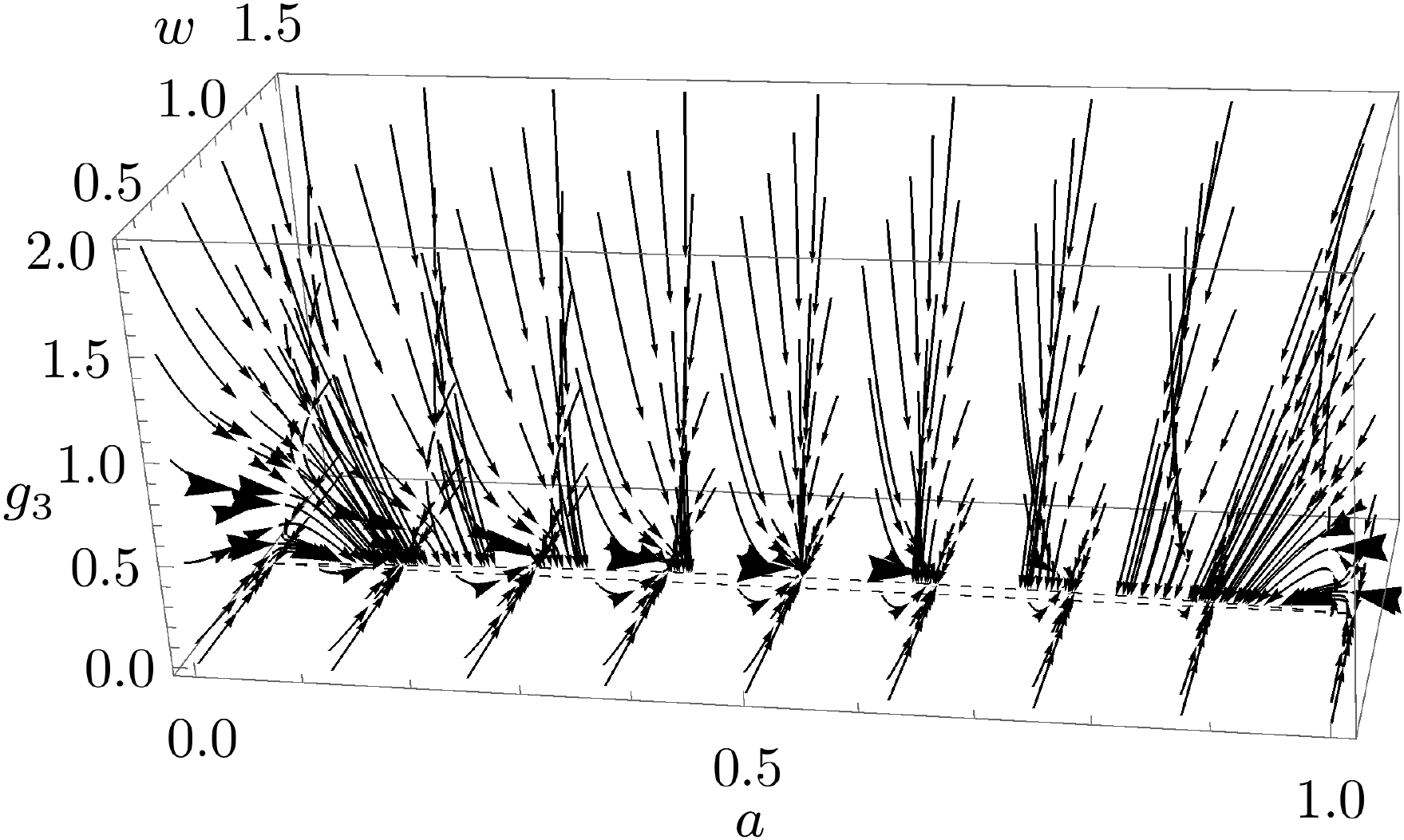} \\
	\includegraphics[width=8.cm]{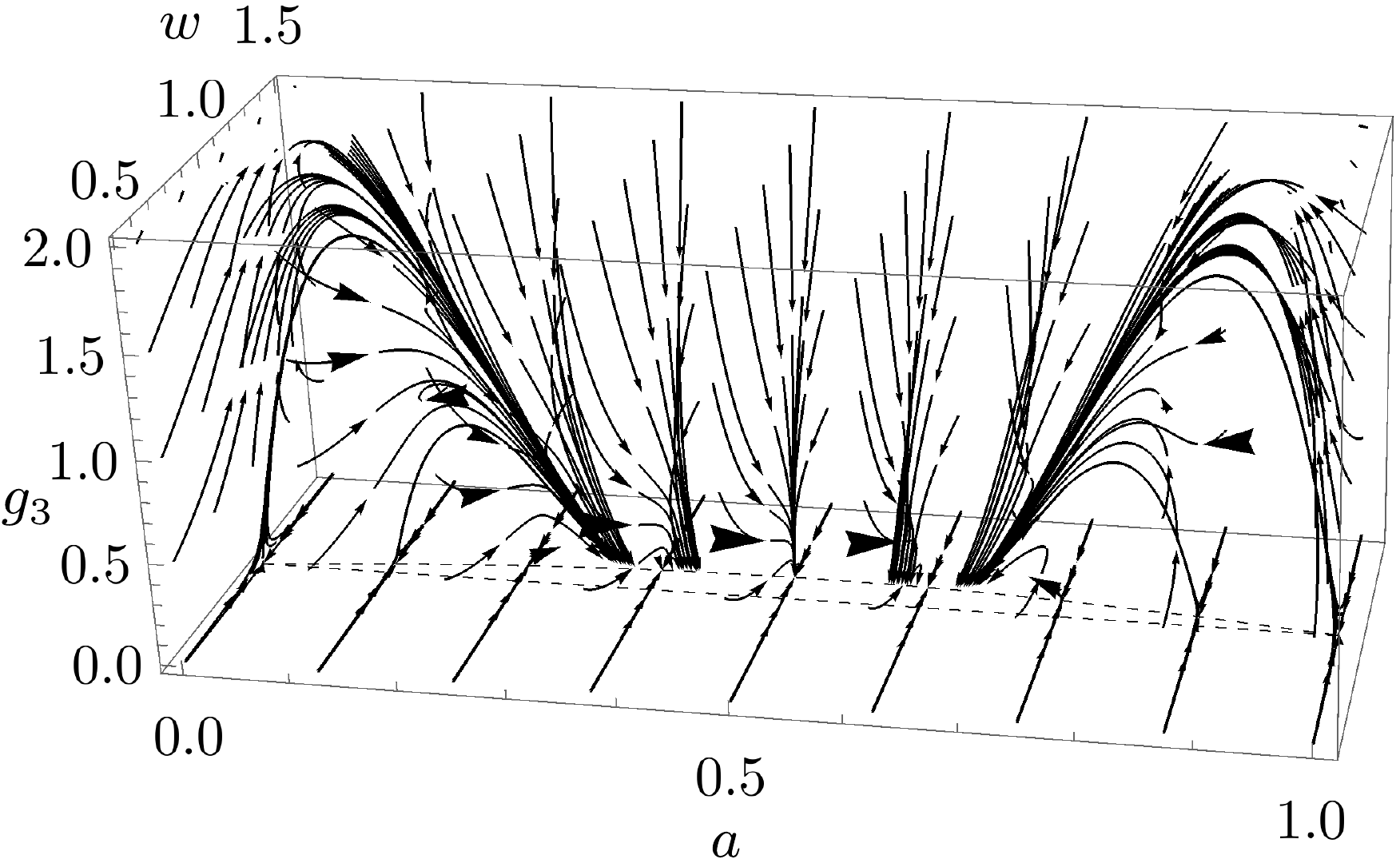}
	\caption{RG flow in the $ (a,w,g_{3}) $ plane for the physical values $ (y,\varepsilon)=(4,1) $
	and for $ \alpha\in\{0,1,\infty\} $ (from top to the bottom). By increasing the value of 
	$ \alpha $, the fixed (dashed) line shifts from $ w = 1 $.} 
	\label{fig:RGflow}
\end{figure}

To confirm the restricted picture obtained in analytical fashion, we have numerically sought
  fixed-points solutions of $\beta$ functions. Results for the coupling constant $ g_{3}^{*} $ in the
$ (y,\varepsilon) $ plane for three different values of parameter $ \alpha $ are shown
 in Fig.~\ref{fig:g3}. These RG flows are calculated with initial conditions
$ (g_{1},g_{2},u,v) = (1,1,1,1) $ for the cNS charges. Varying initial conditions might, 
of course, change the structure of the RG flow, but the universal quantities have to remain 
unchanged (for positive initial values). In the case of purely transversal random force 
$ \alpha = 0 $, only three stable fixed points have been found FP0, FPII and FPIV, what is in 
 accordance with our analytical results. For the physically relevant
 values $ (y,\varepsilon) = (4,1) $ the system belongs to the universality class of passive scalar
advected by the compressible turbulent flow. By increasing the value of $ \alpha $, the existence 
of another fixed point FPV  emerges. The region of stability for FPV gets larger with an increasing 
$ \alpha $. In the limit $ \alpha \rightarrow \infty $ (pure longitudinal random force
scenario) the boundary between FPIV and FPV does not
cross the physical point $ (y,\varepsilon) = (4,1) $. Therefore, we do not observe any crossover 
between universality classes by changing the structure of the velocity field random force. 
\begin{figure}
	\includegraphics[width=6.5cm]{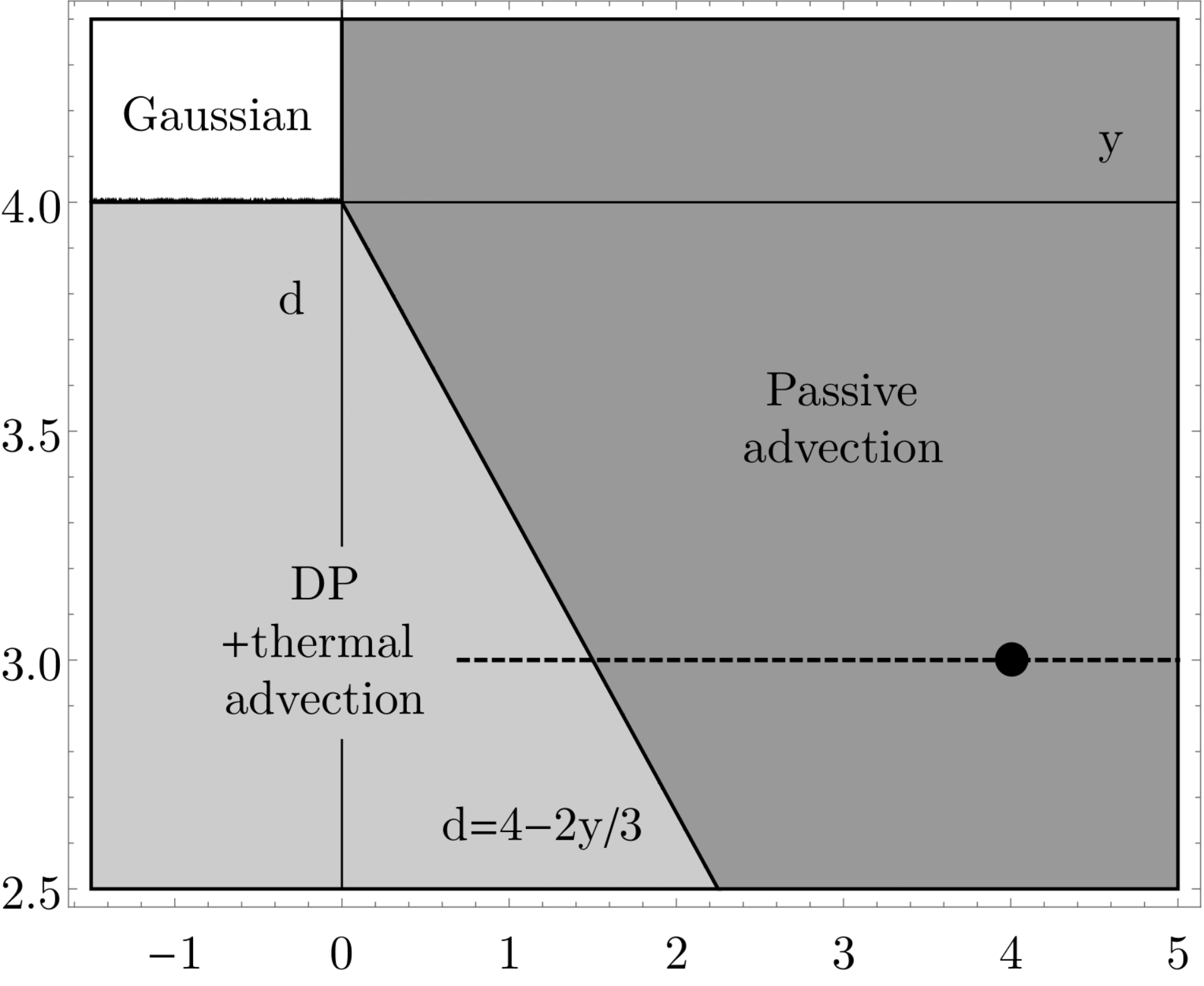}
	\includegraphics[width=6.5cm]{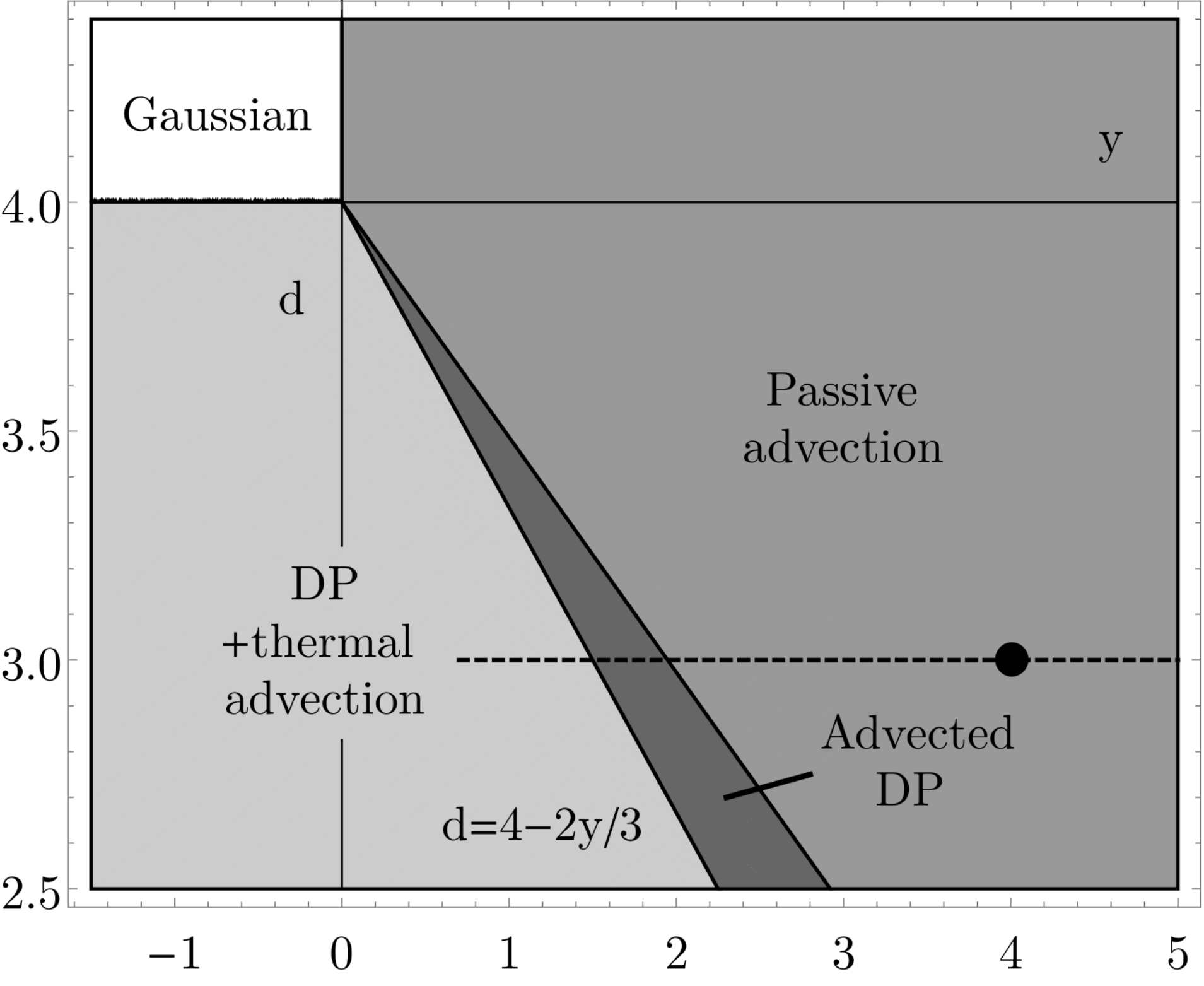}
	\includegraphics[width=6.5cm]{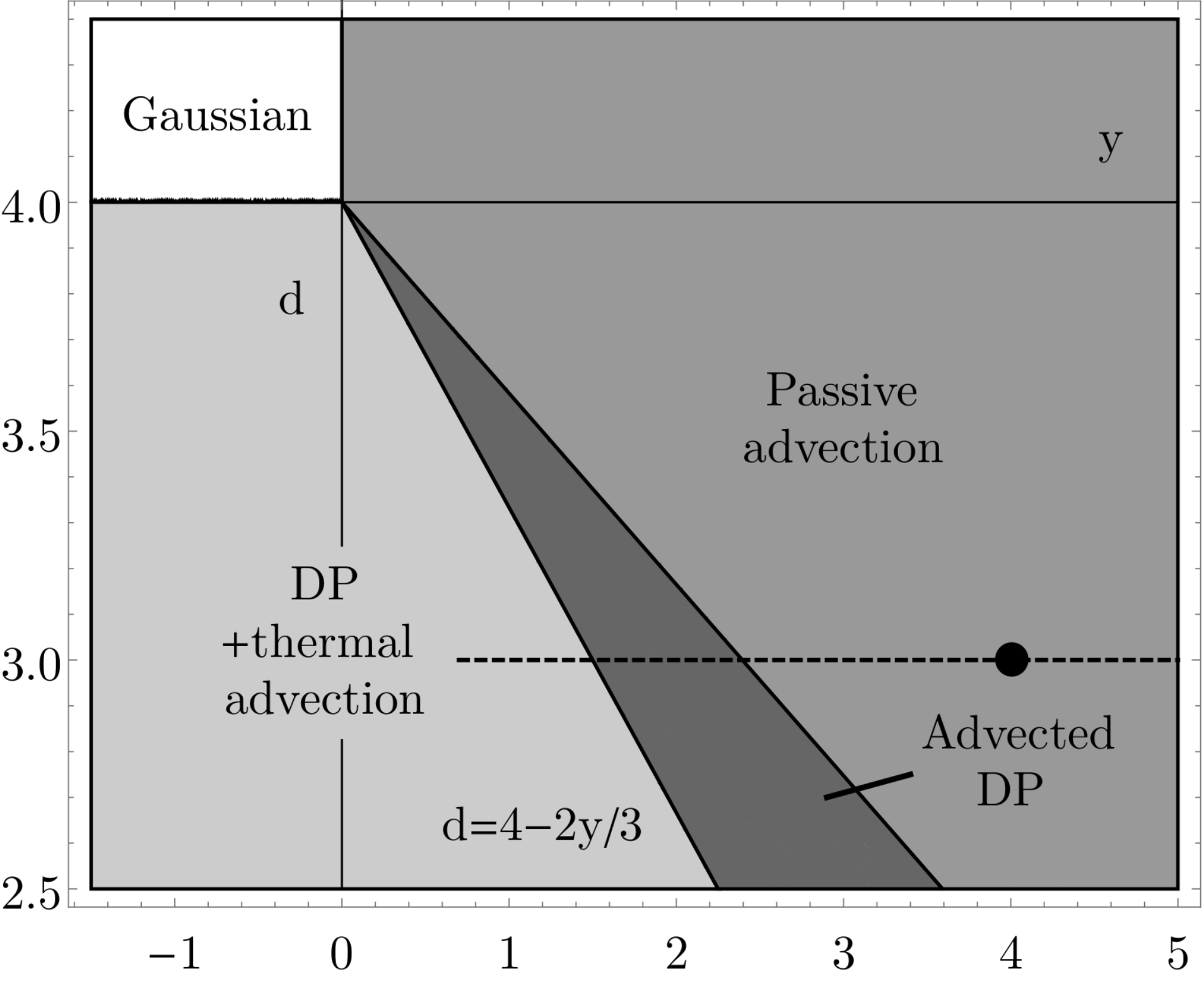}
	\caption{ Phase portraits in the $ (d,y)- $ plane for the
	following values  $ \alpha \in \{0,1,\infty\} $
	(from top to the bottom). For $ \alpha > 0 $ four regimes are present and in the limit
	$ \alpha \rightarrow 0 $ advected DP regime vanishes. The case $d=3\,  (\varepsilon = 1) $ is
	denoted with the dashed line, so that the crossover with growing $\alpha$ between
	distinct regimes is clearly visible. The most relevant point $ (d,y)=(3,4) $ (denoted by the black dot)
	belongs to the universality class of passive advection. 
	} 
	\label{fig:PD}
\end{figure}
A numerical calculation of the RG flow for the physical values of parameters 
$ (y,\varepsilon)=(4,1) $ can be seen in Fig.~\ref{fig:RGflow}. We have again performed a
 calculation for three different values of $ \alpha $. For $ \alpha = 0 $ one immediately observes
that there is an entire line of fixed points with $ g_{3}^{*} = 0, \ w^{*} = 1 $ and $ a^{*} $ not
fixed, denoted by dashed line. Moreover, the RG flow is symmetric around the plane $ a^{*} = 1/2 $ and the 
further from the center the flow begins, the more attracted towards the center it is.
The large value $\alpha$ takes, the stronger attraction is observed.
 This can be particularly seen in the 
case of $ \alpha = 1 $. In addition, the line of fixed points shifts away from the former line to 
$ w^{*}=w^{*}(a^{*}) $. In the case $ \alpha \rightarrow \infty $ even the stability of this
line changes, so that regions around $ a^{*}=0 $ and $ a^{*}=1 $ become unstable.
\begin{table*}
	\centering
	\def\arraystretch{1.75}
	\begin{tabular}{ | c || c c c c | c c c | }
		\hline \hline
		FP$\!\!	\biggl/\!\! \gamma^*|\text{exp.} $ & \multicolumn{2}{c}{$ \gamma_{\psi}^{*},\gamma_{\psi'}^{*} $} & $ \gamma_{\tau}^{*} $ & $ \gamma_{D}^{*} $ & $ \tilde{z} $ & $ \Theta $ & $ \delta, \ \delta' $ \\
		\hline\hline
		FP0 & \multicolumn{2}{c}{$ 0 $} & $ 0 $ & $ 0 $ & $ 1 $ & $ 0 $ & $ 1 - \frac{1}{4}\varepsilon $ \\
		
		FPI & \multicolumn{2}{c}{$ -\frac{1}{12} \varepsilon $} & $ -\frac{1}{4} \varepsilon $ & $ \frac{1}{12} \varepsilon $ & $ 1 + \frac{1}{14} \varepsilon $ & $ \frac{1}{12} \varepsilon $ & $ 1 - \frac{1}{4} \varepsilon $ \\
		
		FPII & \multicolumn{2}{c}{$ \gamma_{\psi}(a^{*},\varepsilon) \neq \gamma_{\psi'}(a^{*},\varepsilon) $} & $ -\frac{1}{2}\varepsilon $ & $ \frac{1}{2}\varepsilon $ & $ 1 + \frac{1}{4}\varepsilon $ & $ T(a^{*})\varepsilon $ & $ 1 - \frac{\gamma_{\psi}(a^{*},\varepsilon)}{2} \neq 1 - \frac{\gamma_{\psi'}(a^{*},\varepsilon)}{2} $ \\
		
		FPIII & \multicolumn{2}{c}{$ -0.0603(6) \varepsilon $} & $ -0.5438(1) \varepsilon $ & $ \frac{1}{2} \varepsilon $ & $ 1 + 0.0109(5) \varepsilon $ & $ 0.0219(1) \varepsilon $ & $ 1 - \frac{1}{4} \varepsilon $ \\
		
		FPIV & \multicolumn{2}{c}{$ \gamma_{\psi}^{*}(a^{*},\Delta) \neq \gamma_{\psi'}^{*}(a^{*},\Delta) $} & $ -\frac{1}{3} y $ & $ \frac{1}{3} y $ & $ \frac{2}{2-y/3} $ & $ \Theta^{}(a^{*},\Delta) $ & $ \delta^{}(a^{*},\Delta) \neq {\delta'}^{}(a^{*},\Delta) $ \\
		
		FPV & \multicolumn{2}{c}{$ \gamma_{\psi}^{*}(\Delta) = \gamma_{\psi'}^{*}(\Delta) $} & $ -\frac{1}{3}y + \frac{1}{8}g_{3}^{*}(\Delta) $ & $ \frac{1}{3} y $ & $ \frac{2}{2-y/3} $ & $ \Theta(\Delta) $ & $ \delta(\Delta) = \delta'(\Delta) $ \\
		
		$ \underset{\alpha \rightarrow 0}{\text{FPIV}} $ & \multicolumn{2}{c}{$ 0 $} & $ -\frac{1}{3}y $ & $ \frac{1}{3}y $ & $ \frac{2}{2-y/3} $ & $ 0 $ & $ \frac{4-\varepsilon}{2(2-y/3)}  $ \\
		
		$ \underset{\alpha \rightarrow 0}{\text{FPV}} $ & \multicolumn{2}{c}{$ \frac{1}{30} (2y - 3 \varepsilon) $} & $ -\frac{1}{5}(y + \varepsilon) $ & $ \frac{1}{3} y $ & $ \frac{2}{2-y/3} $ & $ -\frac{3(2y-3\varepsilon)}{5(2-y/3)} $ & $ \frac{30+y-9 \epsilon}{15 (2 - y/3)} $ \\
		\hline
		FPVI & \multicolumn{2}{c}{$ 0 $} & $ -\frac{1}{3}\varepsilon $ & $ \frac{1}{3}\varepsilon $ & $ 1 + \frac{1}{6} \varepsilon $ & $ 0 $ & $ 1 - \frac{1}{12}\varepsilon $  \\
		
		FPVII & \multicolumn{2}{c}{$ -\frac{1}{30} \varepsilon $} & $ -\frac{2}{5} \varepsilon $ & $ \frac{1}{3} \varepsilon $ & $ 1 + \frac{1}{6} \varepsilon $ & $ \frac{1}{30}\varepsilon $ & $ 1 - \frac{1}{10} \varepsilon $ \\
		
		FPVIII & \multicolumn{2}{c}{$ 0 $} & $ -\frac{1}{3} y $ & $ \frac{1}{3} y $ & $ \frac{2}{2-y/3} $ & $ 0 $ & $ \frac{4-\varepsilon}{2(2-y/3)} $ \\
		
		FPIX & \multicolumn{2}{c}{$ \frac{1}{30} (2y - 3 \varepsilon) $} & $ -\frac{1}{5}(y + \varepsilon) $ & $ \frac{1}{3} y $ & $ \frac{2}{2-y/3} $ & $ -\frac{3(2y-3\varepsilon)}{5(2-y/3)} $ & $ \frac{30+y-9 \epsilon}{15 (2 - y/3)} $ \\
		\hline \hline		
	\end{tabular}
	\caption{Anomalous dimensions and  critical exponents for various fixed points 
	with shorthand notation
	$ \Delta=\{y,\varepsilon,\alpha\} $. Some gamma functions are not displayed due 
	to their cumbersome structure. Non-universality affects only anomalous dimensions for
	DP fields and critical exponents that originate from them. Exponents $ \delta $ and 
	$ \delta' $ differ only if the non-universality is present for $ a^{*} \neq 1/2 $. Corresponding
	values for fixed point FPX are not displayed since they are identical to FPI.}
	\label{tab:CS}
\end{table*}

For convenience, we have constructed
 a schematic phase portrait with regions of stability
 in Fig.~\ref{fig:PD} in the $ (d,y) $ plane. Different 
regions of stability are denoted by different shades of gray. For the physical values of parameters 
$ (d,y) = (3,4) $ the system lies within the regime of the passive scalar advected by the compressible
turbulent flow, where DP interactions are irrelevant. This phase portrait possesses a few differences from
results obtained by previous authors. In \cite{AIK10,AKM11} it was established that in the case of 
incompressible turbulence for the physical values of parameters, the model should belong to the universality
class of passive advection and DP is effectively irrelevant. With the account of the
effect of compressibility,
the region of stability for an advected DP expands and above a certain level of compressibility DP 
 interactions become relevant. This difference might be traced to the fact that in the present 
model the parameter $ \alpha $, which is responsible for the quantitative change of the phase 
portrait, does not generally describe the level of compressibility. Further, in
the previous work \cite{gawedzki00,A06} the parameter $ \alpha $ (in their notation) has to fulfill a certain condition
 involving scaling parameter of the velocity field.
{ \subsection{Critical scaling  } \label{subsec:CritScal} }
In this section we  discuss  universal scaling properties of the DP process advected by 
the fully developed compressible turbulent flow. At the critical point, the total scaling dimension 
of any quantity $ Q $ is given by the relation \cite{Vasiliev,Zinn}
\begin{align}
  \Delta_{Q} = d_{Q}^{k} + \Delta_{\omega} d_{Q}^{\omega} + \gamma_{Q}^{*}, \quad
  \Delta_{\omega} = 2 - \gamma_{D}^{*},
\end{align}
where $ \gamma_{Q}^{*} = \gamma_{Q}(g^{*}) $ is  a fixed point's value. 
Applying  a scaling analysis on definitions
(\ref{eq:CExtp1})-(\ref{eq:CExtp2}) we get the following expressions for critical
exponents~\cite{AK10,HHL08}
\begin{align}
  \Theta &= - \frac{\gamma_{\psi}^{*} + \gamma_{\psi'}^{*}}{\Delta_{\omega}},  
  &\tilde{z} & =  \frac{2}{\Delta_{\omega}} , \label{CS:eq.CE1} \\
  \delta & = \frac{d/2 + \gamma_{\psi}^{*}}{\Delta_{\omega}},
  &\delta' & =  \frac{d/2 +
  \gamma_{\psi'}^{*}}{\Delta_{\omega}}. 
  \label{eq:CE2}
\end{align} 
Anomalous dimensions and critical exponents of the model under consideration can be seen in
 Tab.~\ref{tab:CS}. A few analytical expression were too lengthy, so they are not displayed explicitly.

Let us first begin by discussing a presence of non-universality. It has turned out that the anomalous 
dimensions $ \gamma_{\tau}^{*} $ and $ \gamma_{D}^{*} $ are independent of charges 
$ a^{*}(w^{*}) $ in the case of all fixed points (see Appendix \ref{app:anomalousTD}).
As a result, the non-universality shows up only in anomalous dimensions $ \gamma_{\psi} $
and $ \gamma_{\psi'} $ for regimes FPII and FPIV. The only exception is FPV, 
where $ \gamma_{\tau}^{*} $ depends on $ \alpha $. It should also be  emphasized that the 
result $ \gamma_{D}^{*} = y/3 $ for regimes FPIV, FPV, FPVIII, FPIX and $ \gamma_{\tau}^{*}=-y/3 $ 
for regimes FPIV, FPVIII are in fact exact results. This follows from the fact, that in these cases 
$ \gamma_{D}^{*} $ and $ \gamma_{\tau}^{*} $ are calculated solely from $ \gamma_{\nu}^{*} $
which is known exactly due to non-renormalizability of the non-local part of the random force 
correlator (\ref{eq:vRFcorr}) \cite{AGKL17,Vasiliev}.

Although the rapidity symmetry (\ref{eq:DP_RS}) is generally broken, critical exponents 
describing density of species $ \delta $ and the survival probability $ \delta' $ are identical if
DP is relevant (FPI, FPIII, FPV). This is because anomalous dimensions for DP fields posses the 
following symmetry
\begin{align}
  \gamma_{\psi}(a) = \gamma_{\psi'}(1-a), \label{eq:gammaSymm}
\end{align}
and so they are equal for $ a^{*} = 1/2 $ (see Eq. (\ref{eq:GammaP}) and (\ref{eq:GammaPs})).
However, if the DP is irrelevant (FPII, FPIV) the parameter $ a^{*} $ does not necessarily reach the
fixed point value $ a^{*} = 1/2 $. The region of stability as well as the final value of $ a^{*} $
then depends on its initial value $ a $ in the RG flow and on parameters $ \Delta $ as it has been shown
in~Fig. \ref{fig:RGflow}. For any other fixed point value than $ a^{*} = 1/2 $ critical 
exponents $ \delta $ and $ \delta' $ differ.
\begin{figure}
	\includegraphics[width=8.cm]{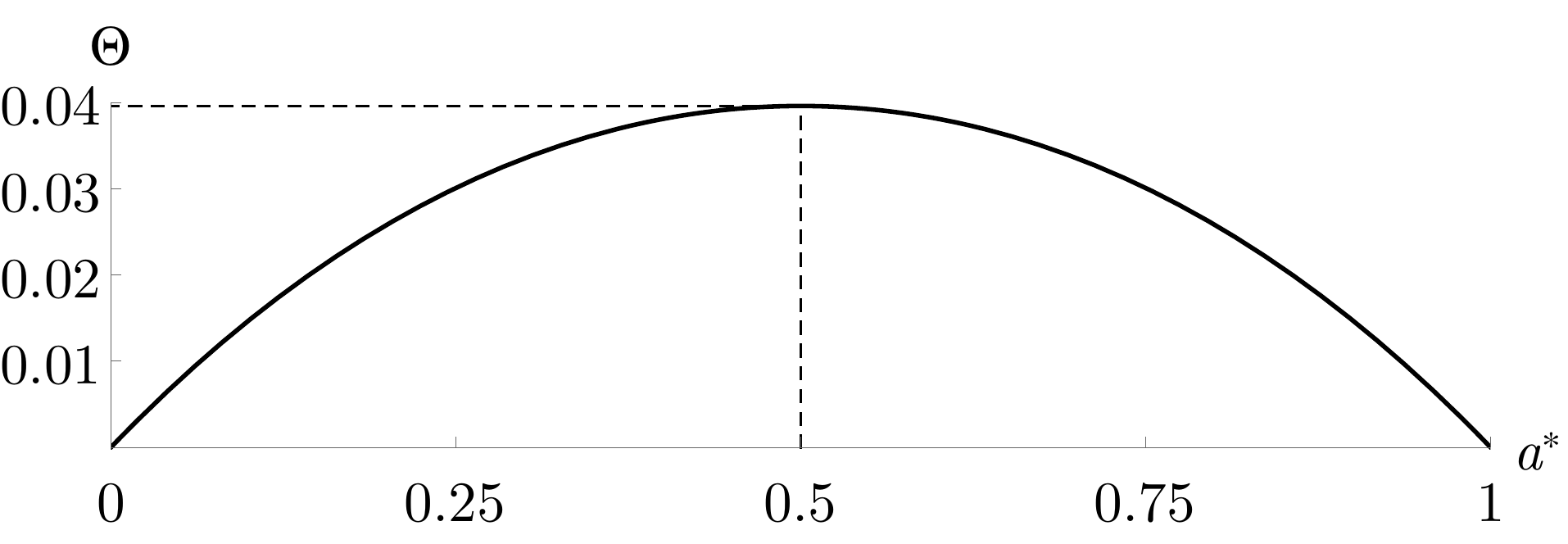}
	\includegraphics[width=8.cm]{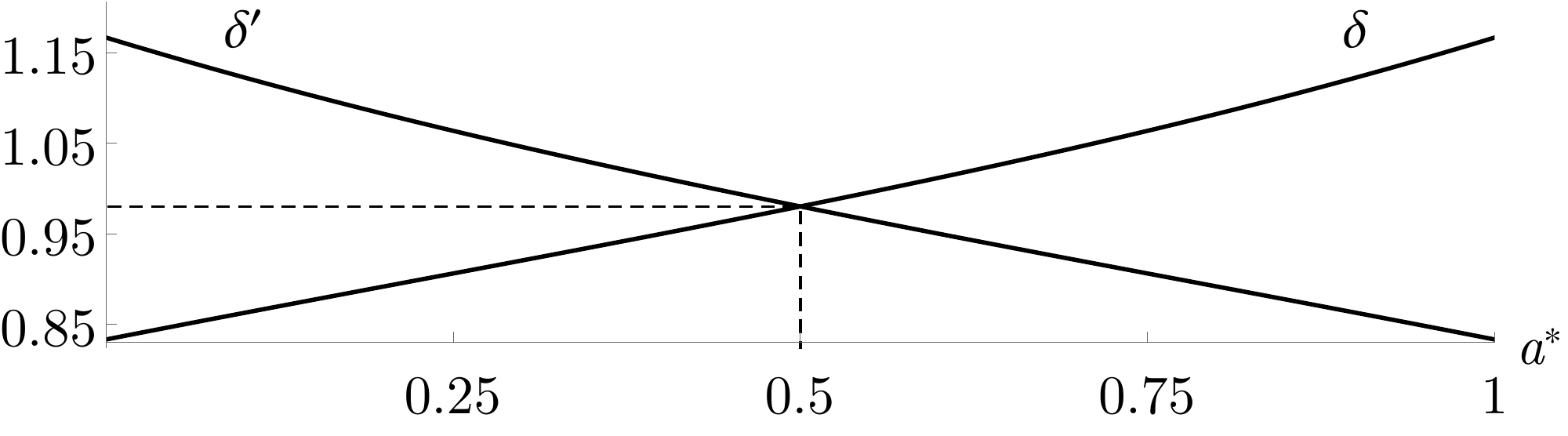}
	\caption{Critical exponents for FPII for $ \varepsilon = 1 $ as a 
	function of $ a^{*} $.}
	\label{fig:expII}
\end{figure}
Scaling properties of the Gaussian and DP fixed point (FP0, FPI) are in agreement with \cite{JT04}. 
Results of FPII and FPIII universality class have not been obtained before, but these fixed
points are unstable. Critical exponents $ \Theta, \delta $ and $ \delta' $ for FPII are
 depicted in Fig.~\ref{fig:expII}. Critical exponent $ \Theta $ is symmetric around $ a^{*} = 1/2 $ 
and exponents $ \delta $ and $ \delta' $ are symmetric within each other with respect to
$ a^{*} = 1/2 $. The final expressions of $ \tilde{z} $ for regimes FPIV and FPV are
exact. Note that in \cite{AIK10} authors do not normalize the definition for
$ \mathcal{R} $ with the expression $ \mathcal{N} $. As a result, a different exponent $ \tilde{z} $ 
 is obtained. For the physical value of parameter $ y = 4 $ we obtain $ R^{2} \sim t^{3} $ 
which is in agreement with the Richardson law $ \dRM R^{2} / \dRM t \sim R^{4/3} $ for turbulent
diffusion \cite{turbo,davidson}. Explicit results for other exponents and anomalous dimensions are in the
case of FPIV and FPV too long for an analytical analysis. We discuss a numerical calculation of
critical exponents later.

Scaling properties of the first two universality classes in the incompressible limit (FPVI, FPVII) 
have not been found in the previous work~\cite{AKM11}. This is due to the fact, that the
incompressible model of NS turbulence does not posses divergence around $ d = 4 $ and 
therefore a fixed point determined solely by $ g_{2} $ (in our notation) does not exist. 
The results of the other two universality classes in the incompressible limit (FPVIII, FPIX) 
are in agreement with the results obtained in~\cite{AKM11}.

A very intriguing result is 
that the universality class of the passive scalar and DP advected by the compressible 
turbulent flow (FPIV,FPV) coalesces with the incompressible limit for $ \alpha \rightarrow 0 $. 
 The reason for this result may be related to the fact that in the limit 
$ \alpha \rightarrow 0 $ the model (\ref{eq:cNS_action}) for the fully developed compressible
turbulence shows an incompressible Kolmogorov $ -5/3 $ energy spectrum \cite{ANU97}.

\begin{figure}
	\includegraphics[width=7cm]{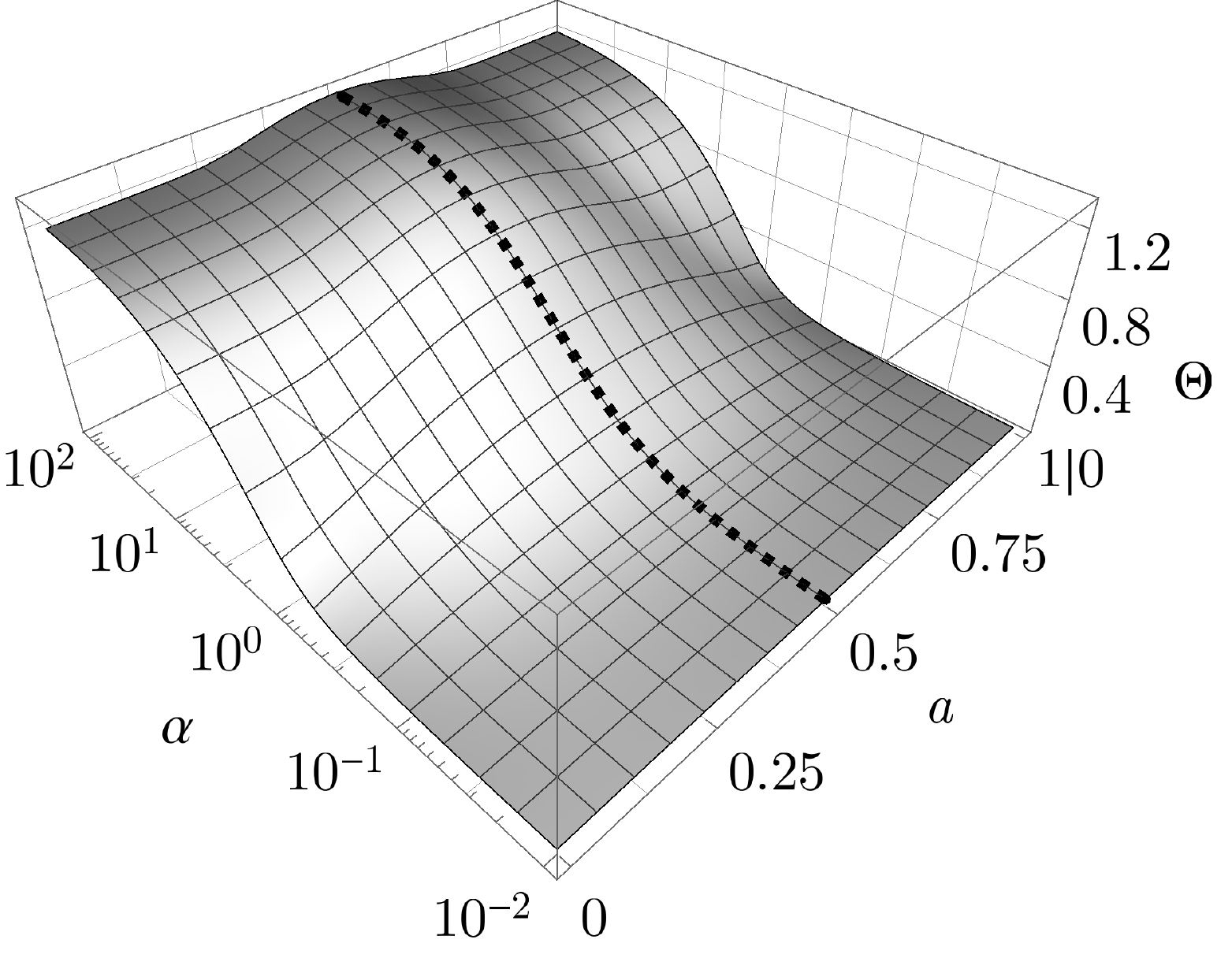} \\
	\includegraphics[width=7cm]{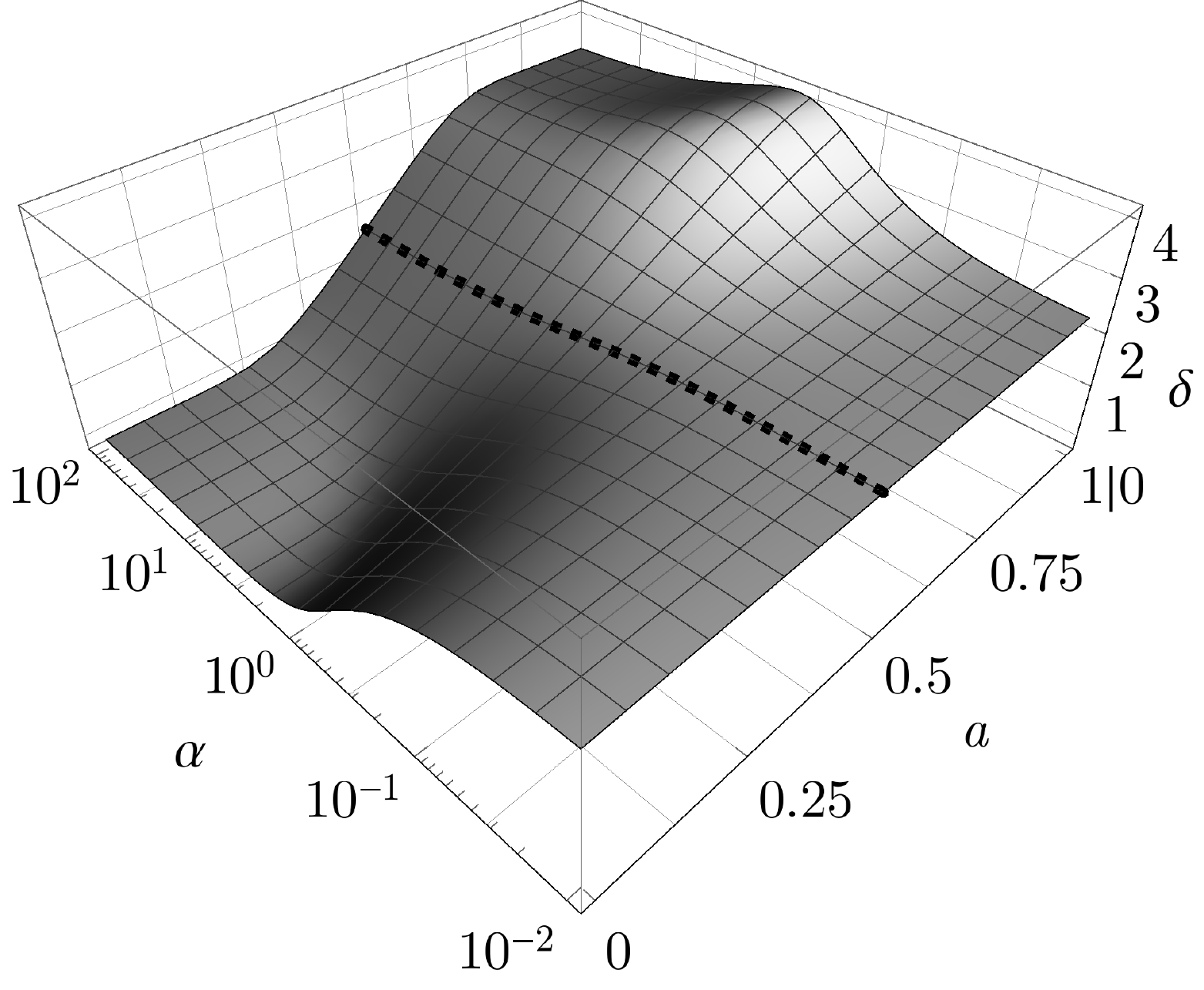} \\
	\includegraphics[width=7cm]{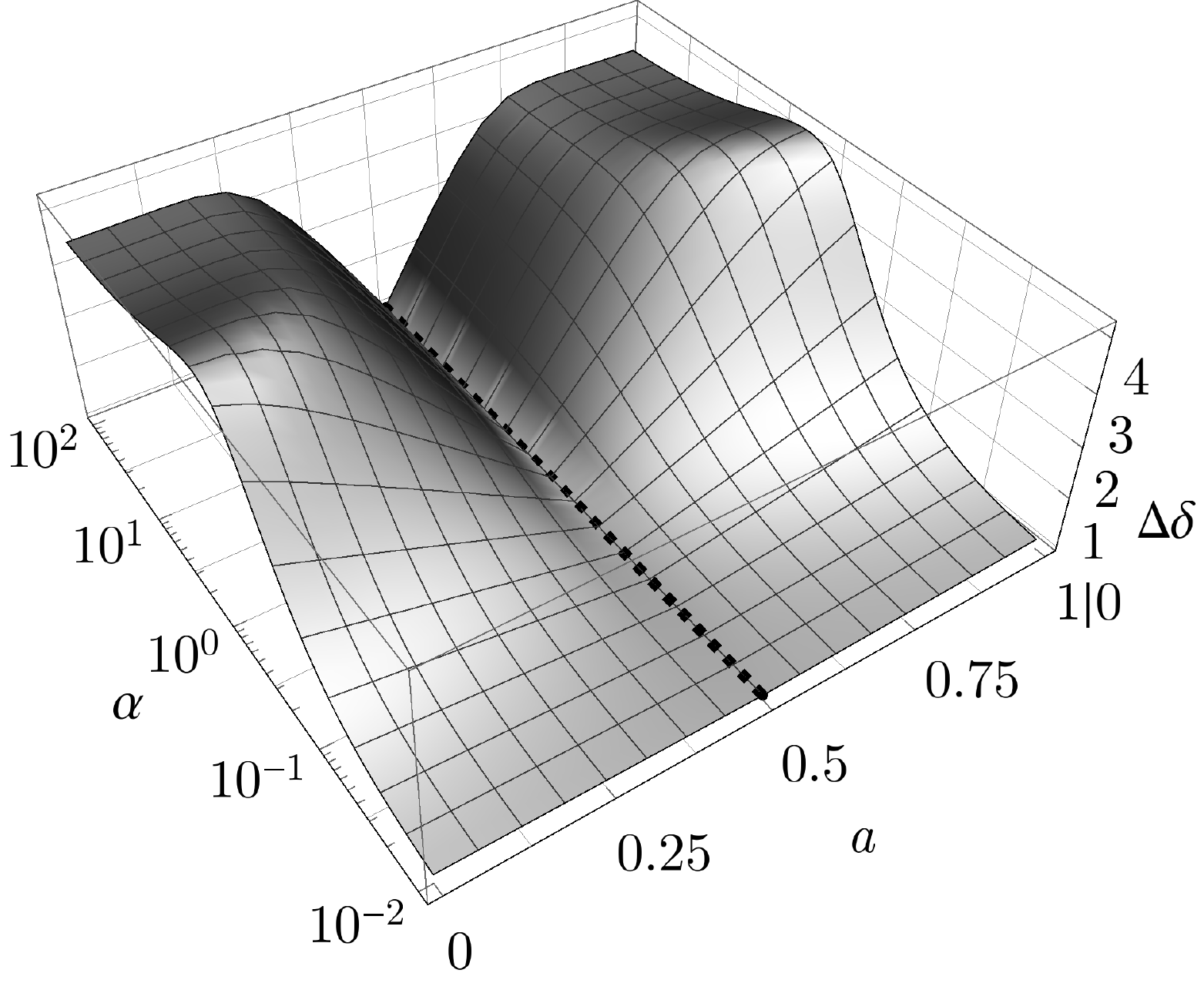}
	\caption{Numerical solution for critical exponents $ \Theta, \delta $ and the
	difference $ \Delta \delta = |\delta - \delta'| $ (from top to the bottom). Result
	for the exponent $ \delta' $ was found horizontally symmetric to $ \delta $ with 
	respect to the plane $ a=1/2 $ (not displayed). For $ \alpha \rightarrow 0 $ exponents
	tend to the incompressible limit and the difference $ \Delta \delta $
	vanishes. }
	\label{fig:expIV}
\end{figure}

We have computed critical exponents numerically for the physical $ (y,\varepsilon) = (4,1) $ as 
a function of $ \alpha $ and the initial value $ a $ in the RG flow. The result for exponents
$ \Theta,\delta $ and the difference $ \Delta\delta = |\delta-\delta'| $ are displayed 
 in Fig.~\ref{fig:expIV}. 
  It is noticeable, that the limit $ \alpha \rightarrow 0 $
 converges to the incompressible case and the results are universal, i.e. independent 
 of the initial value $ a $. For $a=1/2$ we do not observe any substantial change of $\delta$ or
 $\Delta\delta$ as a function of $\alpha$.
  By increasing $ \alpha $ a non-universality with respect to
 the initial value $ a $ emerges. The more $a$ deviates from $a=1/2$, the faster increase
 of $\delta$ and $\Delta\delta$ as function of $\alpha$, mainly in region $\alpha\ge 1$.
 A similar situation is observed in other
 stochastic models, e.g. in case of the stochastic magnetohydrodynamic
turbulence \cite{MHD01}, where the decisive r\^{o}le is played by a forcing decay-parameter.
 
 Increase of the value of the parameter $ \alpha $ also drastically changes 
 values of critical exponents. The exponent $ \Theta $ shows a rapid increase for 
 $ \alpha \gtrapprox 1 $, while exponents $ \delta $ and $ \delta' $ show
 very weak dependence for $ \alpha \lessapprox 1 $.
Moreover, we observe a certain degree of symmetry in critical exponents. For instance,
 the exponent describing a number of particles $ \Theta $ is symmetric with respect to 
 the transformation $ a \leftrightarrow 1-a $, 
i.e. symmetric around $ a=1/2 $. The graph for $ \delta' $ is horizontally symmetric to $ \delta $
with respect to the $ a = 1/2 $ plane. This is due to the symmetry (\ref{eq:gammaSymm}), which 
 follows from the fact that renormalization constants are symmetric with respect to the transformation
\begin{align}
  Z_{i}(a) &= Z_{i}(1-a), \quad \text{for} \ i = 1,2,3; \label{eq:RS1} \\
  Z_{4}(a) &= Z_{5}(1-a), \\
  Z_{1}(a) - a Z_{6}(a) &= (1-a) Z_{6}(1-a). 
  \label{eq:RS2}
\end{align}
The above symmetry was proposed in \cite{AK10} as a result of the time-reversal 
symmetry of the compressible Kraichnan-velocity ensemble. It is unclear whether relations
(\ref{eq:RS1})-(\ref{eq:RS2}) hold also in the present case, since the cNS
model (\ref{eq:cNS_action}) is not time-reversal invariant. A broken time-reversal invariance by the 
incompressible NS velocity ensemble has not been discussed in \cite{AKM11} although
the authors did obtain the same anomalous dimensions $ \gamma_{\psi} $ and $ \gamma_{\psi'} $.

{\section{Conclusions} \label{sec:concl}}

In this paper we have analyzed the critical behavior of the DP process
in the presence of the velocity fluctuations of an ambient environment. 
  It has been shown how a functional representation of the problem can
  be constructed. We have demonstrated that the model is multiplicatively renormalizable, which
  ensures utilization of methods of the field-theoretic renormalization group to 
  obtain relevant information about the large-scale and long-time behavior of the model.
Depending on the dimensionality of the space $ d $, the scaling exponent $y$ of the velocity field
 and pumping parameter $\alpha$,
 we have found eleven universality classes, out of which only four classes have IR stable fixed points and
 thus  are macroscopically relevant. In contrast to previous works \cite{AK10,AHKLM16}, the compressibility is present
 in the model at all stages. A special r\^{o}le is played by the parameter $\alpha$, which
 is associated with the pumping of longitudinal modes into the velocity flow.
 Compressibility changes stability of certain regimes, which can be deduced
 from the fixed-point structure. The pure DP regime of FPI becomes unstable
 and is effectively replaced by the regime of FPIII with the combined effect of DP interactions
 and local stirring of the velocity field.
 We observe that the mutual interplay between the
 nonlinear DP interactions and advection in the compressible flow gives rise
 to a nontrivial regime of FPV. Further, we see that the larger the input of compressible modes
 the larger stability region. Compressible modes thus enhance stabilization of the DP nonlinearities.
 We have also estimated that the effect is not strong enough to affect the physically relevant
 three-dimensional turbulent case.
 
 Additional consequences are obtained from a numerical analysis of the critical exponents.
  First, we see that the 
 parameter $\alpha$ has direct influence on the value of some critical exponents
 (see FPIV and FPV in Tab.~\ref{tab:FP_tab1}).
 The spreading exponent $\tilde{z}$ is not affected by $\alpha$ at all and for all fixed points it is
 expressed in terms of the universal quantities $\eps$ and $y$.
The incompressible limit $u\rightarrow\infty$  and the limit
$v\rightarrow\infty$
have been analyzed and the results are in agreement with previously
obtained results. 
 Non-universality is especially pronounced in the physically most relevant three-dimensional ($d=3$) 
 case with $y=4$. There, an enhancement  of
 the DP process is exhibited in the behavior of the exponent $\Theta$ (see Figs.~\ref{fig:expII} and 
 \ref{fig:expIV}). Although $\alpha$ does not affect the stability of FPV it does affect the critical
 exponents quite heavily. This fact can be explained by
the presence of compressible sinks into which particles are attracted \cite{Bouchaud}. 

Ffuture studies should involve higher-order loop calculations to confirm the physical picture
we have presented here. Moreover, there are
interesting questions regarding possible influence of the percolation process
 on the velocity field. Work on these topics is in progress.

\subsection*{Acknowledgments}
The authors thank to Paolo Muratore-Ginanneschi, Dhrubaditya Mitra and Michal Hnati\v{c} for
 many illuminating and fruitful discussions. The work was supported by VEGA grant No. $1/0345/17$ of
 the Ministry of
Education, Science, Research and Sport of the Slovak Republic,
 the grant of the Slovak Research and Development Agency under the contract No. APVV-16-0186
 and by the Ministry of Education and Science
 of the Russian Federation (the Agreement number 02.a03.21.0008).
\appendix
  {\section{Renormalization constants}   \label{app:const} }
  \subsection{Self-energy diagram}

  In this section we give an explicit example of a typical calculation of the Feynman diagram.
  Let us consider the self-energy diagram
\begin{align}
  I = \raisebox{-0.25cm}{ \includegraphics[width=2cm]{psp2.pdf}}. 
\end{align}
  We choose an external momentum $ p =(\mpp,\Omega) $ to flow through the lower
  propagator $ \langle \psi \psi' \rangle_{0} $ and the internal momentum $ k=(\mk,\omega) $
  flows clockwise in the loop. Using the standard Feynman diagrammatic technique based on
  Eqs.~(\ref{eq:prop1})-(\ref{eq:ver2}), we construct the following algebraic expression for the diagram
\begin{align}
  I(p) & =  \int \frac{\dRM^{d}k}{(2\pi)^{d}}
  \int \frac{ \dRM \omega}{ 2\pi }
  V_{\psi'v_{1}(-\mk)\psi(\mk-\mpp)} 
  \langle \psi \psi' \rangle_{0} (p-k)  \nonumber \\
  & \times V_{\psi'v_{2}(\mk)\psi(\mpp-\mk)} \langle v_{1} v_{2} \rangle_{0} ( k ),
\end{align}
  or in a detailed explicit form
\begin{align}
  I & = (-i)^{2} \int \frac{\dRM^{d}k }{(2\pi)^{d}} 
  \int \frac{ \dRM \omega}{ 2\pi } 
  \frac{-((p-k)_{1} + 
  ak_{1})(ak_{2} - p_{2})}{L( p - k )}  \nonumber \\
  & \ \times \left( P_{12}(\mk) \frac{d_{1}^{f}(\mk)}{|\epsilon_{1}( k )|^{2}} + 
  Q_{12}(\mk) d_{2}^{f}(\mk) \left| \frac{\epsilon_{3}( k )}{R( k )} \right|^{2} \right).
  \label{}
\end{align}
  First of all, let us note that the correction to the vertex function $ \G^{\psi\psi'} $ is
  independent of $ c_{0} $ and therefore we may set $ c_{0} = 0 $ in the above expression. 
  Expanding the terms in brackets,
  the calculation is effectively divided in an evaluation of
  two integrals $ I = I_{P} + I_{Q} $. 
\begin{align}
  I_{P} &= \int \frac{\dRM^{d}k }{(2\pi)^{d}} 
  \int \frac{  \dRM \omega}{2\pi } 
  \frac{T_{12}(\mpp,\mk) P_{12}(\mk) 
  d_{1}^{f}(\mk)}{L(  p - k ) |\epsilon_{1}( k )|^{2}}, \\
  I_{Q} &= \int \frac{ \dRM^{d}k }{(2\pi)^{d}}
  \int \frac{ \dRM \omega}{ 2\pi } 
  \frac{T_{12}(\mpp,\mk) Q_{12}(\mk)
  d_{2}^{f}(\mk)}{L( p - k ) |\epsilon_{2}( k )|^{2}},
\end{align}
  where we have introduced $ T_{12}(\mpp,\mk) = ((p-k)_{1} + ak_{1})(ak_{2} - p_{2}) $. Let us
  first calculate the $ I_{P} $ part. The calculation of the tensor structure yields
\begin{align}
   P_{12}(\mk) T_{12}(\mpp,\mk) &= \left(\frac{(\mpp \cdot \mk)^{2}}{k^{2}} - p^{2}\right).
   \label{eq:nConstPTS}
\end{align}
  Since the tensor structure is already proportional to $ p^{2} $, we can put $ \mpp = \Omega = \tau = 0 $ 
  in the rest of the calculation. The next step is to perform the frequency integration
\begin{align}
  \int \frac{\dRM\omega}{2\pi} \frac{1}{L( - k )|\epsilon_{1}( k )|^{2}} =&
  \frac{1}{2 (1 + w_{0}) \nu_{0}^{2} k^{4}}.
\end{align}
  Then, the expression $ I_{P} $ part is equal to 
\begin{align}
  I_{P} = \frac{1}{2 (1 + w_{0}) \nu_{0}^{2}} \int \frac{\dRM^{d} k}{(2\pi)^{d}}
  \left(\frac{(\mpp \cdot \mk)^{2}}{k^{2}} - p^{2}\right) \frac{d_{1}^{f}(\mk)}{k^{4}}.
\end{align}
  In order to carry out the momentum integration, we need the following formula 
  for isotropic integrals
\begin{align}
  \int \dRM^{d}k\,  k_{i}k_{j} f(k^{2}) = \frac{1}{d} \int \dRM^{d} k \, k^{2} f(k^{2}) 
  \label{eq:nConst_Int1}
\end{align}
  that allows us to perform a valid substitution $ (\mpp \cdot \mk)^{2} \rightarrow p^{2} k^{2} / d $ 
  in Eq.~(\ref{eq:nConstPTS}). We are thus left with simple $ d-$dimensional integrals of two
  types
\begin{align}
  \int_{m}^{\infty} \frac{\dRM^{d} k}{(2\pi)^{d}} \frac{k^{4-d-y}}{k^{4}} &= \overline{S_{d}} 
  \frac{m^{-y}}{y}, \\
  \int_{m}^{\infty} \frac{\dRM^{d} k}{(2\pi)^{d}} \frac{1}{k^{4}} &= \overline{S_{d}} 
  \frac{m^{-\varepsilon}}{\varepsilon},
  \label{eq:nConst_Int2}
\end{align}
  where $ \overline{S_d} \equiv S_{d}/(2\pi)^{d} $ and $ S_{d} = 2\pi^{d/2}/\G(d/2) $ is the 
  surface of a $ d $-dimensional sphere.   The final result then reads
\begin{align}
   \frac{I_{P}}{D_{0}p^{2}} = \frac{(1-d) \overline{S_{d}} }{2 w_{0} (1 + w_{0}) d} 
   \left( \frac{g_{10}m^{-y}}{y} + \frac{g_{20}m^{-\varepsilon}}{\varepsilon} \right).
\end{align}
In a similar way, the
tensor structure of the $ I_{Q} $ part yields
\begin{align}
  Q_{12}(\mk) T_{12}(\mpp,\mk) &= a_{0}(a_{0} - 1) k^{2} + (\mpp \cdot \mk) -
  \frac{(\mpp\cdot \mk)^{2}}{k^{2}},
  \label{eq:nConst_Q1}
\end{align}
  and the frequency integration gives
\begin{align}
  &\int \frac{\dRM\omega}{2\pi} \frac{1}{L( -k )|\epsilon_{2}( k )|^{2}} 
  \approx \frac{1}{2u_{0}(u_{0} + w_{0}) \nu_{0}^{2} k^{4}} \nonumber \\
  &\times \Bigg(1 + \frac{i\Omega + w_{0}(2(\mpp\cdot\mk) - p^{2} - \tau_{0})}{(u_{0} + w_{0}) k^{2}} + 
  \frac{4w_{0}^{2}(\mpp\cdot\mk)^{2}}{(u_{0} + w_{0})^{2} k^{4}} \Bigg), 
  \label{eq:nConst_Q2}
\end{align}  
  where we have already performed the Taylor expansion to the first order 
  in variables $ \Omega,\tau_{0} $, and to the 
  second order in the external momentum $ \mpp $. Multiplying expressions
  (\ref{eq:nConst_Q1}) and (\ref{eq:nConst_Q2}),
  keeping only terms proportional to $ \Omega,p^{2} $ and $ \tau_{0} $ and integrating over 
  the momentum $ \mk $, we finally obtain
\begin{align}
\label{1loopExample}
  &I_{Q} = \frac{\overline{S}_{d}}{2u_{0}(u_{0} + w_{0})^{2}} \left( g_{10} \alpha
  \frac{m^{-y}}{y} + g_{20} \frac{m^{-\varepsilon}}{\varepsilon} \right) \nonumber \\
  & \ \times \Bigg[ p^{2} D_{0} \left( \frac{w_{0} - u_{0}}{d w_{0}} +
  \frac{a_{0}(a_{0}-1)w_{0}}{(u_{0} + w_{0})} \left( \frac{4}{d} - \frac{u_{0}}{w_{0}} - 1 \right)
  \right) \nonumber \\
  & \hspace{0.8cm} + a_{0}(a_{0} - 1) \left( i\Omega - \tau_{0} D_{0} \right) \Bigg].
\end{align}
  
{\subsection{Renormalization constants} \label{subsec:normal}}

Calculation shows that the  general structure of the renormalization constants for the current model in the one-loop approximation 
can be written as follows
\begin{align}
\label{Zstructure}
  Z_{i} = 1 + z_{1}^{(i)}(r) \frac{g_{1}}{y} + z_{2}^{(i)}(r) \frac{g_{2}}{\varepsilon} +
  z_{3}^{(i)}(r) \frac{g_{3}}{\varepsilon},
\end{align}
where $ r = \{u,v,w,a\} $ and all coefficient functions $z_i$ in (\ref{Zstructure}) are analytic functions of the regulators $\varepsilon$ and $y$. It is convenient to 
express contributions of the type (\ref{1loopExample}) in the form of functions of renormalized parameters
with the use of relations (\ref{eq:norm_rel1})-(\ref{mult}). For simplicity, we adopt the normalization-point scheme
with the choice $\mu/m=1$ and calculate the coefficient functions $z_i$ in (\ref{Zstructure}) only at the leading order
of expansion in $\varepsilon$ and  $y$.
The resulting renormalization constants for the DP process are
\begin{align}
  Z_{1} =\ & 1 + \frac{a(1-a)}{2u(u+w)^{2}} G_{1} + 
  \frac{g_{3}}{8\varepsilon}, \\
  Z_{2} =\ & 1 + \frac{(1-d)}{2w(1+w)d} G_{2} +
  \frac{1}{2u(u+w)^{2}} \Bigg[ \frac{w - u}{dw} \nonumber \\
  & +\frac{a(a-1)w}{u + w} \left( \frac{4}{d} - \frac{u}{w} - 1 \right) \Bigg] G_{1}
  + \frac{d - 2}{8d} \frac{g_{3}}{\varepsilon}, \\
  Z_{3} =\ & 1 - \frac{a(a-1)}{2u(u+w)^{2}} G_{1} + \frac{g_{3}}{4\varepsilon}, \\
  Z_{4} =\ & 1 - \left( \frac{ (1 - a)^{2} }{2 uw(u + w)} + \frac{a(a - 1)}{u(u + w)^{2}} \right) 
  G_{1} + \frac{g_{3}}{2\varepsilon}, \\ 
  Z_{5} =\ & 1 - \left( \frac{a^{2} }{2uw(u + w)} + \frac{a(a - 1)}{u(u + w)^{2}} \right)  G_{1}
  + \frac{g_{3}}{2\varepsilon}, \\
  Z_{6} =\ & 1 + \frac{a(1-a)}{2u(u+w)^{2}} G_{1} + \frac{d(d-1)+3}{4ad} \frac{g_{3}}{\varepsilon},  
\end{align}
where we have introduced 
\begin{equation}
  G_{1} = \left(\frac{\alpha g_{1}}{y} + \frac{g_{2}}{\varepsilon}\right), \quad 
  G_{2} = \left(\frac{g_{1}}{y} + \frac{g_{2}}{\varepsilon}\right).
\end{equation}
The remaining Feynman diagrams can be analyzed in a similar fashion.
 { \section{RG functions } \label{app:RG} }
 { \subsection{Anomalous dimensions } \label{app:ADim} }

  Anomalous dimensions are found from the renormalization constants in the following way.
  From Eqs.~(\ref{eq:AD}) and (\ref{Zstructure})the general form of the 
  anomalous dimensions at the one-loop order readily follows
\begin{align}
  \gamma_{i} = - z_{i}^{(1)}(r) g_{1} - z_{i}^{(2)}(r) g_{2} - z_{i}^{(3)}(r) g_{3}\,.
\end{align}
Relations
  between anomalous dimensions are found from Eqs.~(\ref{eq:NC1})-(\ref{eq:NC3})
\begin{align}
  \gamma_{\psi} &= (\gamma_{1} + \gamma_{5} - \gamma_{4})/2, 
  &\gamma_{D}& = \gamma_{2} - \gamma_{1}, 
  \label{eq:gammaRel1} \\
  \gamma_{\psi'} &= (\gamma_{1} - \gamma_{5} + \gamma_{4})/2, 
  &\gamma_{\tau}& = \gamma_{3} - \gamma_{2}, \\
  \gamma_{\lambda} &= (\gamma_{4} + \gamma_{5} - \gamma_{1})/2 -\gamma_{2}, 
  &\gamma_{g_{3}}& = 2 \gamma_{\lambda}, \\
  \gamma_{w} &= \gamma_{D} - \gamma_{\nu} = \gamma_{2} - \gamma_{1} - \gamma_{\nu}, 
  &\gamma_{a}& = \gamma_{6} - \gamma_{1}. 
  \label{eq:gammaRel2}
\end{align}
  The final form of the anomalous dimensions is
\begin{align}
  \gamma_{\psi} =&\ \left(\frac{(a-1) a}{2 u (u+w)^2}+\frac{2 a-1}{2 u w (u+w)}\right) 
  \left(\alpha  g_1+g_2\right) - \frac{g_3}{8},
  \label{eq:GammaP} \\
  \gamma_{\psi'} =&\ \left(\frac{(a-1) a}{2 u (u+w)^2}+\frac{1-2 a}{2 u w (u+w)}\right) 
  \left(\alpha  g_1+g_2\right) - \frac{g_3}{8}, 
  \label{eq:GammaPs} \\
  \gamma_{g_{3}} =&\ \left(\alpha  g_1+g_2\right)\biggl(\frac{3 (a-1) a}{2 u (u+w)^2}+\frac{2 (a-1) a+1}{2 u w (u+w)}
   \nonumber\\
  & - \frac{4 (a-1) a u w+u^2-w^2}{4 u w (u+w)^3} \biggl)  -
  \frac{3 \left(g_1+g_2\right)}{4 w (w+1)} -\frac{3 g_3}{4}, \\
  \gamma_{\tau} =&\ \frac{\left((1-2 a)^2 w^2-u^2\right)}{8 u w (u+w)^3}\left(\alpha  g_1+g_2\right) -
  \frac{3 \left(g_1+g_2\right)}{8 w (w+1)} \nonumber\\
  &-\frac{3 g_3}{16},
  \label{eq:gammaTau} \\
  \gamma_{D} =&\ \frac{\left(u^2-(1-2 a)^2 w^2\right)}{8 u w (u+w)^3}\left(\alpha  g_1+g_2\right) + 
  \frac{3 \left(g_1+g_2\right)}{8 w (w+1)} \nonumber\\
  &+\frac{g_3}{16}, 
  \label{eq:gammaD} \\
  \gamma_{\nu} =&\ \frac{(u-1)}{8 u (u+1)^2}\left(\alpha  g_1+g_2\right) + 
  \frac{\left(3 u^2+8 u+7\right)}{24 (u+1)^2} \nonumber\\
  & \times \left(g_1+g_2\right),
  \label{eq:ADnu} \\
  \gamma_{w} =&\ \left(\frac{u^2-(1-2 a)^2 w^2}{8 u w (u+w)^3}+\frac{1-u}{8 u (u+1)^2}\right) 
  \left(\alpha  g_1 + g_2\right) \nonumber\\
  &+ \left(\frac{3}{8 w (w+1)}-\frac{3 u^2+8 u+7}{24 (u+1)^2}\right)
  \left(g_1+g_2\right) + \frac{g_3}{16}, \\
  \gamma_{a} =&\ \frac{(1-2 a) g_3}{16 a}. 
\end{align}
  where we have included $ \gamma_{\nu} $ for completeness \cite{AGKL17}.    
  
 { \subsection{Beta functions}  \label{app:BFunc}}
  The $\beta$ functions, which express RG flow,  are easily found from Eq.~(\ref{eq:BetaF})
  \begin{align}
	\beta_{g_{3}} & =  -g_{3} \Bigg[ \varepsilon + \biggl(\frac{3 (a-1) a}{2 u (u+w)^2}
	+\frac{2 (a-1) a+1}{2 u w (u+w)} \nonumber\\
	& -\frac{4 (a-1) a u w+u^2-w^2}{4 u w (u+w)^3} \biggl)
	\left(\alpha  g_1+g_2\right) -\frac{3 \left(g_1+g_2\right)}{4 w (w+1)} 
	\nonumber\\
	&- \frac{3 g_3}{4} \Bigg] , \\
	\beta_{w} & =\! -w \Bigg[\!\! \left(\frac{u^2-(1-2 a)^2 w^2}{8 u w (u+w)^3}+
	\frac{1-u}{8 u (u+1)^2}\right) (\alpha  g_1 + g_2)
	 \nonumber\\
	& +
	\left(\frac{3}{8 w (w+1)}-\frac{3 u^2+8 u+7}{24 (u+1)^2}\right) \left(g_1+g_2\right)
	+ \frac{g_3}{16} \Bigg], \\
	\beta_{a} & =  g_3 \frac{2a - 1}{16}.
\end{align}

{\section{Explicit expressions} \label{app:explicit}}
\subsection{Universality of anomalous dimensions $ \gamma_{\tau}^{*} $ and $ \gamma_{D}^{*} $} 
\label{app:anomalousTD}
Let us consider a fixed point with $ \gamma_{w}^{*} = 0 $ and $ w^{*} \neq 0$.  Using 
  relations~(\ref{eq:gammaRel1})-(\ref{eq:gammaRel2}) we derive
\begin{equation}
  \gamma_{\tau}^{*} = -\gamma_{\nu}^{*} + \gamma_{3}^{*} - \gamma_{1}^{*} = - 
  \gamma_{\nu}^{*} - \frac{g_{3}^{*}}{8}, \quad 
  \gamma_{D}^{*} = \gamma_{\nu}^{*}.
  \label{eq:afor}
\end{equation}
  We have seen that $ \gamma_{\nu}^{*} $ is independent of $ a^{*}(w^{*}) $ (see Eq.~(\ref{eq:ADnu}).
  Relations (\ref{eq:afor}) then imply
  that anomalous dimensions $ \gamma_{\tau}^{*} $ and $ \gamma_{D}^{*} $ are unaffected by
  the non-universality in $ a^{*}(w^{*}) $ appearing for fixed points FPII and FPIV. In 
  the case of FPI with $ w^{*} = 0 $ anomalous dimensions (\ref{eq:gammaTau}) and 
  (\ref{eq:gammaD}) are independent of $ a^{*} $. To avoid possible confusion in this case, let us note
  that we can set $ w^{*} = 0 $ after the limit $g_i\rightarrow 0;i=1,2$ has been performed.
{\subsection{FPII} \label{app:FPII}}

The results for fixed point FPII are non-universal with respect to parameter $ a^{*}(w^{*}) $. Although this 
dependence is more instructive for results, it is more difficult for practical calculations. 
Therefore, in  what follows we choose the independent parameter to be $ w^{*} $. The 
relation between $ a^{*} $ and $ w^{*} $, anomalous dimensions and corresponding 
\begin{equation}
  a_{\pm}^{*}(w^{*}) = \frac{1}{2} \left(1\pm\frac{\sqrt{X(w^{*})}}{\sqrt{2} (w^{*})}\right),
  \label{eq:Expressions1}
\end{equation}
where
\begin{equation}
  X(w) = -3 w^{4}-9 w^3-3 w^2+9 w+8 .
\end{equation}
Note that two solutions (\ref{eq:Expressions1}) correspond to two intersections
of the curve (\ref{fig:wa}) at fixed horizontal line $w^*=const.$
If we consider $ a^{*} \in \langle0,1\rangle $, from the Eq.~(\ref{eq:Expressions1})
we find that $ w^{*} $ ranges from $1$ (at $ a^{*} \in\{0,1\} $) to
$1.0518(8) $ at $ a^{*}=1/2 $.

Eigenvalues of the Jacobi matrix are the following
\begin{align}
  \lambda_{5}(w) &= \frac{\left(6 w^4+9 w^3+9 w+16\right)}{6 w (w+1)^3}, \\
  \lambda_{6}(w) &= \frac{\left(9 w^4+24 w^3+w^2-26 w-16\right)}{6 w^3 (w+1)}.
\end{align}
There are two values for critical exponents due to the existence of
two intersections of the curve $a^*=a^*(w^*)$ in Fig.~\ref{fig:wa}
\begin{align}
  \gamma_{\psi\pm}^{*} &= \frac{\left(Y(w^{*}) \pm 4 \sqrt{2} \sqrt{X(w^{*})}\right) }
  {12 (w^{*})^{2} ((w^{*})+1)} \varepsilon, \\
  \gamma_{\psi\pm}^{*} &= \frac{\left(Y(w^{*}) \mp 4 \sqrt{2} \sqrt{X(w^{*})}\right) }
  {12 (w^{*})^{2} ((w^{*})+1)} \varepsilon, \\
  \Theta &= \frac{\left(3 w^{3} + 6 w^{2} - w - 8 \right) }{12 w^{2} (w + 1)} \varepsilon,
\end{align}
where
\begin{align}
  X(w) &= -3 w^{4}-9 w^3-3 w^2+9 w+8, \\
  Y(w) &= -3 w^{4} - 6 w^{3} + w^{2} + 8 w.
\end{align}

{\subsection{FPIV} \label{app:FPIV}}
Similarly to the fixed point FPII from previous section, for fixed point FPIV we have
\begin{align}
  a_{\pm}^{*} = \frac{1}{2} \left( 1 + \frac{\sqrt{X(w^{*})}}{2 \sqrt{2} \alpha  w^2 (y-\epsilon )} 
  \right),
\end{align}
where
\begin{align}
  X & =  \ \alpha  w^2 (y-\epsilon ) \Big(2 \alpha  w^2 (y-\epsilon ) - \left(w^2-1\right) 
  \nonumber \\
  &\times \big(y (4 (\alpha +1)+(\alpha +2) w^2+3 (\alpha +2) w ) \nonumber \\
  &- \epsilon  (2 \alpha +3 w^2+9 w+6)\big)\Big).
\end{align}

Eigenvalues $\lambda_5$ and $\lambda_6$ from Table~\ref{tab:FP_tab1} are
\begin{align}
  \lambda_{5}(w,\Delta) & =  \frac{1}{I} \Big( 9 (w+1) w^3 \epsilon ^2+y^2 \big(-8 (\alpha +1) \nonumber \\
  & +5 (\alpha +2) w^4+10 (\alpha +2) w^3-(\alpha -2) w^2
  \nonumber \\
  &-2 (5 \alpha +8) w\big) 
   + y \epsilon  \big(4 (\alpha +3)-3 (\alpha +7) w^4
  \nonumber \\ 
   &-3 (\alpha +12) w^3 
    +(2 \alpha -3) w^2+2 (\alpha +12) w\big) \Big), 
\end{align}    
and
\begin{align}
  \lambda_{6}(w,\Delta) &= \frac{1}{I} \Big\{ y \big(y [8 (\alpha +1)+2 (\alpha +2) w^4 3 
 (\alpha +2) w^3 + \nonumber \\
  & +3 (\alpha +2) w]-\epsilon  [4 (\alpha +3)+6 w^4+9 w^3 \nonumber \\
  & +9 w] \big) \Big\},
\end{align}
where the expression $I$ is defined as follows
\begin{equation}
 I = 3 w (w+1)^3 [(\alpha +2) y-3 \epsilon].
\end{equation} 
{\subsection{FPIX} \label{app:FPX}}
For the fixed point FPX we have the following eigenvalues
\begin{align}
  \lambda_{5}(y,\varepsilon) &= \frac{A + \sqrt{6B}}{1200 y}, \\
  \lambda_{6}(y,\varepsilon) &= \frac{A - \sqrt{6B}}{1200 y},
\end{align}
where we have used expressions
\begin{align}
  A =&\ 48 y^2 -12 y \left(C-48 \varepsilon \right) + 3 \varepsilon  \left(C+\varepsilon \right), \\
  B =&\ 145408 y^4 + 3 \varepsilon ^3 \left(C+\varepsilon \right) + 304 y^2 \varepsilon 
  \left(11 C+377 \varepsilon \right) \nonumber \\
  &\ - 12 y \varepsilon ^2 \left(53 C+59 \varepsilon \right) - 64 y^3 \left(53 C+3807 \varepsilon \right), \\
  C =&\ \sqrt{176 y^2-48 y \varepsilon +\varepsilon ^2}.
\end{align}

{\subsection{The boundary between FPIV and FPV} \label{subsec:boundary}}

\begin{figure}[b!]
	\includegraphics[width=7.5cm]{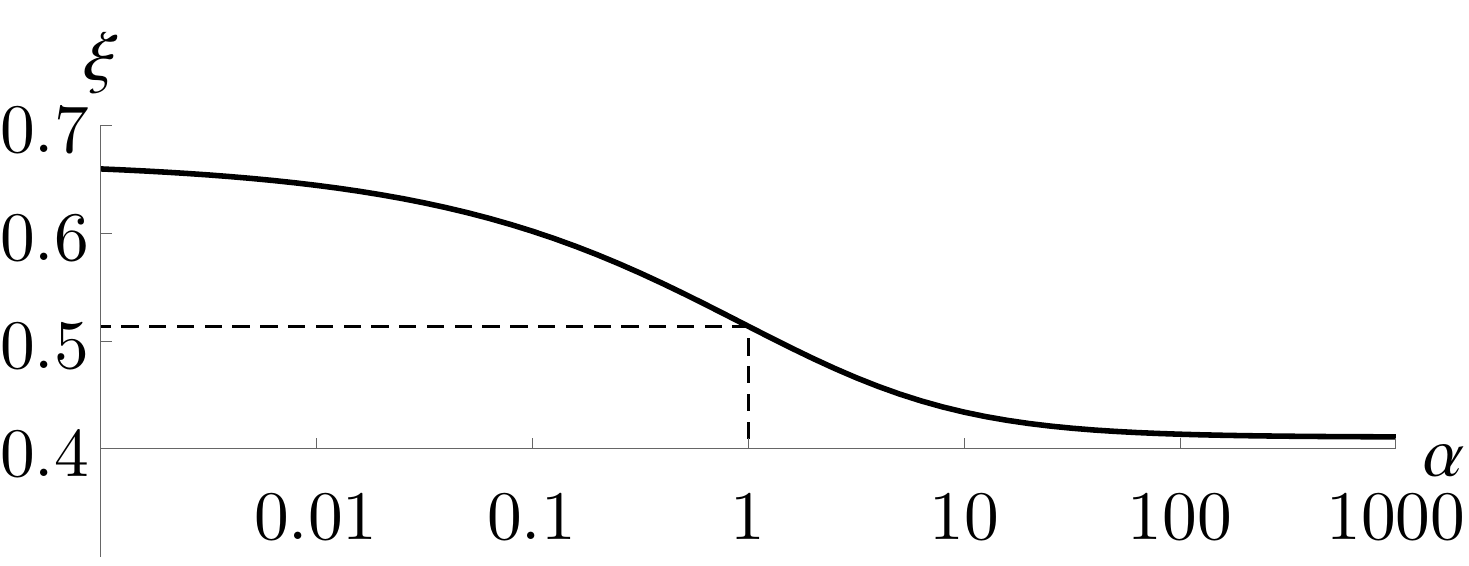} 
	\caption{The parameter $ \xi $ as a function of $ \alpha $. The results are 
	consistent with the numerical solution shown in Fig.~\ref{fig:g3}.} 
	\label{fig:xiAlpha}
\end{figure}

In order to find the boundary between two  non-trivial fixed points FPIV and FPV, we 
  work in the ray scheme \cite{HN95,AHK05}. Since expansion parameters must be proportional,
we relate them as $ y = \frac{1}{\xi} \varepsilon $. Now we have to consider 
values of charges for both fixed points.
According to our analytical and numerical solutions (see Tab.~\ref{tab:FP_tab1}
and Fig.~\ref{fig:g3}), 
charges $ g_{3}^{*} $ and $ a^{*} $ seem to converge at the boundary  between FPIV and FPV.
Hence, we may put $ g_{3}^{*} = 0 $ and $ a^{*} = 1/2 $ during the calculation process. In this case,
 beta functions $ \beta_{g_{3}} $ and $ \beta_{a} $ vanish and 
 coordinate $ w^{*} $ is found as a zero point of the function
\begin{align}
  \beta_{w} & = \frac{\varepsilon}{A} \bigg[ w \left(w^3+3 w^2+w-3\right) (\alpha -3 \xi +2) \nonumber \\
  & + 2 \alpha  (\xi -2) + 6 \xi - 4  \bigg],
\end{align}
 where $ A = 3 \xi  (w+1)^3 (\alpha -3 \xi +2) $. Note that $ \xi < 2/3 $, since in the limit 
$ \alpha \rightarrow 0 $ the boundary between FPIV and FPV is simply $ y = 3\varepsilon/2 $. By
calculating corresponding eigenvalues of the stability matrix we find that
$ \lambda_{7} = 0, \ \lambda_{6} $ is positive in the relevant region and the line 
$ \lambda_{5}(\xi,\alpha)/\varepsilon = 0 $ is plotted  in Fig.~\ref{fig:xiAlpha}.

\bibliographystyle{apsrev}
\bibliography{mybib}

\begin{thebibliography}{69}
\expandafter\ifx\csname natexlab\endcsname\relax\def\natexlab#1{#1}\fi
\expandafter\ifx\csname bibnamefont\endcsname\relax
  \def\bibnamefont#1{#1}\fi
\expandafter\ifx\csname bibfnamefont\endcsname\relax
  \def\bibfnamefont#1{#1}\fi
\expandafter\ifx\csname citenamefont\endcsname\relax
  \def\citenamefont#1{#1}\fi
\expandafter\ifx\csname url\endcsname\relax
  \def\url#1{\texttt{#1}}\fi
\expandafter\ifx\csname urlprefix\endcsname\relax\def\urlprefix{URL }\fi
\providecommand{\bibinfo}[2]{#2}
\providecommand{\eprint}[2][]{\url{#2}}

\bibitem[{\citenamefont{Schmittmann and Zia}(1995)}]{Zia95}
\bibinfo{author}{\bibfnamefont{B.}~\bibnamefont{Schmittmann}} \bibnamefont{and}
  \bibinfo{author}{\bibfnamefont{R.~K.~P.} \bibnamefont{Zia}},
  \emph{\bibinfo{title}{{Statistical mechanics of driven diffusive systems}}},
  vol.~\bibinfo{volume}{17} of \emph{\bibinfo{series}{Phase transitions and
  critical phenomena}} (\bibinfo{publisher}{Academic Press},
  \bibinfo{year}{1995}).

\bibitem[{\citenamefont{Krapivsky et~al.}(2010)\citenamefont{Krapivsky, Redner,
  and Ben-Naim}}]{krapivsky}
\bibinfo{author}{\bibfnamefont{P.~L.} \bibnamefont{Krapivsky}},
  \bibinfo{author}{\bibfnamefont{S.}~\bibnamefont{Redner}}, \bibnamefont{and}
  \bibinfo{author}{\bibfnamefont{E.}~\bibnamefont{Ben-Naim}},
  \emph{\bibinfo{title}{A Kinetic View of Statistical Physics}}
  (\bibinfo{publisher}{Cambridge University Press}, \bibinfo{year}{2010}).

\bibitem[{\citenamefont{T{\"{a}}uber}(2014)}]{Tauber}
\bibinfo{author}{\bibfnamefont{U.~C.} \bibnamefont{T{\"{a}}uber}},
  \emph{\bibinfo{title}{{Critical Dynamics: A Field Theory Approach to
  Equilibrium and Non-Equilibrium Scaling Behavior}}}
  (\bibinfo{publisher}{Cambridge University Press}, \bibinfo{year}{2014}).

\bibitem[{\citenamefont{Amit and Mart\'in-Mayor}(2005)}]{Amit}
\bibinfo{author}{\bibfnamefont{D.~J.} \bibnamefont{Amit}} \bibnamefont{and}
  \bibinfo{author}{\bibfnamefont{V.}~\bibnamefont{Mart\'in-Mayor}},
  \emph{\bibinfo{title}{Field Theory, the Renormalization Group and Critical
  Phenomena}} (\bibinfo{publisher}{World Scientific},
  \bibinfo{address}{Singapore}, \bibinfo{year}{2005}).

\bibitem[{\citenamefont{Zinn-Justin}(1996)}]{Zinn}
\bibinfo{author}{\bibfnamefont{J.}~\bibnamefont{Zinn-Justin}},
  \emph{\bibinfo{title}{{Quantum field theory and critical phenomena}}}
  (\bibinfo{publisher}{Clarendon Press}, \bibinfo{year}{1996}).

\bibitem[{\citenamefont{Zinn-Justin}(2007)}]{ZinnRG}
\bibinfo{author}{\bibfnamefont{J.}~\bibnamefont{Zinn-Justin}},
  \emph{\bibinfo{title}{{Phase transitions and renormalization group}}}
  (\bibinfo{publisher}{Clarendon}, \bibinfo{year}{2007}).

\bibitem[{\citenamefont{Vasil'ev}(2004)}]{Vasiliev}
\bibinfo{author}{\bibfnamefont{A.~N.} \bibnamefont{Vasil'ev}},
  \emph{\bibinfo{title}{{The Field Theoretic Renormalization Group in Critical
  Behavior Theory and Stochastic Dynamics}}} (\bibinfo{publisher}{Chapman
  Hall/CRC, Boca Raton, FL}, \bibinfo{year}{2004}).

\bibitem[{\citenamefont{Henkel et~al.}(2008)\citenamefont{Henkel, Hinrichsen,
  and L{\"{u}}beck}}]{HHL08}
\bibinfo{author}{\bibfnamefont{M.}~\bibnamefont{Henkel}},
  \bibinfo{author}{\bibfnamefont{H.}~\bibnamefont{Hinrichsen}},
  \bibnamefont{and}
  \bibinfo{author}{\bibfnamefont{S.}~\bibnamefont{L{\"{u}}beck}},
  \emph{\bibinfo{title}{{Non-Equilibrium Phase Transitions: Volume 1-Absorbing
  Phase Transitions}}} (\bibinfo{publisher}{Springer}, \bibinfo{year}{2008}).

\bibitem[{\citenamefont{Cardy and Sugar}(1980)}]{Cardy80}
\bibinfo{author}{\bibfnamefont{J.}~\bibnamefont{Cardy}} \bibnamefont{and}
  \bibinfo{author}{\bibfnamefont{R.~L.} \bibnamefont{Sugar}},
  \bibinfo{journal}{J. Phys. A: Math. Gen.} \textbf{\bibinfo{volume}{13}},
  \bibinfo{pages}{L423} (\bibinfo{year}{1980}).

\bibitem[{\citenamefont{Obukhov}(1980)}]{obukhov80}
\bibinfo{author}{\bibfnamefont{S.~P.} \bibnamefont{Obukhov}},
  \bibinfo{journal}{Physic A} \textbf{\bibinfo{volume}{101}},
  \bibinfo{pages}{145} (\bibinfo{year}{1980}).

\bibitem[{\citenamefont{Janssen}(1981)}]{Janssen81}
\bibinfo{author}{\bibfnamefont{H.-K.} \bibnamefont{Janssen}},
  \bibinfo{journal}{Z. Phys. B} \textbf{\bibinfo{volume}{42}},
  \bibinfo{pages}{151} (\bibinfo{year}{1981}).

\bibitem[{\citenamefont{Grassberger}(1982)}]{Grassberger82}
\bibinfo{author}{\bibfnamefont{P.}~\bibnamefont{Grassberger}},
  \bibinfo{journal}{Z. Phys. B} \textbf{\bibinfo{volume}{47}},
  \bibinfo{pages}{365} (\bibinfo{year}{1982}).

\bibitem[{\citenamefont{{\'{O}}dor}(2004)}]{Odor04}
\bibinfo{author}{\bibfnamefont{G.}~\bibnamefont{{\'{O}}dor}},
  \bibinfo{journal}{Phys. Rev. E} \textbf{\bibinfo{volume}{70}},
  \bibinfo{pages}{026119} (\bibinfo{year}{2004}).

\bibitem[{\citenamefont{Janssen and T{\"{a}}uber}(2005)}]{JT04}
\bibinfo{author}{\bibfnamefont{H.-K.} \bibnamefont{Janssen}} \bibnamefont{and}
  \bibinfo{author}{\bibfnamefont{U.~C.} \bibnamefont{T{\"{a}}uber}},
  \bibinfo{journal}{Ann. Phys. (N. Y.)} \textbf{\bibinfo{volume}{315}},
  \bibinfo{pages}{147} (\bibinfo{year}{2005}).

\bibitem[{\citenamefont{Rupp et~al.}(2003)\citenamefont{Rupp, Richter, and
  Rehberg}}]{RRR03}
\bibinfo{author}{\bibfnamefont{P.}~\bibnamefont{Rupp}},
  \bibinfo{author}{\bibfnamefont{R.}~\bibnamefont{Richter}}, \bibnamefont{and}
  \bibinfo{author}{\bibfnamefont{I.}~\bibnamefont{Rehberg}},
  \bibinfo{journal}{Phys. Rev.~E} \textbf{\bibinfo{volume}{67}},
  \bibinfo{pages}{036209} (\bibinfo{year}{2003}).

\bibitem[{\citenamefont{Takeuchi et~al.}(2007)\citenamefont{Takeuchi, Kuroda,
  Chat{\'{e}}, and Sano}}]{TKCS07}
\bibinfo{author}{\bibfnamefont{K.~A.} \bibnamefont{Takeuchi}},
  \bibinfo{author}{\bibfnamefont{M.}~\bibnamefont{Kuroda}},
  \bibinfo{author}{\bibfnamefont{H.}~\bibnamefont{Chat{\'{e}}}},
  \bibnamefont{and} \bibinfo{author}{\bibfnamefont{M.}~\bibnamefont{Sano}},
  \bibinfo{journal}{Phys. Rev. Lett.} \textbf{\bibinfo{volume}{99}},
  \bibinfo{pages}{234503} (\bibinfo{year}{2007}).

\bibitem[{\citenamefont{Sano and Tamai}(2016)}]{Sano2016}
\bibinfo{author}{\bibfnamefont{M.}~\bibnamefont{Sano}} \bibnamefont{and}
  \bibinfo{author}{\bibfnamefont{K.}~\bibnamefont{Tamai}},
  \bibinfo{journal}{Nat. Phys.} \textbf{\bibinfo{volume}{12}},
  \bibinfo{pages}{249} (\bibinfo{year}{2016}).

\bibitem[{\citenamefont{Lemoult et~al.}(2016)\citenamefont{Lemoult, Shi, Avila,
  Avila, and Hof}}]{LSAJAH16}
\bibinfo{author}{\bibfnamefont{G.}~\bibnamefont{Lemoult}},
  \bibinfo{author}{\bibfnamefont{L.}~\bibnamefont{Shi}},
  \bibinfo{author}{\bibfnamefont{S.~V.} \bibnamefont{Avila},
  \bibfnamefont{K.~Jalikop}},
  \bibinfo{author}{\bibfnamefont{M.}~\bibnamefont{Avila}}, \bibnamefont{and}
  \bibinfo{author}{\bibfnamefont{B.}~\bibnamefont{Hof}},
  \bibinfo{journal}{Nature Phys} \textbf{\bibinfo{volume}{12}},
  \bibinfo{pages}{254} (\bibinfo{year}{2016}).

\bibitem[{\citenamefont{Janssen et~al.}(1999)\citenamefont{Janssen, Oerding,
  van Wijland, and Hilhorst}}]{Jan99}
\bibinfo{author}{\bibfnamefont{H.-K.} \bibnamefont{Janssen}},
  \bibinfo{author}{\bibfnamefont{K.}~\bibnamefont{Oerding}},
  \bibinfo{author}{\bibfnamefont{F.}~\bibnamefont{van Wijland}},
  \bibnamefont{and} \bibinfo{author}{\bibfnamefont{H.}~\bibnamefont{Hilhorst}},
  \bibinfo{journal}{Eur. Phys. J. B} \textbf{\bibinfo{volume}{7}},
  \bibinfo{pages}{137} (\bibinfo{year}{1999}).

\bibitem[{\citenamefont{R{\"{o}}ssner and Hinrichsen}(2006)}]{Hin06}
\bibinfo{author}{\bibfnamefont{S.}~\bibnamefont{R{\"{o}}ssner}}
  \bibnamefont{and}
  \bibinfo{author}{\bibfnamefont{H.}~\bibnamefont{Hinrichsen}},
  \bibinfo{journal}{Phys. Rev.~E} \textbf{\bibinfo{volume}{74}},
  \bibinfo{pages}{041607} (\bibinfo{year}{2006}).

\bibitem[{\citenamefont{Hinrichsen}(2007)}]{Hin07}
\bibinfo{author}{\bibfnamefont{H.}~\bibnamefont{Hinrichsen}},
  \bibinfo{journal}{J. Stat. Mech. Theor. Exp.}
  \textbf{\bibinfo{volume}{2007}}, \bibinfo{pages}{P07006}
  (\bibinfo{year}{2007}).

\bibitem[{\citenamefont{Hinrichsen}(2000)}]{Hin00}
\bibinfo{author}{\bibfnamefont{H.}~\bibnamefont{Hinrichsen}},
  \bibinfo{journal}{Adv. Phys.} \textbf{\bibinfo{volume}{49}},
  \bibinfo{pages}{815} (\bibinfo{year}{2000}).

\bibitem[{\citenamefont{Janssen}(1997)}]{Janssen97}
\bibinfo{author}{\bibfnamefont{H.-K.} \bibnamefont{Janssen}},
  \bibinfo{journal}{Phys. Rev.~E} \textbf{\bibinfo{volume}{55}},
  \bibinfo{pages}{6253} (\bibinfo{year}{1997}).

\bibitem[{\citenamefont{Moreira and Dickman}(1996)}]{MorDic96}
\bibinfo{author}{\bibfnamefont{A.~G.} \bibnamefont{Moreira}} \bibnamefont{and}
  \bibinfo{author}{\bibfnamefont{R.}~\bibnamefont{Dickman}},
  \bibinfo{journal}{Phys. Rev. E} \textbf{\bibinfo{volume}{54}},
  \bibinfo{pages}{R3090} (\bibinfo{year}{1996}).

\bibitem[{\citenamefont{Cafiero et~al.}(1998)\citenamefont{Cafiero, Gabrielli,
  and Mu{\~{n}oz}}}]{CGM98}
\bibinfo{author}{\bibfnamefont{R.}~\bibnamefont{Cafiero}},
  \bibinfo{author}{\bibfnamefont{A.}~\bibnamefont{Gabrielli}},
  \bibnamefont{and} \bibinfo{author}{\bibfnamefont{M.~A.}
  \bibnamefont{Mu{\~{n}oz}}}, \bibinfo{journal}{Phys. Rev. E}
  \textbf{\bibinfo{volume}{57}}, \bibinfo{pages}{5060} (\bibinfo{year}{1998}).

\bibitem[{\citenamefont{Vojta and Dickison}(2005)}]{Vojta05}
\bibinfo{author}{\bibfnamefont{T.}~\bibnamefont{Vojta}} \bibnamefont{and}
  \bibinfo{author}{\bibfnamefont{M.}~\bibnamefont{Dickison}},
  \bibinfo{journal}{Phys. Rev. E} \textbf{\bibinfo{volume}{72}},
  \bibinfo{pages}{036126} (\bibinfo{year}{2005}).

\bibitem[{\citenamefont{Vojta and Lee}(2006)}]{Vojta06}
\bibinfo{author}{\bibfnamefont{T.}~\bibnamefont{Vojta}} \bibnamefont{and}
  \bibinfo{author}{\bibfnamefont{M.~Y.} \bibnamefont{Lee}},
  \bibinfo{journal}{Phys. Rev. Lett.} \textbf{\bibinfo{volume}{96}},
  \bibinfo{pages}{035701} (\bibinfo{year}{2006}).

\bibitem[{\citenamefont{Adzhemyan and Antonov}(1998)}]{AdzAnt98}
\bibinfo{author}{\bibfnamefont{L.~T.} \bibnamefont{Adzhemyan}}
  \bibnamefont{and} \bibinfo{author}{\bibfnamefont{N.~V.}
  \bibnamefont{Antonov}}, \bibinfo{journal}{Phys. Rev. E}
  \textbf{\bibinfo{volume}{58}}, \bibinfo{pages}{7381} (\bibinfo{year}{1998}).

\bibitem[{\citenamefont{Antonov}(2000)}]{Ant00}
\bibinfo{author}{\bibfnamefont{N.~V.} \bibnamefont{Antonov}},
  \bibinfo{journal}{Physica D} \textbf{\bibinfo{volume}{144}},
  \bibinfo{pages}{370} (\bibinfo{year}{2000}).

\bibitem[{\citenamefont{Frisch}(1995)}]{Frisch}
\bibinfo{author}{\bibfnamefont{U.}~\bibnamefont{Frisch}},
  \emph{\bibinfo{title}{{Turbulence: the legacy of A.N. Kolmogorov}}}
  (\bibinfo{publisher}{Cambridge University Press}, \bibinfo{year}{1995}).

\bibitem[{\citenamefont{Davidson}(2015)}]{davidson}
\bibinfo{author}{\bibfnamefont{P.~A.} \bibnamefont{Davidson}},
  \emph{\bibinfo{title}{Turbulence: An Introduction for Scientists and
  Engineers (2nd edition)}} (\bibinfo{publisher}{Oxford University Press},
  \bibinfo{year}{2015}).

\bibitem[{\citenamefont{Davidson et~al.}(2013)\citenamefont{Davidson, Kaneda,
  and Sreenivasan}}]{chapters}
\bibinfo{author}{\bibfnamefont{P.~A.} \bibnamefont{Davidson}},
  \bibinfo{author}{\bibfnamefont{Y.}~\bibnamefont{Kaneda}}, \bibnamefont{and}
  \bibinfo{author}{\bibfnamefont{K.~R.} \bibnamefont{Sreenivasan}},
  \emph{\bibinfo{title}{Ten Chapters in Turbulence}}
  (\bibinfo{publisher}{Cambridge University Press}, \bibinfo{year}{2013}).

\bibitem[{\citenamefont{Davidson et~al.}(2011)\citenamefont{Davidson, Kaneda,
  and Moffatt}}]{voyage}
\bibinfo{author}{\bibfnamefont{P.~A.} \bibnamefont{Davidson}},
  \bibinfo{author}{\bibfnamefont{Y.}~\bibnamefont{Kaneda}}, \bibnamefont{and}
  \bibinfo{author}{\bibfnamefont{K.~R.} \bibnamefont{Moffatt},
  \bibfnamefont{K.~Sreenivasan}}, \emph{\bibinfo{title}{A Voyage Through
  Turbulence}} (\bibinfo{publisher}{Cambridge University Press},
  \bibinfo{year}{2011}).

\bibitem[{\citenamefont{Antonov et~al.}(2009)\citenamefont{Antonov, Iglovikov,
  and Kapustin}}]{AIK10}
\bibinfo{author}{\bibfnamefont{N.~V.} \bibnamefont{Antonov}},
  \bibinfo{author}{\bibfnamefont{V.~I.} \bibnamefont{Iglovikov}},
  \bibnamefont{and} \bibinfo{author}{\bibfnamefont{A.~S.}
  \bibnamefont{Kapustin}}, \bibinfo{journal}{J. Phys. A: Math. Gen.}
  \textbf{\bibinfo{volume}{42}}, \bibinfo{pages}{135001}
  (\bibinfo{year}{2009}).

\bibitem[{\citenamefont{Antonov et~al.}(2010)\citenamefont{Antonov, Ignatieva,
  and Malyshev}}]{AIM10}
\bibinfo{author}{\bibfnamefont{N.~V.} \bibnamefont{Antonov}},
  \bibinfo{author}{\bibfnamefont{A.~A.} \bibnamefont{Ignatieva}},
  \bibnamefont{and} \bibinfo{author}{\bibfnamefont{A.~V.}
  \bibnamefont{Malyshev}}, \bibinfo{journal}{Physics of Particles and Nuclei}
  \textbf{\bibinfo{volume}{41}}, \bibinfo{pages}{998} (\bibinfo{year}{2010}).

\bibitem[{\citenamefont{Antonov et~al.}(2011)\citenamefont{Antonov, Kapustin,
  and Malyshev}}]{AKM11}
\bibinfo{author}{\bibfnamefont{N.~V.} \bibnamefont{Antonov}},
  \bibinfo{author}{\bibfnamefont{A.~S.} \bibnamefont{Kapustin}},
  \bibnamefont{and} \bibinfo{author}{\bibfnamefont{A.~V.}
  \bibnamefont{Malyshev}}, \bibinfo{journal}{Theor. Math. Phys.}
  \textbf{\bibinfo{volume}{169}}, \bibinfo{pages}{1470} (\bibinfo{year}{2011}).

\bibitem[{\citenamefont{Dančo et~al.}(2013)\citenamefont{Dančo, Hnatič,
  Lu{\v{c}}ivjansk{\'{y}}, and Mižišin}}]{DHLM13}
\bibinfo{author}{\bibfnamefont{M.}~\bibnamefont{Dančo}},
  \bibinfo{author}{\bibfnamefont{M.}~\bibnamefont{Hnatič}},
  \bibinfo{author}{\bibfnamefont{T.}~\bibnamefont{Lu{\v{c}}ivjansk{\'{y}}}},
  \bibnamefont{and}
  \bibinfo{author}{\bibfnamefont{L.}~\bibnamefont{Mižišin}},
  \bibinfo{journal}{Theor. Math. Phys} \textbf{\bibinfo{volume}{176}},
  \bibinfo{pages}{79} (\bibinfo{year}{2013}).

\bibitem[{\citenamefont{Kraichnan}(1968)}]{Kraichnan68}
\bibinfo{author}{\bibfnamefont{R.}~\bibnamefont{Kraichnan}},
  \bibinfo{journal}{Phys. Fluids} \textbf{\bibinfo{volume}{11}},
  \bibinfo{pages}{945} (\bibinfo{year}{1968}).

\bibitem[{\citenamefont{Falkovich et~al.}(2001)\citenamefont{Falkovich,
  Gaw{\c{e}}dzki, and Vergassola}}]{FGV01}
\bibinfo{author}{\bibfnamefont{G.}~\bibnamefont{Falkovich}},
  \bibinfo{author}{\bibfnamefont{K.}~\bibnamefont{Gaw{\c{e}}dzki}},
  \bibnamefont{and}
  \bibinfo{author}{\bibfnamefont{M.}~\bibnamefont{Vergassola}},
  \bibinfo{journal}{Rev. Mod. Phys.} \textbf{\bibinfo{volume}{73}},
  \bibinfo{pages}{913} (\bibinfo{year}{2001}).

\bibitem[{\citenamefont{Adzhemyan et~al.}(1999)\citenamefont{Adzhemyan,
  Antonov, and Vasil'ev}}]{turbo}
\bibinfo{author}{\bibfnamefont{L.~T.} \bibnamefont{Adzhemyan}},
  \bibinfo{author}{\bibfnamefont{N.~V.} \bibnamefont{Antonov}},
  \bibnamefont{and} \bibinfo{author}{\bibfnamefont{A.~N.}
  \bibnamefont{Vasil'ev}}, \emph{\bibinfo{title}{The Field Theoretic
  Renormalization Group in Fully Developed Turbulence}}
  (\bibinfo{publisher}{Gordon \& Breach}, \bibinfo{address}{London},
  \bibinfo{year}{1999}).

\bibitem[{\citenamefont{Benzi and Nelson}(2009)}]{Benzi09}
\bibinfo{author}{\bibfnamefont{R.}~\bibnamefont{Benzi}} \bibnamefont{and}
  \bibinfo{author}{\bibfnamefont{D.~R.} \bibnamefont{Nelson}},
  \bibinfo{journal}{Physica D} \textbf{\bibinfo{volume}{238}},
  \bibinfo{pages}{2003} (\bibinfo{year}{2009}).

\bibitem[{\citenamefont{Pigolotti et~al.}(2012)\citenamefont{Pigolotti, Benzi,
  Jensen, and Nelson}}]{Pig12}
\bibinfo{author}{\bibfnamefont{S.}~\bibnamefont{Pigolotti}},
  \bibinfo{author}{\bibfnamefont{R.}~\bibnamefont{Benzi}},
  \bibinfo{author}{\bibfnamefont{M.~H.} \bibnamefont{Jensen}},
  \bibnamefont{and} \bibinfo{author}{\bibfnamefont{D.~R.}
  \bibnamefont{Nelson}}, \bibinfo{journal}{Phys. Rev. Lett.}
  \textbf{\bibinfo{volume}{108}}, \bibinfo{pages}{128102}
  (\bibinfo{year}{2012}).

\bibitem[{\citenamefont{Volk et~al.}(2014)\citenamefont{Volk, Mauger, Bourgoin,
  Cottin-Bizonne, Ybert, and Raynal}}]{Volk14}
\bibinfo{author}{\bibfnamefont{R.}~\bibnamefont{Volk}},
  \bibinfo{author}{\bibfnamefont{C.}~\bibnamefont{Mauger}},
  \bibinfo{author}{\bibfnamefont{M.}~\bibnamefont{Bourgoin}},
  \bibinfo{author}{\bibfnamefont{C.}~\bibnamefont{Cottin-Bizonne}},
  \bibinfo{author}{\bibfnamefont{C.}~\bibnamefont{Ybert}}, \bibnamefont{and}
  \bibinfo{author}{\bibfnamefont{F.}~\bibnamefont{Raynal}},
  \bibinfo{journal}{Phys. Rev.~E} \textbf{\bibinfo{volume}{90}},
  \bibinfo{pages}{013027} (\bibinfo{year}{2014}).

\bibitem[{\citenamefont{{De Pietro} et~al.}(2015)\citenamefont{{De Pietro},
  {van Hinsberg}, Biferale, Clercx, Perlekar, and Toschi}}]{depietro15}
\bibinfo{author}{\bibfnamefont{M.}~\bibnamefont{{De Pietro}}},
  \bibinfo{author}{\bibfnamefont{M.~A.~T.} \bibnamefont{{van Hinsberg}}},
  \bibinfo{author}{\bibfnamefont{L.}~\bibnamefont{Biferale}},
  \bibinfo{author}{\bibfnamefont{H.~J.~H.} \bibnamefont{Clercx}},
  \bibinfo{author}{\bibfnamefont{P.}~\bibnamefont{Perlekar}}, \bibnamefont{and}
  \bibinfo{author}{\bibfnamefont{F.}~\bibnamefont{Toschi}},
  \bibinfo{journal}{Phys. Rev.~E} \textbf{\bibinfo{volume}{91}},
  \bibinfo{pages}{053002} (\bibinfo{year}{2015}).

\bibitem[{\citenamefont{{De Dominicis}}(1976)}]{DeDM76}
\bibinfo{author}{\bibfnamefont{C.}~\bibnamefont{{De Dominicis}}},
  \bibinfo{journal}{Le Journal de Physique Colloques}
  \textbf{\bibinfo{volume}{37}}, \bibinfo{pages}{C1} (\bibinfo{year}{1976}).

\bibitem[{\citenamefont{Janssen}(1976)}]{Janssen76}
\bibinfo{author}{\bibfnamefont{H.~K.} \bibnamefont{Janssen}},
  \bibinfo{journal}{Z. Phys. B} \textbf{\bibinfo{volume}{23}},
  \bibinfo{pages}{377} (\bibinfo{year}{1976}).

\bibitem[{\citenamefont{Martin et~al.}(1973)\citenamefont{Martin, Siggia, and
  Rose}}]{MSR}
\bibinfo{author}{\bibfnamefont{P.~C.} \bibnamefont{Martin}},
  \bibinfo{author}{\bibfnamefont{E.~D.} \bibnamefont{Siggia}},
  \bibnamefont{and} \bibinfo{author}{\bibfnamefont{H.~A.} \bibnamefont{Rose}},
  \bibinfo{journal}{Phys. Rev.~A} \textbf{\bibinfo{volume}{8}},
  \bibinfo{pages}{423} (\bibinfo{year}{1973}).

\bibitem[{\citenamefont{Landau and Lifshitz}(1987)}]{Landau_fluid}
\bibinfo{author}{\bibfnamefont{L.~D.} \bibnamefont{Landau}} \bibnamefont{and}
  \bibinfo{author}{\bibfnamefont{E.~M.} \bibnamefont{Lifshitz}},
  \emph{\bibinfo{title}{{Fluid Mechanics: Landau and Lifshitz: Course of
  Theoretical Physics, Volume 6}}} (\bibinfo{publisher}{Elsevier Science},
  \bibinfo{year}{1987}).

\bibitem[{\citenamefont{Monin and Yaglom}(1971)}]{Monin}
\bibinfo{author}{\bibfnamefont{A.~S.} \bibnamefont{Monin}} \bibnamefont{and}
  \bibinfo{author}{\bibfnamefont{A.~M.} \bibnamefont{Yaglom}},
  \emph{\bibinfo{title}{{Statistical Fluid Mechanics Volume 1}}}
  (\bibinfo{publisher}{MIT Press}, \bibinfo{year}{1971}).

\bibitem[{\citenamefont{Antonov and Kapustin}(2010)}]{AK10}
\bibinfo{author}{\bibfnamefont{N.~V.} \bibnamefont{Antonov}} \bibnamefont{and}
  \bibinfo{author}{\bibfnamefont{A.~S.} \bibnamefont{Kapustin}},
  \bibinfo{journal}{J. Phys. A: Math. Gen.} \textbf{\bibinfo{volume}{43}},
  \bibinfo{pages}{405001} (\bibinfo{year}{2010}).

\bibitem[{\citenamefont{Staroselsky et~al.}(1990)\citenamefont{Staroselsky,
  Yakhot, Kida, and Orszag}}]{SYKO90}
\bibinfo{author}{\bibfnamefont{I.}~\bibnamefont{Staroselsky}},
  \bibinfo{author}{\bibfnamefont{V.}~\bibnamefont{Yakhot}},
  \bibinfo{author}{\bibfnamefont{S.}~\bibnamefont{Kida}}, \bibnamefont{and}
  \bibinfo{author}{\bibfnamefont{S.~A.} \bibnamefont{Orszag}},
  \bibinfo{journal}{Phys. Rev. Lett.} \textbf{\bibinfo{volume}{65}},
  \bibinfo{pages}{171} (\bibinfo{year}{1990}).

\bibitem[{\citenamefont{Antonov et~al.}(1997)\citenamefont{Antonov, Nalimov,
  and Udalov}}]{ANU97}
\bibinfo{author}{\bibfnamefont{N.~V.} \bibnamefont{Antonov}},
  \bibinfo{author}{\bibfnamefont{M.~Y.} \bibnamefont{Nalimov}},
  \bibnamefont{and} \bibinfo{author}{\bibfnamefont{A.~A.}
  \bibnamefont{Udalov}}, \bibinfo{journal}{Theor. Math. Phys.}
  \textbf{\bibinfo{volume}{110}}, \bibinfo{pages}{305} (\bibinfo{year}{1997}).

\bibitem[{\citenamefont{Antonov et~al.}(2017)\citenamefont{Antonov, Gulitskiy,
  Kostenko, and Lu{\v{c}}ivjansk{\'{y}}}}]{AGKL17}
\bibinfo{author}{\bibfnamefont{N.~V.} \bibnamefont{Antonov}},
  \bibinfo{author}{\bibfnamefont{N.~M.} \bibnamefont{Gulitskiy}},
  \bibinfo{author}{\bibfnamefont{M.~M.} \bibnamefont{Kostenko}},
  \bibnamefont{and}
  \bibinfo{author}{\bibfnamefont{T.}~\bibnamefont{Lu{\v{c}}ivjansk{\'{y}}}},
  \bibinfo{journal}{Phys. Rev.~E} \textbf{\bibinfo{volume}{95}},
  \bibinfo{pages}{033120} (\bibinfo{year}{2017}).

\bibitem[{\citenamefont{Hnati{\v{c}} et~al.}(2016)\citenamefont{Hnati{\v{c}},
  Honkonen, and Lu{\v{c}}ivjansk{\'{y}}}}]{HHL16}
\bibinfo{author}{\bibfnamefont{M.}~\bibnamefont{Hnati{\v{c}}}},
  \bibinfo{author}{\bibfnamefont{J.}~\bibnamefont{Honkonen}}, \bibnamefont{and}
  \bibinfo{author}{\bibfnamefont{T.}~\bibnamefont{Lu{\v{c}}ivjansk{\'{y}}}},
  \bibinfo{journal}{Acta Phys. Slovaca} \textbf{\bibinfo{volume}{66}},
  \bibinfo{pages}{69} (\bibinfo{year}{2016}).

\bibitem[{\citenamefont{Antonov and Kostenko}(2014)}]{AK14}
\bibinfo{author}{\bibfnamefont{N.~V.} \bibnamefont{Antonov}} \bibnamefont{and}
  \bibinfo{author}{\bibfnamefont{M.~M.} \bibnamefont{Kostenko}},
  \bibinfo{journal}{Phys. Rev. E} \textbf{\bibinfo{volume}{90}},
  \bibinfo{pages}{063016} (\bibinfo{year}{2014}).

\bibitem[{\citenamefont{Adzhemyan et~al.}(1995)\citenamefont{Adzhemyan,
  Nalimov, and Stepanova}}]{ANS95}
\bibinfo{author}{\bibfnamefont{L.~T.} \bibnamefont{Adzhemyan}},
  \bibinfo{author}{\bibfnamefont{M.~Y.} \bibnamefont{Nalimov}},
  \bibnamefont{and} \bibinfo{author}{\bibfnamefont{M.~M.}
  \bibnamefont{Stepanova}}, \bibinfo{journal}{Theor. Math. Phys.}
  \textbf{\bibinfo{volume}{104}}, \bibinfo{pages}{971} (\bibinfo{year}{1995}).

\bibitem[{\citenamefont{Volchenckov and Nalimov}(1996)}]{VN96}
\bibinfo{author}{\bibfnamefont{D.~Y.} \bibnamefont{Volchenckov}}
  \bibnamefont{and} \bibinfo{author}{\bibfnamefont{M.~Y.}
  \bibnamefont{Nalimov}}, \bibinfo{journal}{Theor. Math. Phys.}
  \textbf{\bibinfo{volume}{106}}, \bibinfo{pages}{307} (\bibinfo{year}{1996}).

\bibitem[{\citenamefont{Antonov}(1999)}]{Ant99}
\bibinfo{author}{\bibfnamefont{N.~V.} \bibnamefont{Antonov}},
  \bibinfo{journal}{Phys. Rev.~E} \textbf{\bibinfo{volume}{60}},
  \bibinfo{pages}{6691} (\bibinfo{year}{1999}).

\bibitem[{\citenamefont{Antonov}(2006)}]{A06}
\bibinfo{author}{\bibfnamefont{N.~V.} \bibnamefont{Antonov}},
  \bibinfo{journal}{J. Phys. A: Math. Gen.} \textbf{\bibinfo{volume}{39}},
  \bibinfo{pages}{7825} (\bibinfo{year}{2006}).

\bibitem[{\citenamefont{Antonov et~al.}(2016)\citenamefont{Antonov,
  Hnati{\v{c}}, Kapustin, Lu{\v{c}}ivjansk{\'{y}}, and
  Mi{\v{z}}i{\v{s}}in}}]{AHKLM16}
\bibinfo{author}{\bibfnamefont{N.~V.} \bibnamefont{Antonov}},
  \bibinfo{author}{\bibfnamefont{M.}~\bibnamefont{Hnati{\v{c}}}},
  \bibinfo{author}{\bibfnamefont{A.~S.} \bibnamefont{Kapustin}},
  \bibinfo{author}{\bibfnamefont{T.}~\bibnamefont{Lu{\v{c}}ivjansk{\'{y}}}},
  \bibnamefont{and}
  \bibinfo{author}{\bibfnamefont{L.}~\bibnamefont{Mi{\v{z}}i{\v{s}}in}},
  \bibinfo{journal}{Phys. Rev.~E} \textbf{\bibinfo{volume}{93}},
  \bibinfo{pages}{012151} (\bibinfo{year}{2016}).

\bibitem[{\citenamefont{Vasil'ev}(1998)}]{Vasiliev_blue}
\bibinfo{author}{\bibfnamefont{A.~N.} \bibnamefont{Vasil'ev}},
  \emph{\bibinfo{title}{{Functional Methods in Quantum Field Theory and
  Statistical Physics}}} (\bibinfo{publisher}{Gordon and Breach Science
  Publishers}, \bibinfo{address}{Amsterdam}, \bibinfo{year}{1998}).

\bibitem[{\citenamefont{Symanzik}(1973)}]{Sym73}
\bibinfo{author}{\bibfnamefont{K.}~\bibnamefont{Symanzik}},
  \bibinfo{journal}{Lett. Nuovo Cimento} \textbf{\bibinfo{volume}{8}},
  \bibinfo{pages}{771} (\bibinfo{year}{1973}).

\bibitem[{\citenamefont{Schloms and Dohm}(1989)}]{Schloms89}
\bibinfo{author}{\bibfnamefont{R.}~\bibnamefont{Schloms}} \bibnamefont{and}
  \bibinfo{author}{\bibfnamefont{V.}~\bibnamefont{Dohm}},
  \bibinfo{journal}{Nucl. Phys. B} \textbf{\bibinfo{volume}{328}},
  \bibinfo{pages}{639} (\bibinfo{year}{1989}).

\bibitem[{\citenamefont{Wolfram\mbox{ }Research}(2012)}]{wolfram}
\bibinfo{author}{\bibnamefont{Wolfram\mbox{ }Research}},
  \emph{\bibinfo{title}{Mathematica, Version 9.0}}
  (\bibinfo{address}{Champaign, Illinois}, \bibinfo{year}{2012}).

\bibitem[{\citenamefont{Gaw{\c{e}}dzki and Vergassola}(2000)}]{gawedzki00}
\bibinfo{author}{\bibfnamefont{K.}~\bibnamefont{Gaw{\c{e}}dzki}}
  \bibnamefont{and}
  \bibinfo{author}{\bibfnamefont{M.}~\bibnamefont{Vergassola}},
  \bibinfo{journal}{Physica D} \textbf{\bibinfo{volume}{138}},
  \bibinfo{pages}{63} (\bibinfo{year}{2000}).

\bibitem[{\citenamefont{Hnatich et~al.}(2001)\citenamefont{Hnatich, Honkonen,
  and Jurcisin}}]{MHD01}
\bibinfo{author}{\bibfnamefont{M.}~\bibnamefont{Hnatich}},
  \bibinfo{author}{\bibfnamefont{J.}~\bibnamefont{Honkonen}}, \bibnamefont{and}
  \bibinfo{author}{\bibfnamefont{M.}~\bibnamefont{Jurcisin}},
  \bibinfo{journal}{Phys. Rev. E} \textbf{\bibinfo{volume}{64}},
  \bibinfo{pages}{056411} (\bibinfo{year}{2001}).

\bibitem[{\citenamefont{Bouchaud and Georges}(1990)}]{Bouchaud}
\bibinfo{author}{\bibfnamefont{J.-P.} \bibnamefont{Bouchaud}} \bibnamefont{and}
  \bibinfo{author}{\bibfnamefont{A.}~\bibnamefont{Georges}},
  \bibinfo{journal}{Phys. Rep.} \textbf{\bibinfo{volume}{195}},
  \bibinfo{pages}{127} (\bibinfo{year}{1990}).

\bibitem[{\citenamefont{Honkonen and Nalimov}(1995)}]{HN95}
\bibinfo{author}{\bibfnamefont{J.}~\bibnamefont{Honkonen}} \bibnamefont{and}
  \bibinfo{author}{\bibfnamefont{M.~Y.} \bibnamefont{Nalimov}},
  \bibinfo{journal}{Z. Phys. B} \textbf{\bibinfo{volume}{99}},
  \bibinfo{pages}{297} (\bibinfo{year}{1995}).

\bibitem[{\citenamefont{Adzhemyan et~al.}(2005)\citenamefont{Adzhemyan,
  Honkonen, Kompaniets, and Vasil'ev}}]{AHK05}
\bibinfo{author}{\bibfnamefont{L.~T.} \bibnamefont{Adzhemyan}},
  \bibinfo{author}{\bibfnamefont{J.}~\bibnamefont{Honkonen}},
  \bibinfo{author}{\bibfnamefont{M.~V.} \bibnamefont{Kompaniets}},
  \bibnamefont{and} \bibinfo{author}{\bibfnamefont{A.~N.}
  \bibnamefont{Vasil'ev}}, \bibinfo{journal}{Phys. Rev.~E}
  \textbf{\bibinfo{volume}{71}}, \bibinfo{pages}{036305}
  (\bibinfo{year}{2005}).

\end{thebibliography}

\end{document}